\documentclass[preprint,times]{aastex62}

\usepackage{color}
\usepackage{natbib}
\usepackage{amsmath}
\usepackage{grffile}    
\tightenlines

\newcommand \bahat      {\hat{\bf a}}
\newcommand \bB         {{\bf B}}

\newcommand \bBhat      {\hat{\bf B}}

\newcommand \bE         {{\bf E}}

\newcommand \bH         {{\bf H}}
\newcommand \bJ         {{\bf J}}

\newcommand \bxhat      {\hat{\bf x}}
\newcommand \byhat      {\hat{\bf y}}
\newcommand \bzhat      {\hat{\bf z}}

\newcommand \beq        {\begin{equation}}
\newcommand \beqa	{\begin{eqnarray}}

\newcommand \cm         {\,{\rm cm}}

\newcommand \ed         {{\rm (ed)}}
\newcommand \eeq	{\end{equation}}
\newcommand \eeqa	{\end{eqnarray}}

\newcommand \eV 	{\,{\rm eV}}
\newcommand \falign     {f_{\rm align}}
\newcommand \fFe        {f_{\rm Fe}} 
\newcommand \fvFe       {f_{\rm sv,Fe}} 
\newcommand \Feinc      {{\rm Fe\,inc}}
\newcommand \gm         {\,{\rm g}}
\newcommand \GHz        {\,{\rm GHz}}
\newcommand \gtsim	{\gtrsim}		 
\newcommand \Ha 	{{\rm H}}

\newcommand \HH	        {{\rm H}_2}

\newcommand \K  	{\,{\rm K}}

\newcommand \Lmin       {L_{\rm min}}

\newcommand \ltsim	{\lesssim}		 

\newcommand \nH         {n_{\rm H}}

\newcommand \NH         {N_{\rm H}}

\newcommand \CDEtwo     {CDE2}
\newcommand \pc  	{\,{\rm pc}}
\newcommand \poro       {{\cal P}}       

\newcommand \Ad        {{\rm Ad}}       
\newcommand \Adhgl     {{\rm Ad,hgl}}   
\newcommand \Astrodust  {Astrodust}     
\newcommand \astrodust  {astrodust}     

\newcommand \xtimes     {{\!\,\times\!\,}}

\newcommand \mm         {\,{\rm mm}}

\newcommand \aeff       {a_{\rm eff}}

\newcommand \Cabs       {C_{\rm abs}}
\newcommand \Cran       {C_{\rm ran}}
\newcommand \Cpol       {C_{\rm pol}}

\newcommand \sil        {{\rm sil}}

\newcommand \deltataupk {\Delta\tau_{9.7\mu{\rm m}}}

\newcommand{\oldtext}[1]{}

\newcommand{\figwidth}{8.0cm}

\countdef\decade=200
\advance\decade by \year
\countdef\hours=201
\advance\hours by \time
\divide\hours by 60
\countdef\mins=202
\advance\mins by \hours
\multiply\mins by 60
\multiply\hours by 100
\countdef\miltime=203
\advance\miltime by \hours
\advance\miltime by \time
\advance\miltime by -\mins
\renewcommand\today{\number\decade.\number\month.\number\day.\number\miltime}

\begin{document}

\title{%
        {\bf 
         The Dielectric Function of ``Astrodust'' and
         Predictions for Polarization in the 3.4$\micron$ and 10$\micron$
         Features
}
	}

\author[0000-0002-0846-936X]{B. T. Draine}
\affiliation{Department of Astrophysical Sciences,
  Princeton University, Princeton, NJ 08544-1001, USA}
\author[0000-0001-7449-4638]{Brandon S. Hensley}
\affiliation{Department of Astrophysical Sciences,
  Princeton University, Princeton, NJ 08544-1001, USA}
\affiliation{Spitzer Fellow}

\correspondingauthor{B. T. Draine}
\email{draine@astro.princeton.edu, bhensley@astro.princeton.edu}

\begin{abstract}
  The dielectric function of interstellar dust material
  is modeled using observations of extinction and polarization in the infrared,
  together with estimates for the mass of interstellar dust.
  The ``astrodust'' material
  is assumed to be a mix of amorphous silicates and
  other materials, including hydrocarbons producing an absorption
  feature at $3.4\micron$.
  The detailed shape of the $10\micron$ polarization profile depends on
  the assumed porosity and grain shape, 
  but the $10\micron$ spectropolarimetric data
  are not yet good enough to clearly favor one shape over another,
  nor to constrain the porosity.
  The expected $3.4\micron$ feature polarization is consistent with existing 
  upper limits,
  provided the $3.4\micron$ absorption is preferentially located in
  grain surface layers; a separate
  population of non-aligned carbonaceous grains is not required.
  We predict the $3.4\micron$ polarization feature to be
  $(\Delta p)_{3.4\mu{\rm m}}/p(10\micron)\approx 0.016$,
  just below current upper limits.
  Polarization by the same grains at submm wavelengths is also calculated.
\end{abstract}

\let\svthefootnote\thefootnote
\let\thefootnote\relax\footnote{\textcopyright 2020.  All rights reserved.}
\let\thefootnote\svthefootnote

\section{Introduction
         \label{sec:intro}}

Calculations of absorption, scattering, and emission by interstellar
dust particles require assumptions concerning the shapes and sizes of the
grains, as well as the dielectric function of the grain material.
We have information on the amount of different
elements in the grains, as well as observations of the extinction,
emission, and polarization properties of the dust at many wavelengths
\citep{Hensley+Draine_2020b};
these provide some spectroscopic clues to the composition,
but at present the detailed composition of the dust remains uncertain.

Some models 
\citep[e.g.][]{Mathis+Rumpl+Nordsieck_1977,
               Draine+Lee_1984,
               Weingartner+Draine_2001a,
             Clayton+Wolff+Sofia+etal_2003,
             Zubko+Dwek+Arendt_2004,
             Draine+Fraisse_2009,
             Compiegne+Verstraete+Jones+etal_2011,
             Jones+Fanciullo+Kohler+etal_2013,
             Siebenmorgen+Voshchinnikov+Bagnulo_2014,
             Kohler+Ysard+Jones_2015,
             Siebenmorgen+Voshchinnikov+Bagnulo+Cox_2017,
             Fanciullo+Guillet+Boulanger+Jones_2017,
             Guillet+Fanciullo+Verstraete+etal_2018} 
have idealized interstellar dust as consisting
of two or more distinct 
populations, generally taken to be silicate-rich and carbon-rich materials.

The present paper considers a new model
for dust in the diffuse interstellar medium
(ISM): the material making up the bulk of the dust is idealized as
a mixture of different constituents.
For brevity, we refer to this mixed material as {\it \astrodust}.

The interstellar dust population
includes ``stardust'' particles that were condensed
in stellar outflows with a variety of compositions, including
silicates, graphite, SiC, Al$_2$O$_3$ and other condensates
\citep{Nittler+Ciesla_2016}.
However, in our view the bulk of the
interstellar dust material was formed in the cold
ISM, by accretion of atoms on grain surfaces 
in the presence of far-UV radiation;
the resulting material is the product of UV photolysis
\citep[see, e.g.,][]{Draine_2009b}.
The grain size distribution evolves as the result of coagulation
in low speed grain-grain collisions, and fragmentation in higher speed
collisions.
In addition, violent events such as blast waves driven by supernovae 
occasionally
create hostile conditions where much of the solid material may be
returned to the gas phase by
sputtering, or by vaporization in high velocity grain-grain collisions.

Because we envision the bulk of the grain material as having been grown
in the ISM, with large grains having been assembled by coagulation, we idealize
grains larger than $\sim$0.02$\micron$ (i.e., the bulk of the interstellar
dust mass) as having a mixed composition that is
independent of grain size.  
In addition to these ``\astrodust'' particles, we envision
the dust population as also including 
one or more populations of nanoparticles that individually have distinct
composition, such as polycyclic aromatic hydrocarbon (PAH) particles,
or nanosilicate particles.
Such single-composition nanoparticles may be the result of fragmentation of
larger grains in grain-grain collisions.

We estimate below that silicates make up about half of the grain mass.
\Astrodust\ is the silicate-bearing grain material, but it
includes other compounds,
including hydrocarbon material with
an absorption feature at 3.4$\micron$.
Astrodust material (including voids,
if present) will be characterized
by an {\it effective} dielectric function $\epsilon_\Ad$.
We will estimate
$\epsilon_\Ad$ by requiring that material with this dielectric function
reproduce
astronomical observations.
We consider different assumptions concerning the
grain shape and porosity.

There have been previous efforts to characterize the
shapes of the silicate-bearing grains.
Starlight polarization at optical wavelengths requires that some of the
grains be both appreciably nonspherical and substantially aligned
\citep[e.g.][]{Kim+Martin_1995b}.
Because polarization is often observed in the $10\micron$ feature
\citep{Smith+Wright+Aitken+etal_2000,Whittet_2011}, 
it is clear that the silicate-bearing grains must be significantly
nonspherical and aligned.
\citet{Draine+Lee_1984} and \citet{Lee+Draine_1985} 
argued that the polarization profile of the 10$\micron$
feature towards the Becklin-Neugebauer (BN) object 
favored oblate grain shapes, and
\citet{Aitken+Smith+Roche_1989} reached the same conclusion.
However, the grains in the diffuse ISM may differ from those
obscuring the BN object.
Some recent studies focusing on submm polarization
\citep[e.g.,][]{Siebenmorgen+Voshchinnikov+Bagnulo+Cox_2017,
                Guillet+Fanciullo+Verstraete+etal_2018}
favored prolate shapes for the grains in the diffuse ISM,
but did not address the 10$\micron$ silicate feature.

\begin{figure}[b]
\begin{center}
\includegraphics[angle=0,width=7.0cm,clip=true,
                 trim=0.5cm 5.0cm 0.5cm 2.5cm]
{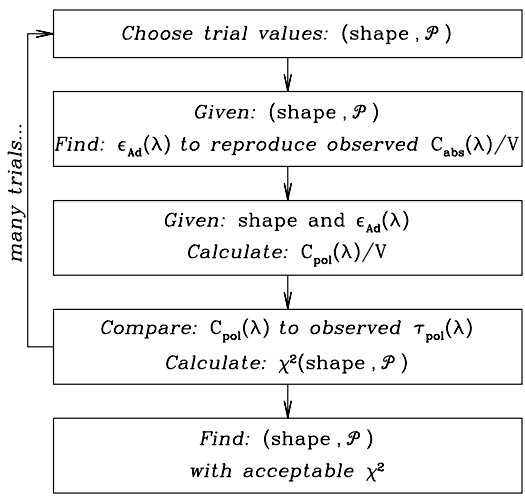}
\caption{\footnotesize\label{fig:flowchart}
Approach used to determine the shape and dielectric function $\epsilon_\Ad$ of
the silicate-bearing ``\astrodust'' grains.
$\poro$ is the porosity
of the \astrodust\ grains (see text).}
\end{center}
\vspace*{-0.3cm}
\end{figure}

The present paper uses the most accurate available observations of 
infrared absorption and polarization by dust, 
summarized in \S\ref{sec:observed absorption}, 
to try to constrain the shape, porosity,
and dielectric function $\epsilon_\Ad$ of the 
silicate-bearing grains in the diffuse ISM.
Our approach (outlined in Figure \ref{fig:flowchart})
consists of testing different hypotheses for
shape and porosity by using the observed interstellar absorption 
(\S\ref{sec:observed absorption})
and estimates of the volume of \astrodust\ material
(see \S\ref{sec:abundance})
to solve for a self-consistent effective dielectric 
function $\epsilon_\Ad(\lambda)$
(see \S\ref{sec:dielectric func}).
Having found $\epsilon_\Ad(\lambda)$, we then calculate the 
10$\micron$ polarization;
comparison of the model polarization to published
spectropolarimetry of the 10$\micron$ silicate feature
(\S\ref{sec:silicate polarization}) favors some shapes over others.
Predicted polarization at other wavelengths 
($3.4\micron$ feature; far-infrared and submm)
is discussed in \S\ref{sec:alignment}.

Our results are discussed in \S\ref{sec:discuss},
and summarized in
\S\ref{sec:summary}.
Various technical points are examined in Appendices \ref{app:mag dipole} --
\ref{app:ferromagnetic}.

\section{Motivation for a One-Component Model}

\begin{figure}[b]
\begin{center}
\includegraphics[angle=0,width=10.0cm,
                 clip=true,trim=0.5cm 5.1cm 0.5cm 2.5cm]
{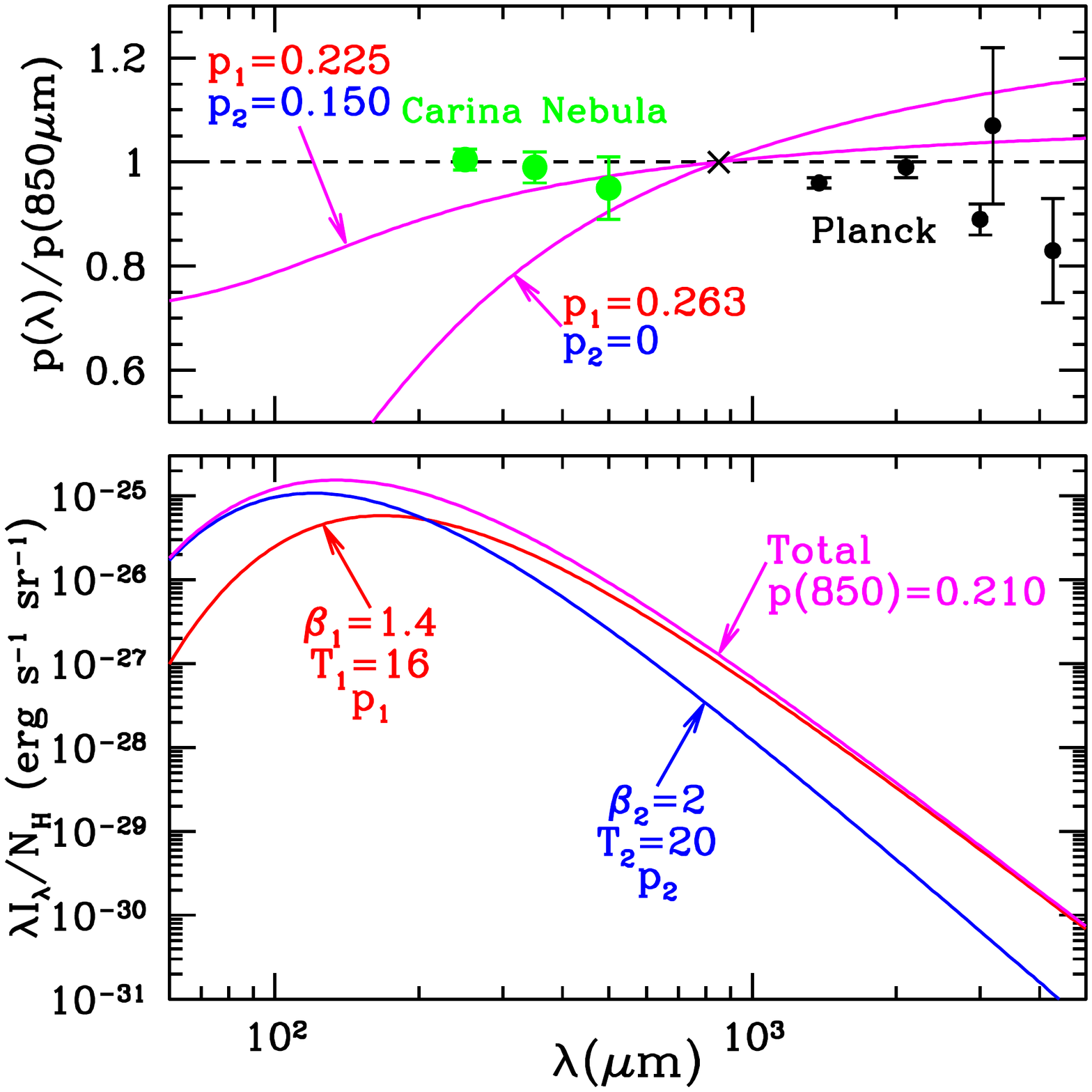}\\
\end{center}
\caption{\footnotesize \label{fig:2 component model}
         Upper panel:
         Symbols: polarization, relative to $p(850\micron)$, measured by 
         {\it Planck} \citep{Planck_int_results_xxii_2015} for the general ISM,
         and
         BLASTPol 
         \citep{Shariff+Ade+Angile+etal_2019}
         for the Carina Nebula.
         Curves: two-component toy model (see text), 
         with two different choices for
         the component polarizations $p_1$ and $p_2$.
         Even with $p_1/p_2=1.5$, the wavelength-dependence of $p(\lambda)$
         appears to be inconsistent with observations.
         Lower panel: emission spectra for the 
         two-component toy model (see text).
         The toy model emission spectrum peaks near $140\micron$, 
         and is broadly consistent with $\lambda\geq 60\micron$
         observations of the diffuse ISM.}
\end{figure}

Many models of interstellar dust have postulated the existence of two
major components: silicate dust and carbonaceous dust, with the two grain types
having similar size distributions, and similar total volumes of material
\citep[e.g.,][]{Mathis+Rumpl+Nordsieck_1977}.  
Such models appeared to be compatible 
with observations of wavelength-dependent 
extinction and starlight polarization,
and seemed to be favored because of nondetection of polarization in the
$3.4\micron$ CH feature on sightlines where the $10\micron$ silicate feature is
appreciably polarized \citep{Chiar+Adamson+Whittet+etal_2006},
which could be explained if for some reason the carbonaceous grains
are not aligned.

At far-infrared and submm wavelengths, 
the fractional polarization $p$ of the thermal emission
from a single grain is expected to be nearly independent of wavelength.
Because the two dust components are likely to be heated to
different temperatures by starlight, 
to have different wavelength-dependences for their
far-infrared (FIR) and submm opacities, and to have
different shapes and degrees of alignment,
two-component models naturally predict that the
fractional polarization of the FIR and submm emission will vary with wavelength
\citep{Draine+Fraisse_2009}.
To illustrate this,
Figure \ref{fig:2 component model} shows the 
polarization fraction $p(\lambda)$ at $\lambda > 60\micron$
for a simple two-component toy model, 
where the two components have slightly different temperatures,  
opacities $\propto \nu^\beta$ with different values of
$\beta$, and polarization fractions $p_1$ and $p_2$.  
The toy model shown has $T_1=16\K$ and $T_2=20\K$,
and $\beta_1=1.4$ and $\beta_2=2$.  
If the two components contribute approximately equally at $200\micron$,
the overall emission for this model approximates the observed 
spectral energy distribution (SED) of the diffuse ISM
\citep[see, e.g.,][]{Hensley+Draine_2020b}.
However, unless the two components each have identical 
fractional polarizations
(i.e., $p_1=p_2$), the overall polarization fraction 
will be frequency dependent.  
The extreme case where only component 1 is polarized 
is illustrated in Figure \ref{fig:2 component model}, resulting in
a total polarization fraction that varies by a factor of $\sim$$1.7$
from $250\micron$ to $3\mm$.
We also show an example with $p_1/p_2=1.5$.
For this example, $p(\lambda)$ varies by a factor of $\sim$$1.13$ from
$250\micron$ to $3\mm$.

{\it Planck} found the fractional polarization to be essentially constant for
$\lambda \geq 850\micron$ \citep{Planck_int_results_xxii_2015}.
BLASTPol
\citep{Ashton+Ade+Angile+etal_2018,Shariff+Ade+Angile+etal_2019}
measured the polarization fraction to be nearly
constant for $250\micron\leq \lambda \leq 850\micron$ in
selected brighter areas (see Figure \ref{fig:2 component model}).
The toy model examples shown in Figure \ref{fig:2 component model}
are inconsistent 
with these observations.
In order to make a two-component model work, one requires that either (1) the
two components have very similar SEDs, 
or (2) the two components have nearly identical fractional polarizations.  
If the two grain types are actually quite
dissimilar (e.g, silicate grains and carbon grains),
one would expect the two components to have different SEDs, {\it and} different
polarized fractions. 

If, however, a {\it single}
grain type dominates the FIR-submm emission, then it is
natural for the polarization fraction to be nearly frequency-independent at
long wavelengths.
This is the type of model considered here.

\section{\label{sec:observed absorption}
         Infrared Absorption by Interstellar Dust
         }

The observed extinction produced by dust in the ISM is
discussed by \citet{Hensley+Draine_2020b}.  At wavelengths $\lambda
\ltsim 30\micron$ the extinction is obtained from the observed
attenuation on sightlines to stars.  In the near-IR ($1-8\micron$) numerous
sightlines have been studied, using observations from the ground and space,
including the IRAC camera on Spitzer Space Telescope and {\it WISE} 
\citep[e.g.,][]{Indebetouw+Mathis+Babler+etal_2005,
                Schlafly+Meisner+Stutz+etal_2016}.
At longer wavelengths, sightlines with very large dust columns are
required to accurately measure the extinction.  The 8--30$\micron$ 
extinction curve that we use here is based on a reanalysis of
archival Spitzer IRS data for the star
Cyg OB2-12 
\citep{Hensley+Draine_2020a}.

\begin{figure}[t]
  \begin{center}
    \includegraphics[width=12.cm,angle=270,clip=true,
                     trim=0.5cm 0.5cm 0.5cm 0.5cm]
                    {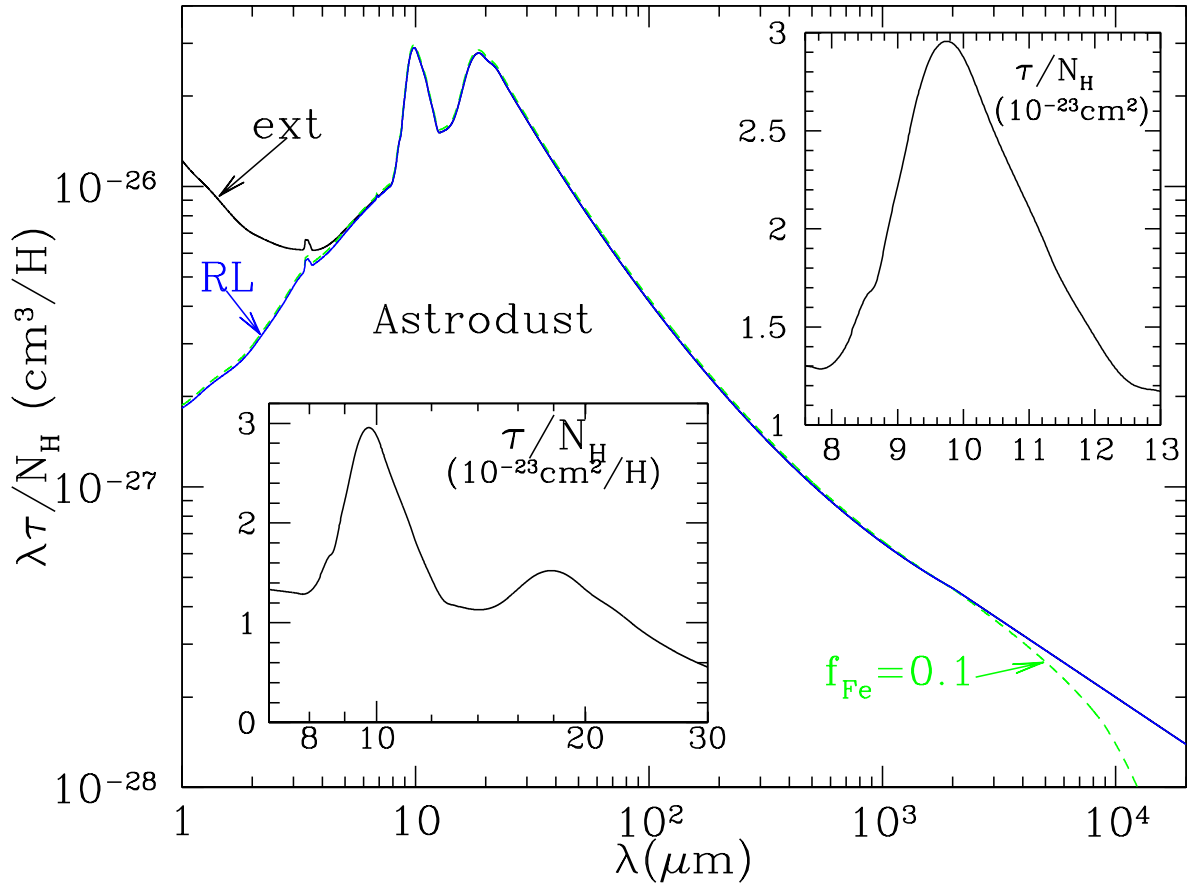}
\caption{\label{fig:obs_ext}\footnotesize
  Solid black curve: $\lambda\tau_{\rm ext}/\NH$ for $\lambda>1\micron$,
  where $\tau_{\rm ext}/\NH$ is the observed extinction cross section
  per H nucleon at intermediate and high galactic latitudes
  at wavelength $\lambda$ \citep[from][]{Hensley+Draine_2020a}.
  Extinction in the PAH features is not included.
  The blue curve (RL) shows our estimate for $\lambda\tau_{\rm abs}/\NH$
 {\it if} the grains were all in the Rayleigh limit.
  The green broken curve shows the reduced level of electric dipole absorption 
  required if there is
  magnetic dipole absorption provided by a fraction 
  $f_{\rm Fe}=0.1$ of the Fe in metallic inclusions.
  Insets show details of the 9.7 and 18$\micron$ silicate features.}
  \end{center}
\end{figure}

At wavelengths $\lambda\gtsim 50\micron$, the
attenuation is very weak and has not been measured directly, 
but the absorption produced by
interstellar dust can be inferred from the observed thermal emission.
The resulting extinction cross
section per H for \astrodust\ in the general diffuse
ISM at intermediate to high galactic latitudes, $\tau_{\rm ext}/\NH$,
is shown in Figure \ref{fig:obs_ext} for $\lambda > 1 \micron$
\citep{Hensley+Draine_2020a,Hensley+Draine_2020b}.

\begin{figure}[t]
\begin{center}
\includegraphics[width=10.0cm,angle=270,clip=true,
                 trim=0.5cm 0.5cm 0.5cm 0.5cm]
                 {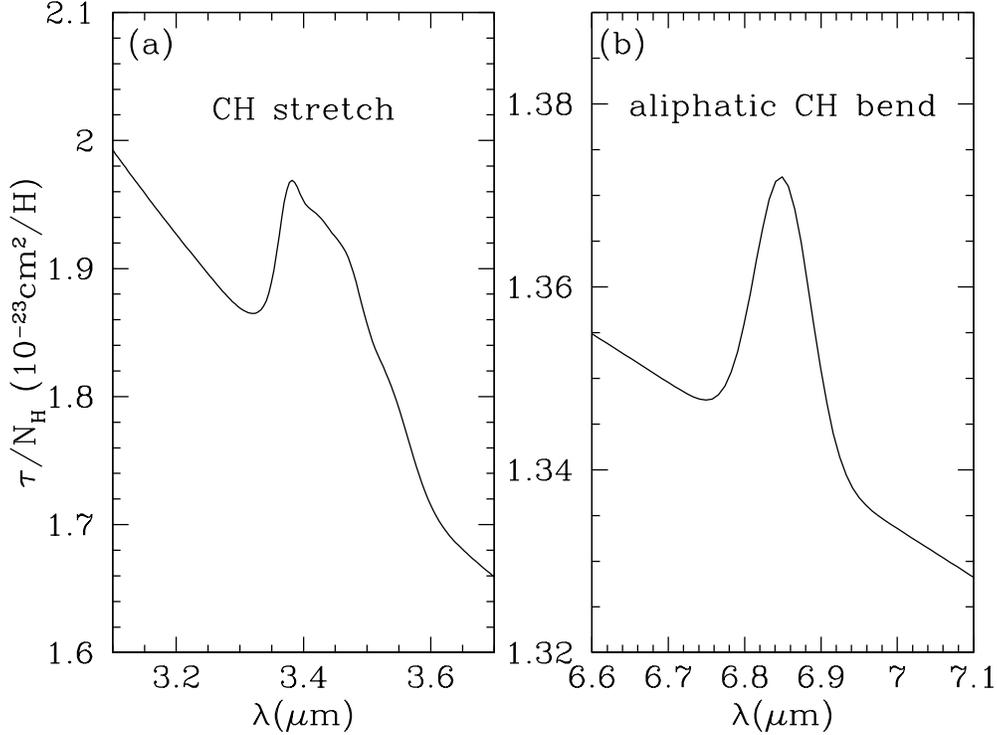}
\caption{\label{fig:CH_features}\footnotesize
   \Astrodust\ hydrocarbon absorption 
   features (see text) at 3.4$\micron$ and $6.85\micron$ based on
   sightline to Cyg OB2-12 \citep{Hensley+Draine_2020a}.
   The $3.29\micron$ aromatic absorption feature has been removed.
   }
\end{center}
\end{figure}

PAH nanoparticles are expected to contribute infrared extinction features
at 3.3, 6.2, and 7.7$\micron$; 
absorption features at these wavelengths are observed
\citep{Schutte+vanderHucht+Whittet+etal_1998,
Chiar+Tielens+Whittet+etal_2000,
Hensley+Draine_2020a} and have been subtracted from the
observed extinction to obtain
the \astrodust-only extinction shown in Figure \ref{fig:obs_ext}.
In addition to the prominent silicate
features at 9.7$\micron$ and 18$\micron$, the astrodust extinction includes
absorption features at $3.4\micron$ and $6.85\micron$ due to aliphatic hydrocarbons; these are shown in Fig.\ \ref{fig:CH_features}.
The interstellar $3.4\micron$ CH absorption feature 
is composed of subfeatures contributed by aromatic, aliphatic, 
and diamond-like hydrocarbons \citep{Chiar+Tielens+Adamson+Ricca_2013,
Hensley+Draine_2020a}.
PAH nanoparticles are expected to
contribute to the ``aromatic CH'' absorption
component at  $3.29\micron$,
and may well be responsible for the bulk of the $3.29\micron$ absorption
feature.
Because we don't know what fraction of the $3.29\micron$ 
extinction feature
to attribute to aromatic material in 
the \astrodust\ component, we obtain a lower bound on the
\astrodust\ absorption by subtracting the entire $3.29\micron$ aromatic
component from the observed extinction, leaving only the contributions from
aliphatic or diamondlike hydrocarbons in the profile shown  
in Fig.\ \ref{fig:CH_features}a.

In this work we seek to estimate the complex dielectric function for
interstellar dust.  When particles are small compared to the
wavelength (i.e., in the Rayleigh limit), extinction is dominated by
absorption; 
the absorption cross section per grain volume 
$C_{\rm abs}/V$ is then directly related to the complex dielectric function
$\epsilon_\Ad(\lambda)$, and is independent of the grain size $a$.

According to models that reproduce
the observed reddening in diffuse regions 
\citep[e.g.,][]{Mathis+Rumpl+Nordsieck_1977,
             Weingartner+Draine_2001a}
the extinction at wavelengths $\lambda \gtsim 3\micron$ in the diffuse ISM
is dominated by {\it absorption}: scattering makes
only a minor contribution.
For example, the $R_V=3.1$ grain 
model of \citet{Weingartner+Draine_2001a} has
albedo $<0.05$ at $\lambda>7\micron$.\footnote{%
   In molecular cloud cores, observations of scattered light
   (``coreshine'') at $3.6\micron$ and $4.5\micron$ provide evidence
   for grain growth, with estimated
   maximum grain radii of $\sim$0.5--0.7$\micron$ for
   4 of the cores studied by \citet{Steinacker+Andersen+Thi+etal_2015}, and
   even larger grains in other cores.
   However, the present study is limited to grains in the diffuse ISM.}
At wavelengths $\lambda \geq 8\micron$ where scattering is minimal,
it may be safely assumed that interstellar grains
are in the Rayleigh limit $a\ll\lambda$.
At shorter wavelengths, we take the ``Rayleigh limit'' 
absorption per H to be a wavelength-dependent fraction of the extinction.
Figure \ref{fig:obs_ext} shows our estimate for
the dust absorption per H nucleon in the Rayleigh limit. 


The complex dielectric function $\epsilon_\Ad(\omega)$ characterizes the
response of the grain material
to applied electric fields, and therefore does not include
any {\it magnetic} absorption.
If a fraction $\fFe$
of the iron is present in the form of metallic Fe inclusions,
these will provide strong magnetic absorption at frequencies
$\nu < 100\GHz$, or wavelengths $\lambda > 3\mm$
\citep{Draine+Hensley_2013}, and will contribute to the
thermal emission at these frequencies.
If the metallic Fe fraction $\fFe>0$, we estimate the magnetic absorption
as described in Appendix \ref{app:mag dipole}.
The magnetic dipole contribution to the absorption is then subtracted from
the observed opacity to yield the {\it electric} contribution to the
absorption.
Figure \ref{fig:obs_ext} shows our estimate for the electric dipole
absorption at long wavelengths for $\fFe=0$ and 0.1.
For $\fFe\ltsim 0.2$,
the magnetic contribution to absorption is significant only at very 
long wavelengths ($\lambda \gtsim 5\mm$, $\nu \ltsim 60 \GHz$).

\section{\label{sec:abundance}
         Abundance of Materials in Interstellar Grains}

In diffuse regions that have been probed by ultraviolet spectroscopy, we can
estimate the total grain mass by summing up the atoms and ions observed in the
gas phase and subtracting this sum
from our best estimate for the total elemental
abundances in the diffuse ISM.
With reasonable assumptions for the mass densities of the
materials containing the missing atoms, 
we can then estimate the volume of solid material. 


\citet{Jenkins_2009} studied variations in gas-phase abundances,
finding that the gas-phase abundances
in a single cloud can be predicted using a single ``depletion parameter''
$F_\star$
that characterizes the overall level of depletion
(sequestration of elements into solid grains) in that cloud.
In Jenkins's sample, 
the median sightline had $F_\star\approx0.5$, which we take
to be representative of diffuse \ion{H}{1}.\footnote{
   A representative density for the diffuse ISM in the solar neighborhood
   is
   $\NH/h\approx 0.5\cm^{-3}$, where $\NH\approx3\times10^{20}\cm^{-2}$ is the
   half-thickness and $h\approx 200\pc$ is the scale height.
   Jenkins found $F_\star\approx0.5$
   for sightlines with $\langle \nH\rangle = 0.3\cm^{-3}$.
   }
%
\begin{table}[t]
{\footnotesize
\begin{center}
\caption{\label{tab:abund}
         Elemental Abundances in Gas and Dust (ppm)
         \citep{Hensley+Draine_2020b}}
\begin{tabular}{|ccc|c|}
\hline
$X$ & $(X/\Ha)_{\rm ISM}$ & $(X/\Ha)_{\rm gas}$ & $(X/\Ha)_{\rm dust}$ \\
\hline
C & 324         & 198     & 126     \\
O & 682         & 434     & 248     \\
Mg&  52.9       & 7.1     &  46     \\
Al&  3.48       & 0.07    &   3.4   \\
Si&  44.6       & 6.6     &  38.0   \\
S &  17.2       & 9.6     &   7.6   \\
Ca&  3.25       & 0.07    &   3.2   \\
Fe&  43.7       & 0.9     &  43     \\
Ni&  2.09       & 0.04    &   2.1   \\
\hline
\end{tabular}
\end{center}
}
\end{table}
\begin{table}[ht]
{\footnotesize
\begin{center}
\caption{\label{tab:composition} Nominal Composition of \astrodust\ Grains}

\begin{tabular}{|l|c|c|}
\hline
\multicolumn{1}{|r|}{$\fFe=$}        & 0   & 0.10 \\
\hline
\multicolumn{1}{|c}{species} & \multicolumn{2}{c|}{(ppm relative to H)}\\
\hline
Mg$_{1.3}$(Fe,Ni)$_{0.3}$SiO$_{3.6}$ 
($\rho=3.41\gm\cm^{-3}$)             & 35.4 & 35.4 \\
(Fe,Ni) metal 
($\rho=7.9\gm\cm^{-3}$)              &  0   &  4.5 \\
(Fe,Ni)$_3$O$_4$ 
($\rho=5.15\gm\cm^{-3}$)             &  8.9 &  7.4 \\
(Fe,Ni)S
($\rho=4.84\gm\cm^{-3}$)             &  7.6 &  7.6 \\
CaCO$_3$
($\rho=2.71\gm\cm^{-3}$)             &  3.2 &  3.2 \\
Al$_2$O$_3$
($\rho=4.02\gm\cm^{-3}$)             &  1.7 &  1.7 \\
SiO$_2$
($\rho=2.20\gm\cm^{-3}$)             &  2.8 &  2.8 \\
C (in hydrocarbons, $\rho\approx2\gm\cm^{-3}$) & 83.  & 83. \\
C in PAH nanoparticles               & 40.  & 40.  \\
\hline
O unaccounted for                    & 66.  & 70. \\
\hline
\multicolumn{1}{|c}{volume estimates} & & \\
\hline
$V_\sil [10^{-27}\cm^3\Ha^{-1}]$     & 2.31 & 2.31 \\
$V_{\rm mix} [10^{-27}\cm^3\Ha^{-1}$ & 1.29 & 1.18 \\
$V_{\rm car} [10^{-27}\cm^3\Ha^{-1}]$ & 0.86& 0.86 \\
$V_{\rm Fe} [10^{-27}\cm^3\Ha^{-1}]$ &  0.  & 0.053 \\
\hline
$V_\Ad(1-\poro)\,\,[10^{-27}\cm^3\Ha^{-1}]$ 
                                     & 4.46 & 4.40 \\
$\rho_\Ad/(1-\poro)\,\,[\gm\cm^{-3}]$& 3.42 & 3.43 \\
\hline
\multicolumn{1}{|c}{mass estimates} & & \\
\hline
Astrodust: & &  \\
\hspace*{3em}silicate mass/H mass    & 0.0047 & 0.0047 \\
\hspace*{3em}(Fe,Fe$_3$O$_4$,FeS)/H mass & 0.0027 & 0.0026 \\
\hspace*{3em}carbon mass/H mass      & 0.0010 & 0.0010 \\
\hspace*{3em}(Al$_2$O$_3$,CaCO$_3$,SiO$_2$)/H mass & 0.0007 & 0.0007 \\
\hspace*{3em}total astrodust mass/H mass & 0.0092 & 0.0091 \\ 
\hline
PAH mass/H mass                      & 0.0005 & 0.0005 \\ 
Total grain mass/H mass              & 0.0097 & 0.0096 \\
\hline
\end{tabular}\\
\end{center}
}
\end{table}
Estimates for the gas-phase
abundances $(X/\Ha)_{\rm gas}$
are given in Table \ref{tab:abund} for 
$X=$ C, O, Mg, Si, S, Al, Ca, Fe, and Ni in diffuse regions with 
$F_\star=0.5$.\footnote{%
   \citet{Jenkins_2009} did not discuss the depletion of Al and Ca.
   We take the depletions of Al and Ca to be similar to
   Fe, which slightly underestimates the solid-phase abundances as
   Al and Ca are usually more strongly depleted than Fe
   \citep[see, e.g.,][]{Welty+Hobbs+Lauroesch+etal_1999}.}
Table \ref{tab:abund} also lists 
interstellar
abundances $(X/\Ha)_{\rm ISM}$ recommended by
\citet{Hensley+Draine_2020b}; using these values
we obtain $(X/\Ha)_{\rm dust}=(X/\Ha)_{\rm ISM}-(X/\Ha)_{\rm gas}$
recommended by \citet{Hensley+Draine_2020b}, shown again
in Table \ref{tab:abund}.

We model the interstellar dust population as separate populations of
(1) astrodust grains, containing the bulk of the dust mass; 
and (2) $a\ltsim 0.01\micron$ nanoparticles, including PAHs.
We assume the \astrodust\ material to be a mixture of different
components: 
(1) silicates, 
(2) a mixture of other compounds of Fe, Si, S, Al, Ca, etc.,
(3) carbonaceous material, and, possibly,
(4) metallic Fe-Ni inclusions.

Let $V_\Ad$ be the \astrodust\ volume/H.
If \astrodust\ grains include a volume fraction $\poro$ of vacuum,
then
\beq
(1-\poro) V_\Ad = V_{\rm sil}  + V_{\rm mix} + V_{\rm car} 
+ V_\Feinc
~,
\eeq
where $V_{\rm sil}$, $V_{\rm mix}$, $V_{\rm car}$, and $V_\Feinc$, are, 
respectively,
the solid volume per H of the silicate material, 
other non-silicate materials,
carbonaceous material, and
the metallic Fe-Ni inclusions.\footnote{%
   Given its chemical similarity to Fe, we take Ni to be
   distributed with the Fe in the ratio Ni:Fe $\approx$ 1:20.}

\subsection{Mg and Si}

The elemental composition of the interstellar silicate material 
remains uncertain,
but evidence points to a Mg-rich composition intermediate between
pyroxenes and olivines.
Based on the profile of the $10\micron$ silicate feature,
\citet{Min+Waters+deKoter+etal_2007} favored a model with
overall silicate composition Mg$_{1.32}$Fe$_{0.10}$SiO$_{3.45}$.
Studies of the extinction profiles of the 10 and 18$\micron$ features
toward $\zeta$Oph led \citet{Poteet+Whittet+Draine_2015} to
favor a silicate mixture with
overall composition Mg$_{1.48}$Fe$_{0.32}$SiO$_{3.79}$.
\citet{Fogerty+Forrest+Watson+etal_2016} fit the silicate profile
toward Cyg OB2-12
with a silicate mixture having overall composition
Mg$_{1.37}$Fe$_{0.18}$Ca$_{0.002}$SiO$_{3.55}$.
We adopt a nominal silicate composition
Mg$_{1.3}$(Fe,Ni)$_{0.3}$SiO$_{3.6}$, 
for which we estimate a density 
$\rho=3.41\gm\cm^{-3}$.\footnote{%
   Estimated from the densities of 
   enstatite MgSiO$_3$ ($3.19\gm\cm^{-3}$),
   ferrosilite FeSiO$_3$ ($4.00\gm\cm^{-3}$),
   forsterite Mg$_2$SiO$_4$ ($3.21\gm\cm^{-3}$),
   and fayalite Fe$_2$SiO$_4$ ($4.39\gm\cm^{-3}$),
   assuming 32.5\% of the Si to be in MgSiO$_3$,
   7.5\% in FeSiO$_3$,
   48.8\% in Mg$_2$SiO$_4$,
   and 11.3\% in Fe$_2$SiO$_4$.}






%

For the adopted nominal composition Mg$_{1.3}$(Fe,Ni)$_{0.3}$SiO$_{3.6}$
and the abundances $(X/\Ha)_{\rm dust}$ in Table \ref{tab:abund},
Mg is the limiting constituent.
The interstellar $9.7\micron$ silicate absorption feature is strong; 
we will assume 100\% of the Mg in grains to be in silicates.
We estimate the amorphous silicate volume to be
\beqa
V_{\rm sil} &=&2.32\xtimes10^{-27}\cm^3\Ha^{-1}
~.
\eeqa

\subsection{Fe}

The fraction $\fFe$ of the solid-phase Fe that is in metallic form is
difficult to constrain directly, because metallic Fe lacks spectral features
in the IR.  However, metallic Fe is ferromagnetic, and would generate thermal
magnetic dipole emission with unusual spectral and polarization
characteristics \citep{Draine+Hensley_2013}.  We treat $\fFe$ as an unknown
parameter, but we consider it likely to be small in the local ISM, 
$\fFe\ltsim 0.1$.

For the assumed silicate composition
Mg$_{1.3}$(Fe,Ni)$_{0.3}$SiO$_{3.6}$, 
$\sim$25\% of
the (Fe,Ni) is in the silicate material; thus
$\fFe\ltsim0.75$.
The volume contributed by Fe-Ni 
metallic inclusions (density $\rho=7.9\gm\cm^{-3}$) is
\beq
V_\Feinc = 5.3\fFe\xtimes10^{-28}\cm^3\Ha^{-1}
~~.
\eeq

Candidate materials for the remaining Fe include
oxides (FeO, Fe$_3$O$_4$, Fe$_2$O$_3$, ...),
carbides (Fe$_3$C, ...), and
sulfur compounds (FeS, FeS$_2$, FeSO$_4$, ...).
Nondetection of absorption features at 22$\micron$ and 16$\micron$ imposes
upper limits on the abundances of FeO and Fe$_3$O$_4$ toward
$\zeta$Oph \citep{Poteet+Whittet+Draine_2015} and
Cyg OB2-12 \citep{Hensley+Draine_2020a}
but still allows a
significant fraction of the Fe to be in each of these species.
At this time the chemical form of $\sim$75\%
of the Fe is unknown.

\subsection{Carbon}

About 126 ppm of C is missing from gas
with $F_\star=0.5$ 
(see Table \ref{tab:abund}).  
We estimate that $\sim$40 ppm of C is contained in PAH nanoparticles that
account for the observed PAH emission features, as well as the
prominent 2175\AA\ extinction feature.\footnote{%
   For the average Drude profile from \citet{Fitzpatrick+Massa_1986} and
   $\NH/E(B-V)=8.8\times10^{21}\cm^{-2}{\rm\,mag}^{-1}$
   \citep{Lenz+Hensley+Dore_2017}, the carrier of the 2175\AA\ feature
   has an oscillator strength per H $n_X f_X/\nH = 6.1\times10^{-6}$ 
   \citep[see][]{Draine_1989a}.
   For an oscillator strength per C atom $f_{\rm C}=0.16$ (the value for small
   graphite spheres), this corresponds to $n_{\rm C}/\nH\approx40\,$ppm.
   }
We assume the remainder of the solid-phase carbon to be in the astrodust, 
with 3\,ppm in CaCO$_3$ and the remainder in
hydrocarbons .
Assuming the hydrocarbon material to have C:H $\sim$ 2:1, and C mass
density $\rho_{\rm C}=2\gm\cm^{-3}$, the hydrocarbon 
volume in the silicate-bearing grains is 
\beqa
V_{\rm hc}&\,=\,& 8.3\times10^{-28}\cm^3\Ha^{-1}
~~.
\eeqa

\subsection{Other}

As noted above, there may be a significant amount of
Fe that is not in silicates and not metallic Fe.
S, Al, and Ca are also present in the grains, although with abundances
small compared to Mg, Si, and Fe (see Table \ref{tab:abund}).
To estimate the volume $V_{\rm mix}$ of non-silicate, non-carbonaceous,
material (not including metallic Fe),
we suppose that
the small fraction ($\sim$10\%) of the 
Si that is not in silicates is in SiO$_2$ 
($\rho=2.2\gm\cm^{-3}$),\footnote{SiC
   could also contain some of the Si that is not in silicates.
   Based on nondetection of the SiC 11.3$\micron$ feature toward WR\,98a
   and WR\,112,
   \citet{Chiar+Tielens_2006} obtained an upper limit of 4\% on the fraction
   of Si atoms that are incorporated into SiC, although
   \citet{Min+Waters+deKoter+etal_2007} found 7\% of the Si atoms to
   be in SiC for their best-fitting model.
   }
the Ca is in CaCO$_3$ ($\rho=2.71\gm\cm^{-3}$), 
the Al is in Al$_2$O$_3$ ($\rho=4.0\gm\cm^{-3}$),
the S is in (Fe,Ni)S ($\rho=4.8\gm\cm^{-3}$)
and the remaining (Fe,Ni) atoms are in a mixture of oxides
(Fe$_2$O$_3$, Fe$_3$O$_4$, FeO) with overall ratio O:(Fe,Ni)::4:3
and density as for magnetite Fe$_3$O$_4$ ($\rho=5.2\gm\cm^{-3}$) .
This mixture contributes a volume per H
\beqa
V_{\rm mix} &\,=\,& 1.29(1-0.78\fFe)\xtimes10^{-27}\cm^3\Ha^{-1}
~.
\eeqa
The \astrodust\ volume is
\beqa \label{eq:Vad F*=0.5}
V_\Ad &\,=\,& 
\frac{4.46\xtimes10^{-27}\cm^3\Ha^{-1}}{(1-\poro)} 
\times \left(1-0.22\fFe\right)
~.
\eeqa
The fraction of the solid volume contributed by 
the metallic Fe nanoparticles is
\beq \label{eq:fvfe}
\fvFe \equiv \frac{V_\Feinc}{(1-\poro)V_\Ad}
\approx
\frac{0.12\fFe}{1-0.22\fFe}
~.
\eeq
For $\fFe\ltsim0.1$, the volume fraction of metallic Fe is small,
$\fvFe\ltsim 0.01$.
The fraction of the solid volume contributed by the
silicate material itself is
\beq
\frac{V_{\rm sil}}{V_{\rm sil}+ V_{\rm mix}+V_{\rm car} +V_\Feinc}
\approx
\frac{0.52}{1-0.22\fFe}
~.
\eeq

\subsection{The Oxygen Problem}

\citet{Jenkins_2009} pointed out that in diffuse regions with average
depletions, there does not
appear to be a plausible candidate material to incorporate the oxygen that
is missing from the gas \citep{Whittet_2010}.
In diffuse regions, H$_2$O ice is not detected, and cannot account for the
missing oxygen.
The adopted \astrodust\ composition accounts for only
\beqa
\left(\frac{\rm O}{\rm H}\right)_{\rm dust}
&\,\approx\,& (183-54\fFe){\rm ppm} 
~,
\eeqa
Comparing to Table \ref{tab:abund}, we see that $\sim$66\,ppm of oxygen
-- 10\% of the total -- 
appears to be unaccounted for in regions where $F_\star=0.5$.
Upper limits on the $3.1\micron$ feature imply that the missing oxygen
cannot be attributed to H$_2$O ice in submicron grains
\citep{Poteet+Whittet+Draine_2015}.
The ``missing oxygen'' remains unexplained.

\subsection{Dust Volume per H}

Table \ref{tab:composition} lists the contributions of each of the materials
to the grain volume.
The density of the \astrodust\ mixture is 
$\rho_\Ad=3.42(1-\poro)\gm\cm^{-3}$ (for $\fFe=0$).  
With the abundances in Table \ref{tab:abund}, the dust/H mass ratio
is $0.0097$.

\subsection{Elemental Abundances in the Diffuse ISM}

Many discussions of dust abundances in the ISM implicitly assume that
the local ISM is vertically well-mixed, 
with the same overall elemental abundances 
both near the midplane as well as
in the more diffuse ISM above
and below the plane.
Because elements such as Mg, Si, and Fe are heavily depleted even in diffuse
regions characterized by depletion parameter $F_\star\approx0.5$
(see Table \ref{tab:abund}),
this would lead to the expectation that the dust/gas ratio would
be nearly the same in all diffuse H\,I and H$_2$ clouds, because the 
additional depletion taking place in dense regions adds very little
Mg, Si, and Fe to the dust.

Using H Lyman\,$\alpha$ and $\HH$ Lyman and Werner band lines
to measure $\NH=N(\Ha)+2N(\HH)$
on sightlines to O and B stars,
\citet{Bohlin+Savage+Drake_1978} found
$\langle E(B-V)/\NH\rangle = 1.7\times10^{-22}\cm^2\Ha^{-1}$
and
\citet{Diplas+Savage_1994} obtained
$\langle E(B-V))/\NH\rangle = (2.03\pm0.12)\times10^{-22}\cm^2\Ha^{-1}$.
For many years these were taken to be canonical values for the
local ISM.

However, more recent studies of
high latitude H\,I gas have found significantly less reddening per H:
$\langle E(B-V)/\NH\rangle \approx (1.12\pm0.1)\times10^{-22}\cm^2\Ha^{-1}$
\citep{Lenz+Hensley+Dore_2017}
and $(1.1\pm0.2)\times10^{-22}\cm^2\Ha^{-1}$
\citep{Nguyen+Dawson+Miville-Deschenes+etal_2018}.
We take this to indicate that the abundances relative to H of the
dust-forming elements 
vary, with enhanced abundances near the midplane.  Because gravity causes
dust grains to drift systematically toward the midplane, an enhancement in
the abundances of the dust-forming elements is anticipated, although the
expected magnitude of the enhancement is highly uncertain, 
depending on the competition
between gravitationally-driven settling versus mixing by
vertical motions, including turbulence.

Our knowledge of the abundances of the dust-forming elements derives from
measurements of elemental abundances in stellar atmospheres, which
reflect the abundances of dust and gas in the gas cloud where the star formed.
Because star formation occurs preferentially in dense regions near the
midplane, the abundances in a stellar atmospheres are, effectively, the
mid-plane abundances in the ISM at the time the star formed.
The dust solid volume per H 
$V_{\rm Ad}(1-\poro)=4.46 \times 10^{-27}\cm^3\Ha^{-1}$
in Table \ref{tab:composition} is based on measured abundances in stars,
and therefore should give the dust volume per H in midplane regions.

Our estimate of the far-infrared opacity per H is based on observations of
\ion{H}{1}-correlated FIR and submm emission at 
intermediate and high galactic latitudes.
As discussed above,
the dust-to-gas ratio in this gas is expected to
be lower than in the mid-plane regions.

Let $V_{\rm Ad,hgl}$ be the volume of \astrodust\ grains per H nucleon
in diffuse \ion{H}{1} at intermediate or high galactic latitudes
(i.e., away from the midplane).
With $E(B-V)/\NH\approx 1.1\times10^{-22}\cm^2\Ha^{-1}$ 
in high-latitude gas only $\sim$$2/3$ of 
$E(B-V)/\NH \approx 1.9\times 10^{-22}\cm^2\Ha^{-1}$ near the mid-plane,
we adopt
\beq \label{eq:Vsbg for DISM}
V_{\Adhgl}
=
\frac{3.0\xtimes10^{-27}}{(1-\poro)}\left(1-0.22\fFe\right)\cm^3\,\Ha^{-1}
~~,
\eeq
i.e., about $2/3$ of the value given in Table \ref{tab:composition}.

\renewcommand{\figwidth}{9.0cm}
\begin{figure}[ht]
\begin{center}
\includegraphics[width=\figwidth,angle=270,clip=true,
                 trim=0.5cm 0.5cm 0.5cm 0.5cm]
{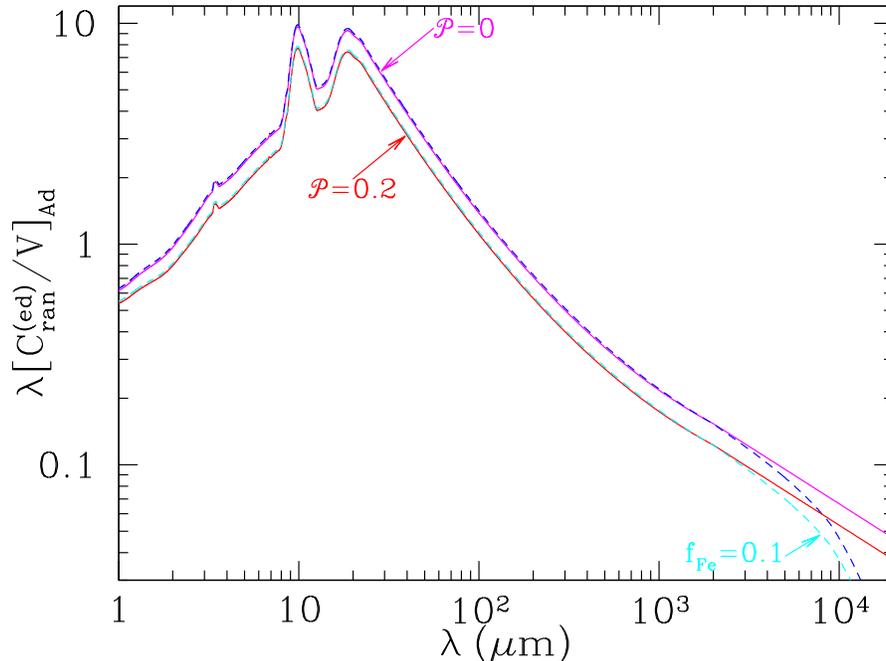}
\vspace*{-0.3cm}
\caption{\label{fig:cabs}\footnotesize
         Assumed $\lambda[\Cran^\ed/V]_{\rm Ad}$, where 
         $[\Cran^\ed/V]_{\rm Ad}$ is the absorption cross section 
         per unit grain volume for randomly-oriented \astrodust\
         grains in the electric dipole limit,
         for porosity $\poro=0$, and 0.2
         and $f_{\rm Fe}=0$ and 0.1 .
         }
\end{center}
\vspace*{-0.3cm}
\end{figure}

\subsection{Astrodust Absorption Per Unit Grain Volume vs.\ $\lambda$}

Except for a small contribution to the mid-IR extinction 
from PAH nanoparticles, we require that \astrodust\ grains reproduce the
entire observed interstellar extinction
for $\lambda > 1\micron$.
Because the porosity of \astrodust\ is not known, 
we consider a range of porosities $0\leq\poro\leq0.9$.

The \astrodust\ cross section per unit volume
\beq
(C/V)_\Ad=\frac{(\tau/\NH)_{\rm hgl}}{V_{\rm Ad,hgl}}
~~.
\eeq
Figure \ref{fig:cabs} shows our estimate for $[\Cabs^\ed/V]_\Ad$,
the absorption per volume in the electric dipole limit, 
at $\lambda\geq1\micron$,
for two values of $\poro$
and two values of $\fFe$.
Note that varying 
$\fFe$ from 0 to 0.1 only changes $\Cabs^\ed/V$ noticeably for 
$\lambda \gtsim 5\mm$ ($\nu\ltsim60\GHz$).


\section{\label{sec:dielectric func}
         Effective Dielectric Function}

We seek a self-consistent {\it effective} complex dielectric function 
$\epsilon_\Ad(\lambda)$
for \astrodust\ that satisfies the
observational constraints.
This effective dielectric function $\epsilon_\Ad(\lambda)$ characterizes 
the overall response of the grain interior, including the silicate
material, vacuum, and other materials present within an assumed 
spheroidal or ellipsoidal
surface.
The derived $\epsilon_\Ad(\lambda)$
depends on
three separate assumptions:
\begin{enumerate}
\item The grain shape (we consider
either spheroids with specified axis ratio $b/a$,
or certain continuous distributions of ellipsoids);
\item $\poro=$ fraction of the \astrodust\ grain volume
that is occupied by vacuum;
\item $\fFe=$ fraction of the interstellar Fe that is
  in the form of metallic Fe inclusions.
\end{enumerate} 
The derived $\epsilon_\Ad(\lambda)$ is intended
to describe the material in the \astrodust\ grains
at wavelengths
from microwaves to X-rays.
We separate the dielectric function into two components:
\beq
\epsilon_\Ad -1 =  (\epsilon_\Ad^{\rm ox}-1) + (\epsilon_\Ad^{\rm ir}-1)
~~,
\eeq
where the ``optical through X-ray'' component
${\rm Im}(\epsilon_\Ad^{\rm ox})$ accounts for
absorption at $\lambda < 1\micron$, and
${\rm Im}(\epsilon_\Ad^{\rm ir})$ describes absorption at 
$\lambda \geq 1\micron$.
We ``solve'' for $\epsilon_\Ad^{\rm ir}$
by adjusting it to comply with observational constraints
on absorption at wavelengths $\lambda > 1\micron$.

The procedure for obtaining ${\rm Im}\left(\epsilon_\Ad^{\rm ox}\right)$
is described in Appendix \ref{app:optical-uv}.
In brief, we obtain the ``optical through X-ray'' dielectric function 
$\epsilon_{\rm mat}^{\rm ox}$ for
the \astrodust\ material, with no voids and no Fe inclusions,
using general considerations for how 
absorptive \astrodust\ material is thought to
be at wavelengths $\lambda<1\micron$.  This dielectric function provides a
level of absorption at optical wavelengths that appears to be consistent with
the observed absorption of starlight by interstellar dust,
and includes a rapid rise in the
absorption shortward of $\sim$$0.15\micron$ 
due to the onset of interband absorption.
The adopted ${\rm Im}\left(\epsilon_\Ad^{\rm ox}\right)$ is consistent with
oscillator strength sum rules for the assumed dust elemental composition.
For each adopted $\poro$ and
$\fFe$, we then use effective medium theory 
to obtain $\epsilon_\Ad^{\rm ox}$ from
$\epsilon_{\rm mat}^{\rm ox}$.

To model the infrared absorption, we take
\beq \label{eq:oscillator model}
\epsilon_\Ad^{\rm ir}(\omega) -1 = 
\sum_{k=1}^N 
\frac{S_k}{\left[
1-\left(
\omega/\omega_{k}
\right)^2
-i\gamma_k
\left(\omega/\omega_{k}\right)
\right]}
~~~,
\eeq
using
$N=3000$ damped Lorentz oscillators, with
resonant frequencies $\omega_k$ distributed 
between $\omega_1=2\pi c/1\micron$ and $\omega_N=2\pi c/5\cm$,
and suitably chosen fractional widths $\gamma_k$
\citep[see][]{Draine+Hensley_2020a}.
The $\omega_k$ are uniformly-distributed in $\ln(\omega)$ between
$\lambda_1=1\micron$ and $\lambda_{2000}=100\micron$, with a smooth
transition to wider spacing in $\ln(\omega)$ at longer wavelengths.
The broadening parameters are taken to be
\beqa \label{eq:gamma_k}
\gamma_k &=& C\left(\frac{\omega_{j-1}}{\omega_{j}}-1\right) ~~~,~~~
j=\max(2,k)
~~~.
\eeqa
For $C=10$
the overlap between resonances yields
a dielectric function that is sufficiently smooth for our purposes,
with individual oscillators having sufficiently narrow resonances
to be able to represent a dielectric function that varies relatively
rapidly in the vicinity of the $9.7\micron$ Si-O profile.

With the frequencies and widths of the resonances specified,
the strengths $S_k$ remain to be determined.
Individual $S_k$ are allowed to be negative -- it is only necessary that
${\rm Im}(\epsilon_\Ad)>0$ after summing over the contributions from all of the
oscillators plus ${\rm Im}(\epsilon^{\rm ox})$.
The $S_k$ are obtained by requiring that the resulting dielectric
function $\epsilon_\Ad$ reproduce the $[C_{\rm ran}^\ed/V]_\Ad$
consistent with the observed extinction and emission -- see
Figure \ref{fig:cabs}.

We consider spheroidal grains 
with a wide range of possible
axis ratios, from prolate shapes with $b/a$ as small as $1/3$, to
oblate shapes with $b/a=3$.
We also consider ellipsoidal grains
with two different continuous distributions of ellipsoidal shapes:
\begin{enumerate}
\item The shape distribution proposed by \citet{Ossenkopf+Henning+Mathis_1992},
referred to as
``CDE2'' by \citet{Fabian+Henning+Jager+etal_2001}.
\item The
``externally-restricted CDE'' (ERCDE) distribution 
\citep{Zubko+Mennella+Colangeli+Bussoletti_1996} with
$L_{\rm min}=0.05$, i.e., restricted to
shape factors $0.05\leq L\leq 0.9$.
\end{enumerate}
Properties of these two shape distributions are discussed in
\citet{Draine+Hensley_2020a}.
Although the continuous distribution of ellipsoids (CDE) 
discussed by \citet{Bohren+Huffman_1983} is often used
\citep[e.g.,][]{
Rouleau+Martin_1991,
Alexander+Ferguson_1994,
Min+Hovenier+deKoter_2003,
Min+Hovenier+Waters+deKoter_2008}
we do not employ it here, because we do not consider it to be realistic.
The Bohren-Huffman CDE
distribution includes
a large fraction
of extremely elongated or flattened shapes -- fully 10\% of
Bohren-Huffman CDE ellipsoids
have axis ratios $a_3/a_1>20$, and 1\% have $a_3/a_1>100$
\citep{Draine+Hensley_2020a}.
Such extreme elongations seem to us to be unlikely, therefore the
present study considers only the ERCDE and CDE2 shape distributions.

\begin{figure}[t]
\begin{center}
\includegraphics[width=8.0cm,angle=0,clip=true,
                 trim=0.5cm 0.5cm 0.5cm 0.5cm]
{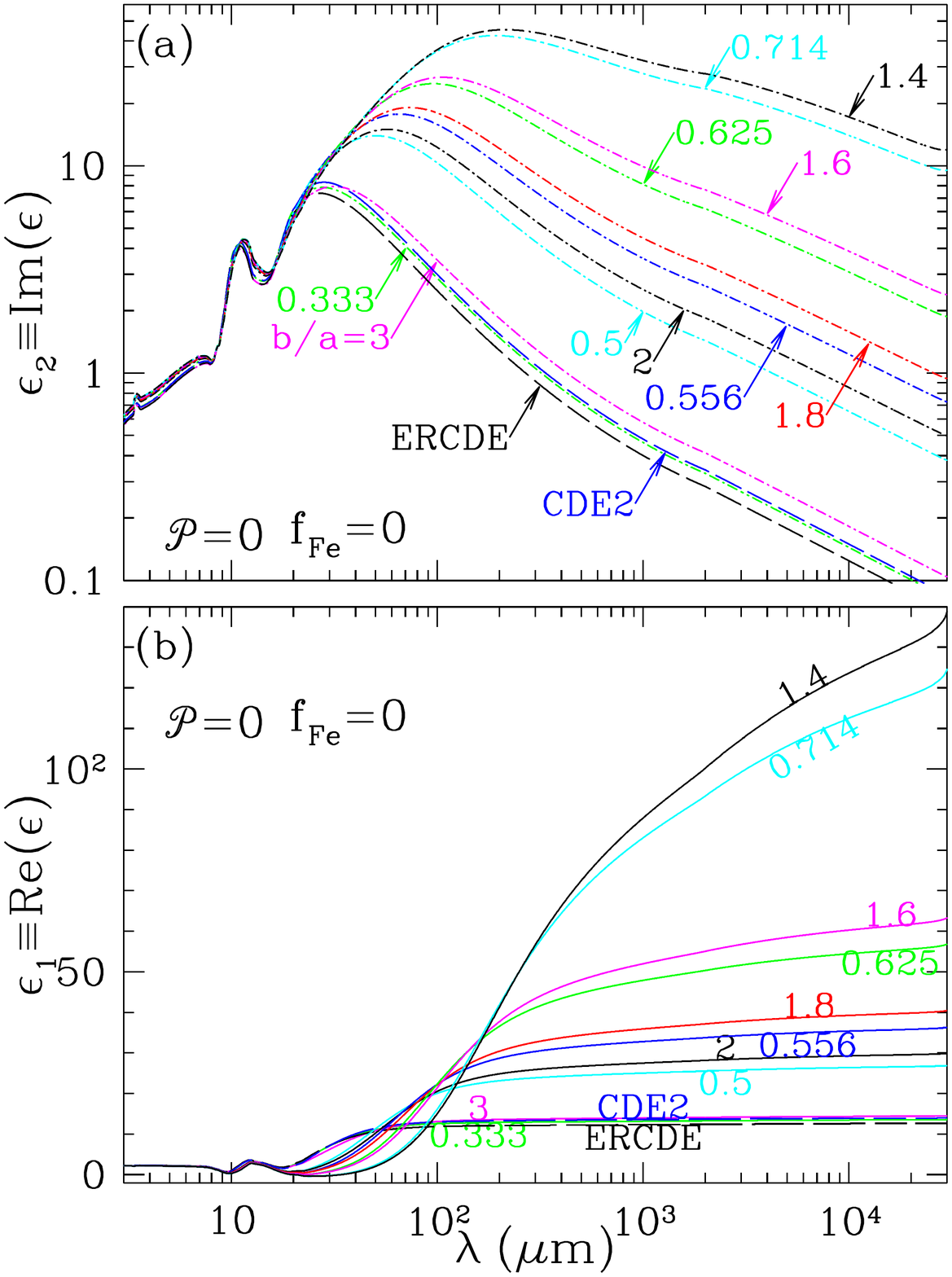}
\includegraphics[width=8.0cm,angle=0,clip=true,
                 trim=0.5cm 0.5cm 0.5cm 0.5cm]
{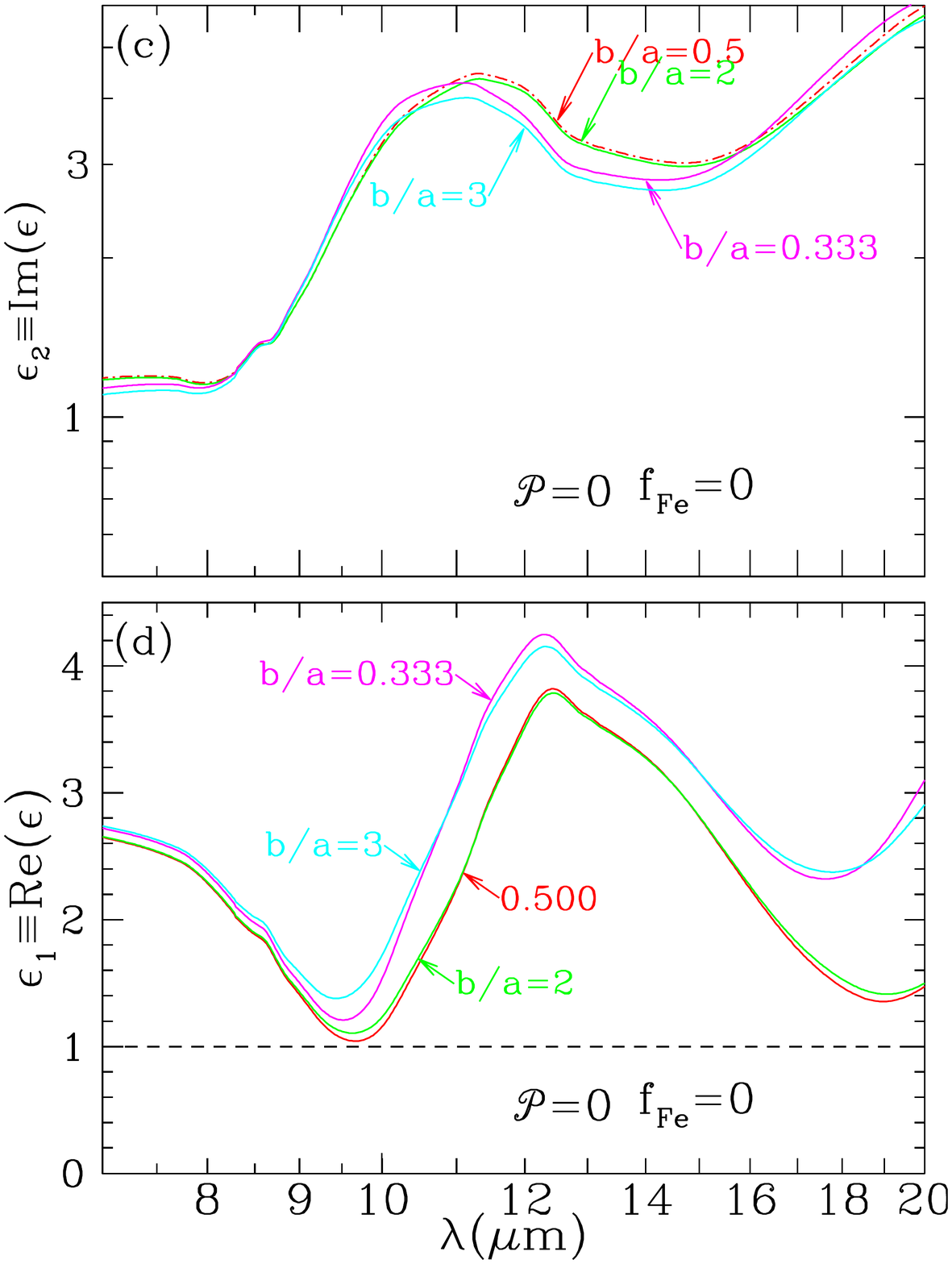}
\caption{\label{fig:epsilon_vary_b/a}\footnotesize
         (a,b): ${\rm Im}(\epsilon)$ and ${\rm Re}(\epsilon)$.
         (c,d): expanded view of 7--20$\micron$ region.
         Results are shown for 
         $\poro=0$
         spheroids with selected axis ratios $b/a$, for
         the ERCDE shape distribution with $\Lmin=0.05$,
         and for the \CDEtwo\ shape distribution.
         Results shown here assume no Fe nanoparticles
         ($\fFe=0$).
         }
\end{center}
\vspace*{-0.3cm}
\end{figure}
Let $\bahat_j$ be the principal axes of the moment of inertia
tensor, with $\bahat_1$ corresponding to the largest moment of inertia.
The cross sections depend on the orientation of the grain relative to the
electric field $\bE$ of the incident wave.
In the Rayleigh limit, 
the absorption cross section for
randomly-oriented grains is
\beqa \label{eq:Cran}
\Cran^{\rm (ed)}(\lambda) &=& 
\frac{1}{3}\left[
\Cabs^{\rm(ed)}(\bE\parallel\bahat_1)+
\Cabs^{\rm(ed)}(\bE\parallel\bahat_2)+
\Cabs^{\rm(ed)}(\bE\parallel\bahat_3)\right]
~~~.
\eeqa
Eq.\ (\ref{eq:Cran}) is exact for
$a/\lambda\rightarrow 0$, and is
an excellent approximation for the submicron grain sizes and
wavelengths $\lambda\gtsim8\micron$ of interest in this study.
For spheroids with symmetry axis $\bahat$, this becomes
\beqa
\Cran^{\rm(ed)}(\lambda) &=&
\frac{1}{3}
\left[\Cabs^{\rm(ed)}(\bE\parallel\bahat)+
2\Cabs^{\rm(ed)}(\bE\perp\bahat)\right]
~~~.
\eeqa
For a given spheroidal or ellipsoidal shape, the 
$\Cabs^{\rm(ed)}(\bE\parallel\bahat_j,\lambda)$
are computed using the well-known solutions to Maxwell's equations
in the limit $a/\lambda\ll 1$
\citep[see, e.g.,][]{Bohren+Huffman_1983,Draine+Lee_1984}.
For shape distributions we use analytic averages 
$\langle\Cran^{\rm(ed)}/V\rangle$ over
the shape distribution obtained 
by \citet{Fabian+Henning+Jager+etal_2001} for the CDE2
and by \citet{Zubko+Mennella+Colangeli+Bussoletti_1996} for the ERCDE
\citep[see Equations 27 and 28 of][]{Draine+Hensley_2020a}. 

As illustrated in Fig.\ \ref{fig:flowchart} 
for each trial shape or shape distribution, 
each trial porosity $\poro$, and each
trial value of $\fFe$, we follow the procedure outlined by
\citet{Draine+Hensley_2020a} and
solve iteratively for the $N=3000$ unknown oscillator strengths $S_k$
such that the model $[\Cran^\ed(\lambda)/V]$ accurately reproduces
the ``observed'' $[\Cran^\ed(\lambda)/V]_\Ad$
(see Fig.\ \ref{fig:cabs})
at $\lambda>1\micron$.
We use the Fortran implementation of the Levenberg-Marquardt algorithm
in the {\tt minpack} library \citep{Garbow+Hillstrom+More_1980}.

Each choice of $\poro$, and grain shape 
(or shape distribution)
gives a different $\epsilon(\lambda)$. 
Figure \ref{fig:epsilon_vary_b/a} shows how 
$\epsilon(\lambda)$ depends upon the assumed grain shape, for
the case of $\poro=0$.

\begin{figure}[t]
\begin{center}
\includegraphics[width=8.0cm,angle=0,clip=true,
                 trim=0.5cm 0.5cm 0.5cm 0.5cm]
{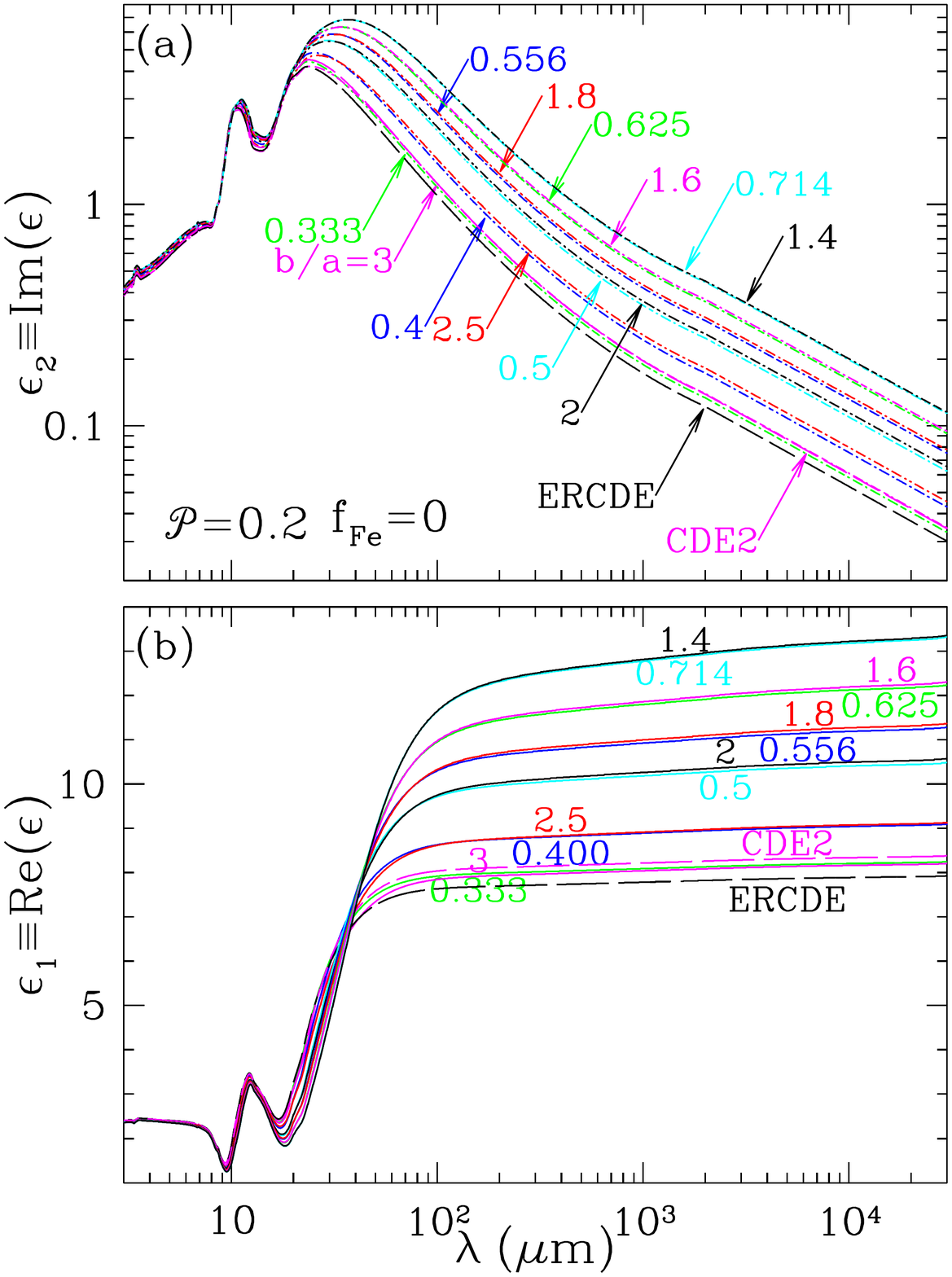}
\includegraphics[width=8.0cm,angle=0,clip=true,
                 trim=0.5cm 0.5cm 0.5cm 0.5cm]
{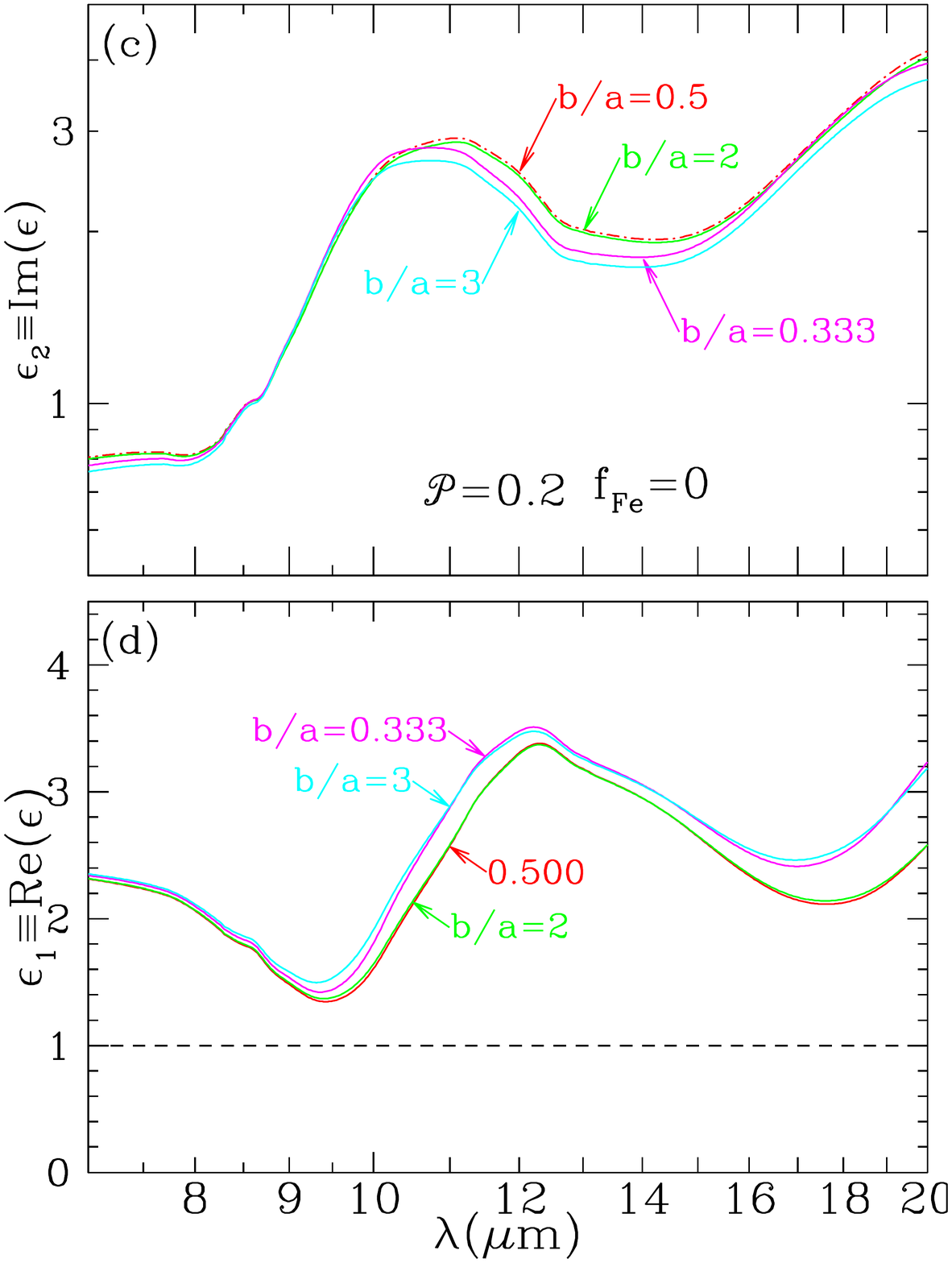}
\caption{\label{fig:epsilon_vary_b/a_P0.20}\footnotesize
         Same as Fig.\ \ref{fig:epsilon_vary_b/a}
         but for $\poro=0.2$.
         Compared to the $\poro=0$ case,
         $\epsilon_2$ is generally lower,
         and $\epsilon_1$ at long wavelengths is reduced.
         }
\end{center}
\vspace*{-0.3cm}
\end{figure}
\begin{figure}
\begin{center}
\includegraphics[width=8.0cm,angle=0,clip=true,
                 trim=0.5cm 0.5cm 0.5cm 0.5cm]
{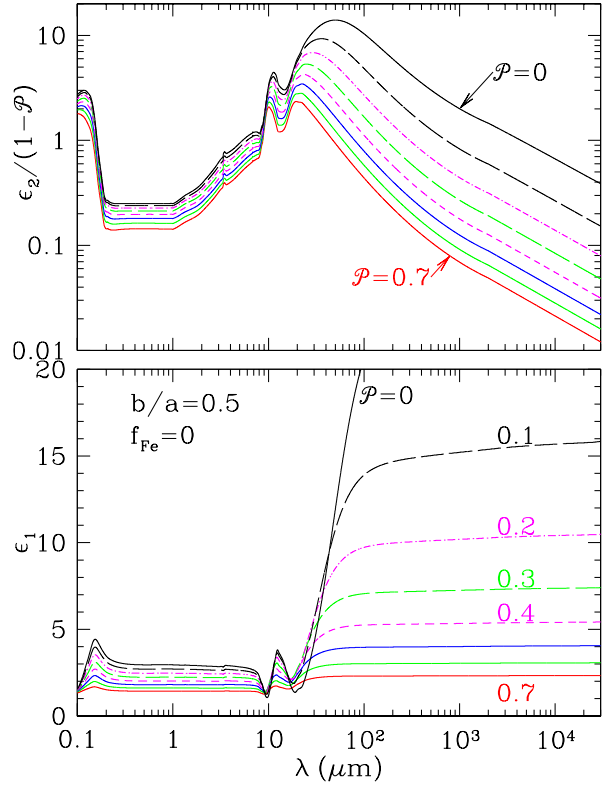}
\caption{\footnotesize \label{fig:epsilon_vary_poro}
         ${\rm Im}(\epsilon)$ and ${\rm Re}(\epsilon)$ for $b/a=0.5$, for
         various porosities $\poro$.}
\end{center}
\end{figure}
Because \astrodust\ grains must provide a substantial
opacity at far-infrared wavelengths, 
$\epsilon_2$ must be relatively large in the far-infrared -- for
example, for $\poro=0$ and $b/a=1.8$, 
we find $\epsilon_2(100\micron)\approx 20$,
and low frequency dielectric constant 
$\epsilon_1(\omega\rightarrow0)\approx40$.

What values of $\epsilon$ are physically plausible?
Consider some examples of strong dielectrics.
Crystalline alumina (Al$_2$O$_3$) has
$\epsilon_1(\omega\!\rightarrow\!0)=11.5$ along the crystal axis, magnetite
(Fe$_3$O$_4$) has $\epsilon_1(\omega\!\rightarrow\!0)\approx 20$, and
barium oxide (BaO) has $\epsilon_1(\omega\!\rightarrow\!0)\approx 34$
\citep{Young+Frederikse_1973}: large values do occur for some minerals.
For $\poro=0$ and $0.6 \ltsim b/a < 1.8$, 
the values found here for \astrodust\
are outside this range; while not physically impossible,
such models seem unrealistic.

Porosity, by increasing the grain volume, 
allows the required absorption to be provided by a
more moderate 
dielectric function.
Figure \ref{fig:epsilon_vary_b/a_P0.20} shows $\epsilon$ if the porosity
is taken to be $\poro=0.2$.  For $b/a=2$, we now find
$\epsilon_2(100\micron)\approx 1$
and
$\epsilon_1(\omega\rightarrow0)\approx8$.
Figure \ref{fig:epsilon_vary_poro} shows results for $b/a=0.5$ and porosities
$\poro$ from 0 to 0.7 .

\begin{figure}
  \begin{center}
    \includegraphics[angle=0,width=10.0cm,
      clip=true,trim=0.5cm 5.0cm 0.5cm 2.5cm]
                    {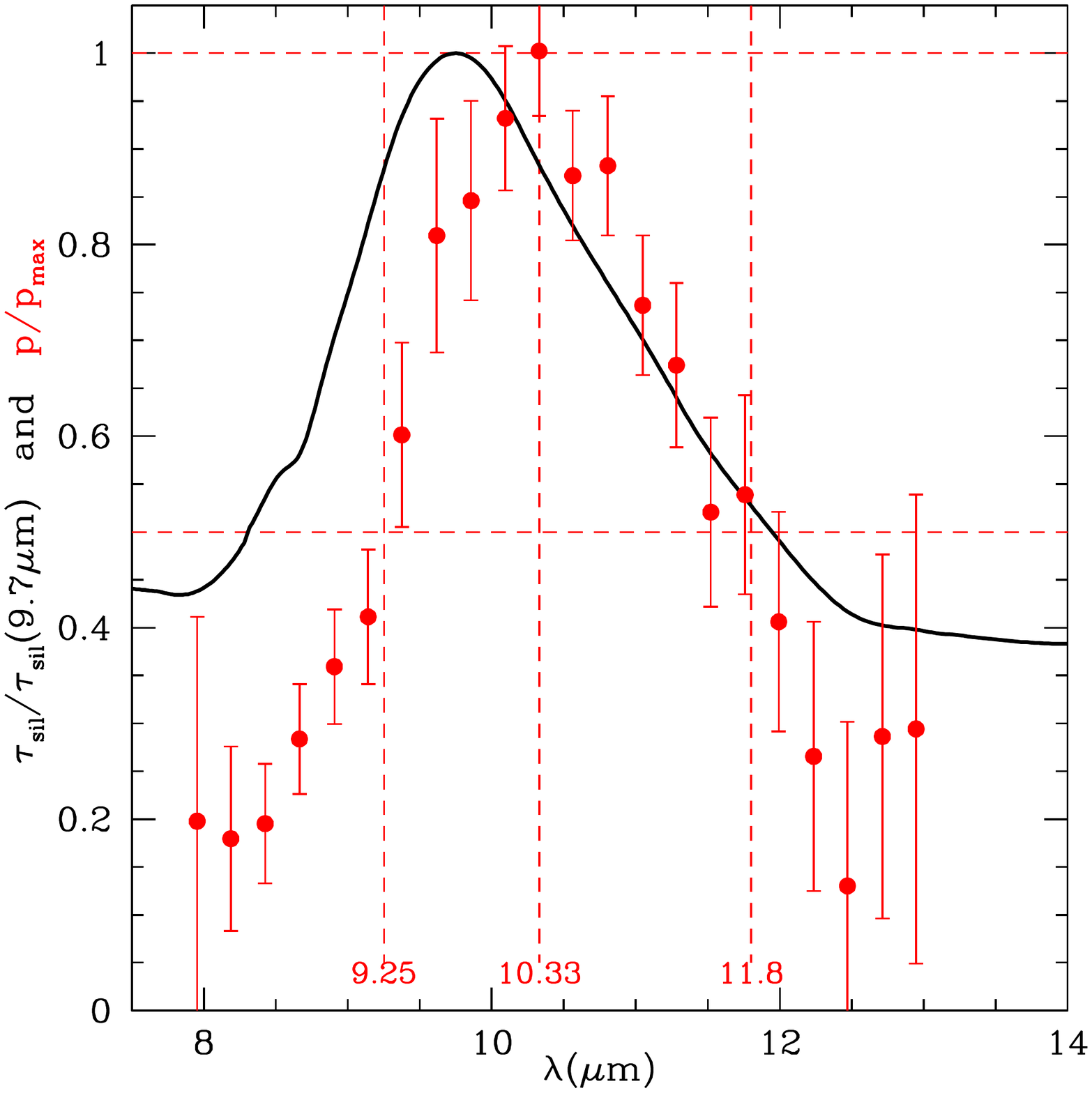}
    \caption{\label{fig:polprof}\footnotesize
      Points: Normalized polarization $p(\lambda)/p_{\rm max}$ toward WR48A and
      WR112=AFGL2014 \citep{Wright+Aitken+Smith+etal_2002}.
      Curve: Extinction toward Cyg OB2-12
      \citep{Hensley+Draine_2020a}.
    }
  \end{center}
\end{figure}

\section{\label{sec:silicate polarization}
         Polarization Near the Silicate Resonances}
\subsection{Polarization Cross Section in the Electric Dipole Limit}

By construction,
the above dielectric functions all give the same
$\lambda > 1\micron$ opacity as a function of wavelength.
However, they differ in
polarization cross sections.
The polarization cross section in the Rayleigh limit ($a\ll\lambda$)
is:
\beqa \label{eq:Cpol_ell}
\Cpol^{\rm(ed)}(\lambda) &\equiv& 
\frac{1}{4}
\left[\Cabs^{\rm(ed)}(\bE\parallel\bahat_2)
+\Cabs^{\rm(ed)}(\bE\parallel\bahat_3)\right]
-\frac{1}{2}\Cabs^{\rm(ed)}(\bE\parallel\bahat_1) ~~~{\rm for~ellipsoids,}
\\ \label{eq:Cpol_obl}
&=& \frac{1}{2}
\left[\Cabs^{\rm(ed)}(\bE\perp\bahat)-
\Cabs^{\rm(ed)}(\bE\parallel\bahat)\right] ~~~{\rm for~oblate~spheroids,}
\\ \label{eq:Cpol_pro}
&=& \frac{1}{4}\left[\Cabs^{\rm(ed)}(\bE\parallel\bahat)-
\Cabs^{\rm(ed)}(\bE\perp\bahat)\right] ~~~{\rm for~prolate~spheroids,}
\eeqa
where we assume each grain to be spinning around the principal
axis of largest moment of inertia $\bahat_1$ (i.e., $\bahat_1\parallel\bJ$,
where $\bJ$ is the angular momentum).
For spheroids with specified axis ratio $b/a$, we calculate $\Cpol$ using
standard formulae \citep[e.g.,][]{Bohren+Huffman_1983,Draine+Lee_1984}.
For the ERCDE and CDE2 shape distributions, we calculate 
$\Cpol^{\rm(ed)}(\lambda)$
using Eq.\ (31) and (33) from \citet{Draine+Hensley_2020a}.
Each choice of grain shape and $\poro$ leads to
different predictions for $\Cpol^{\rm(ed)}(\lambda)$.
By comparing the predicted $\Cpol^{\rm(ed)}(\lambda)$
to the observed polarization profile, we hope to narrow
the domain of allowed values of $\poro$ and grain shape.

Figure \ref{fig:polprof} shows the observed 8--13$\micron$
polarization profile in the ISM
\citep{Wright+Aitken+Smith+etal_2002}.
There is a significant offset between
the polarization profile (peaking near $\sim$$10.3\micron$) and
the extinction profile (peaking near $9.75\micron$).
Such an offset is theoretically expected, because $\Cran^{\rm(ed)}(\lambda)$ 
and $\Cpol^{\rm(ed)}(\lambda)$
depend differently on $\epsilon(\lambda)$.

Figure \ref{fig:Cpol} shows the theoretical $[\Cpol^{\rm(ed)}/(1-\poro)V]_\Ad$
from 7--40$\micron$,
for spheroids with porosity $\poro=0.2$, and various axis ratios $b/a$, 
as well as for the ERCDE and \CDEtwo\ shape distributions.
By construction, these models have identical absorption
profiles $\Cran^{\rm(ed)}(\lambda)/(1-\poro)V_\Ad$.
As expected, spheroids with more extreme axis ratios
have larger $[\Cpol^{\rm(ed)}/V]_\Ad$.
The \CDEtwo\ and ERCDE shape distributions have 
$[\Cpol^{\rm(ed)}(\lambda)/V]_\Ad$ that are
similar to oblate spheroids with $b/a\approx 3$.

\subsection{Grain Shape}

\begin{figure}[t]
\begin{center}
\includegraphics[width=8.0cm,angle=0,clip=true,
                 trim=0.5cm 0.0cm 0.5cm 0.0cm]
{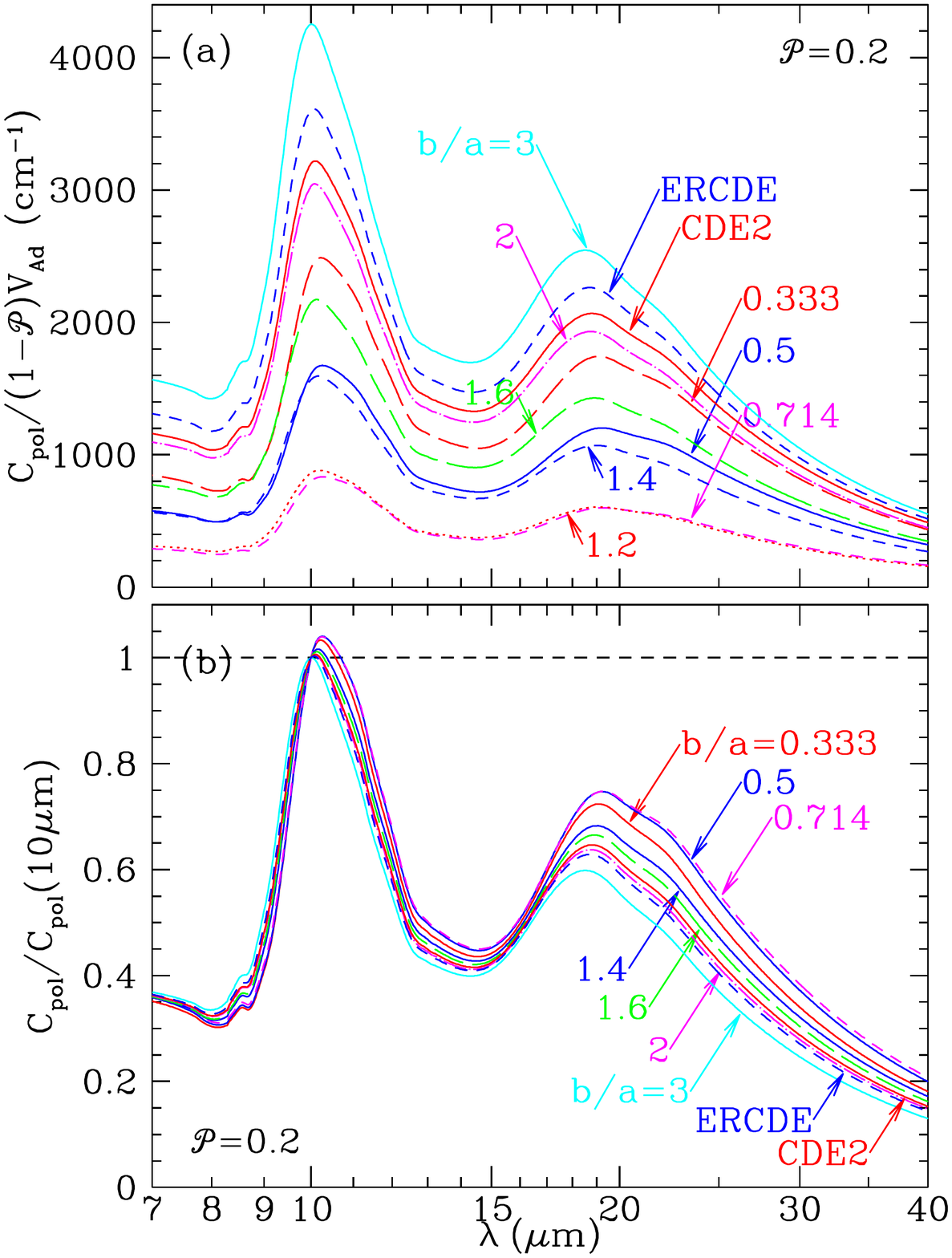}
\includegraphics[width=8.0cm,angle=0,clip=true,
                 trim=0.5cm 0.0cm 0.5cm 0.0cm]
{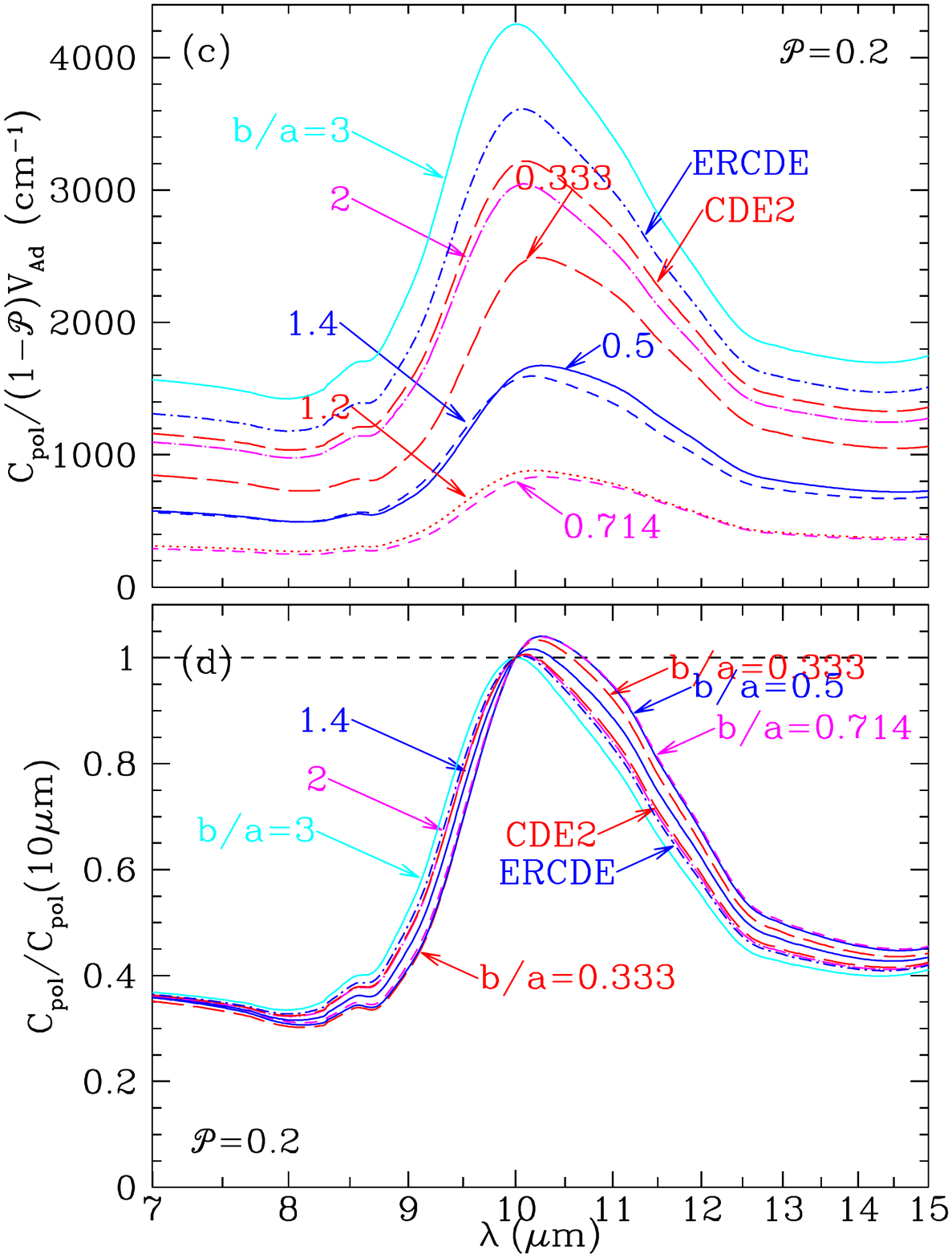}
\caption{\label{fig:Cpol}\footnotesize
         (a) $\Cpol^{\rm(ed)}/(1-\poro)V_\Ad$ and 
         (b) normalized profile 
         $\Cpol^{\rm(ed)}/\Cpol^{\rm(ed)}(10\micron)$ for
         spheroids with different axis ratios $b/a$,
         for the ERCDE distribution with $\Lmin=0.05$,
         and for the \CDEtwo\ distribution,
         for dielectric functions that reproduce
         $\Cabs(\lambda)$ with
         $\poro=0.2$.
         The polarization profile in (b) and (d)
         is seen to be sensitive to the assumed grain shape.
         Polarization for the \CDEtwo\ and ERCDE shape 
         distributions is similar
         to oblate spheroids with $b/a\approx3$.
         }
\end{center}
\vspace*{-0.3cm}
\end{figure}

In addition to affecting the overall magnitude of $\Cpol^{\rm(ed)}/V$,
the grain shape also affects the
{\it shape} of the polarization profile.
Figures \ref{fig:Cpol}b and \ref{fig:Cpol}d
show normalized polarization profiles.
Varying the grain shape from oblate to
prolate systematically shifts the $10\micron$ and $18\micron$ 
polarization profiles to
longer wavelengths.
Figure \ref{fig:Cpol}d shows that the short-wavelength side of the 
$10\micron$ profile
shifts by $\sim$$0.19\micron$ as $b/a$ varies from 3 to $1/3$.
Observations of the $10\micron$ polarization profile therefore provide a
way to constrain the grain shape.

\begin{figure}[h]
\begin{center}
\includegraphics[width=8.5cm,angle=0,clip=true,
                 trim=0.5cm 5.0cm 0.5cm 2.0cm]
{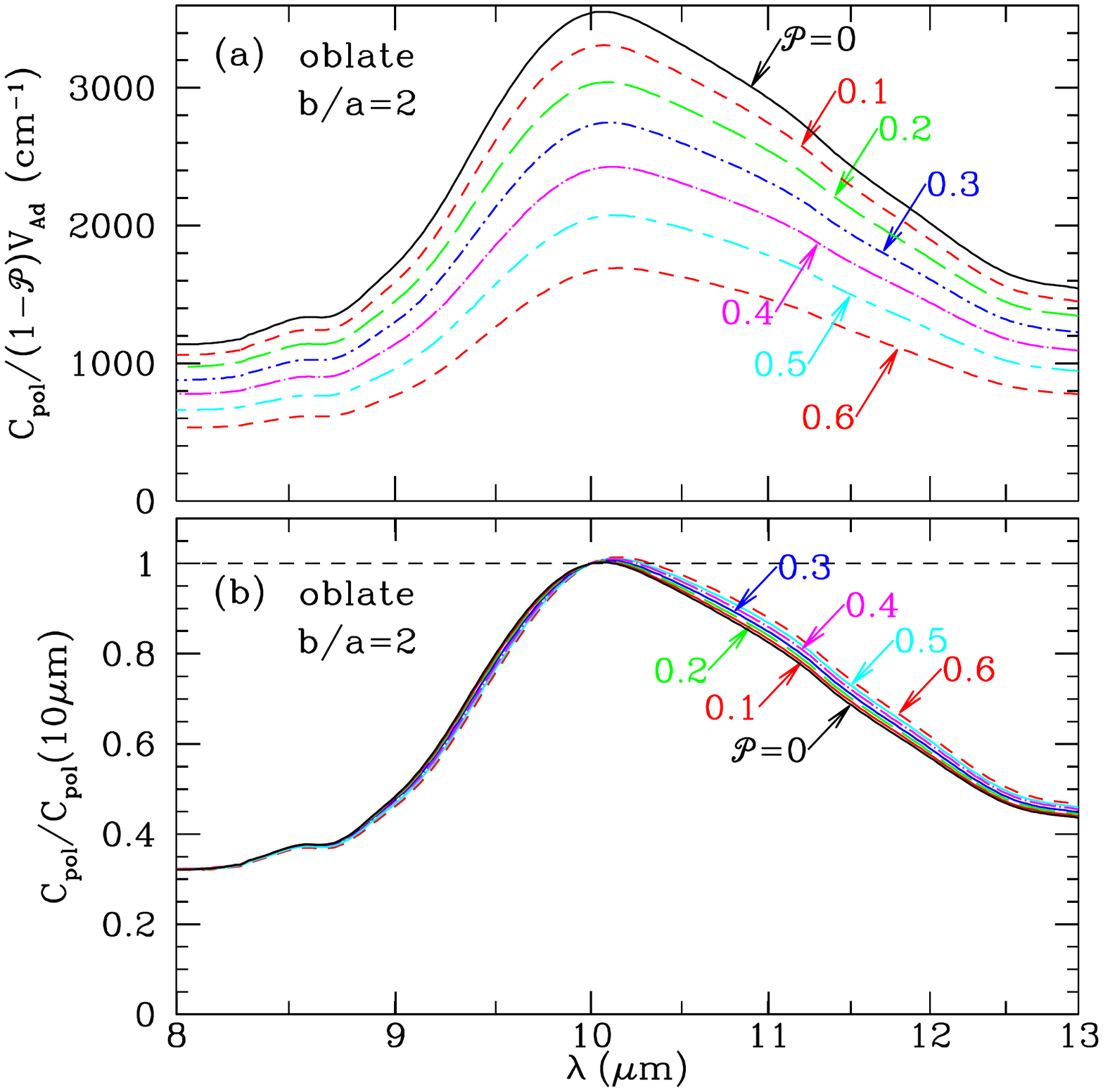}
\includegraphics[width=8.5cm,angle=0,clip=true,
                 trim=0.5cm 5.0cm 0.5cm 2.0cm]
{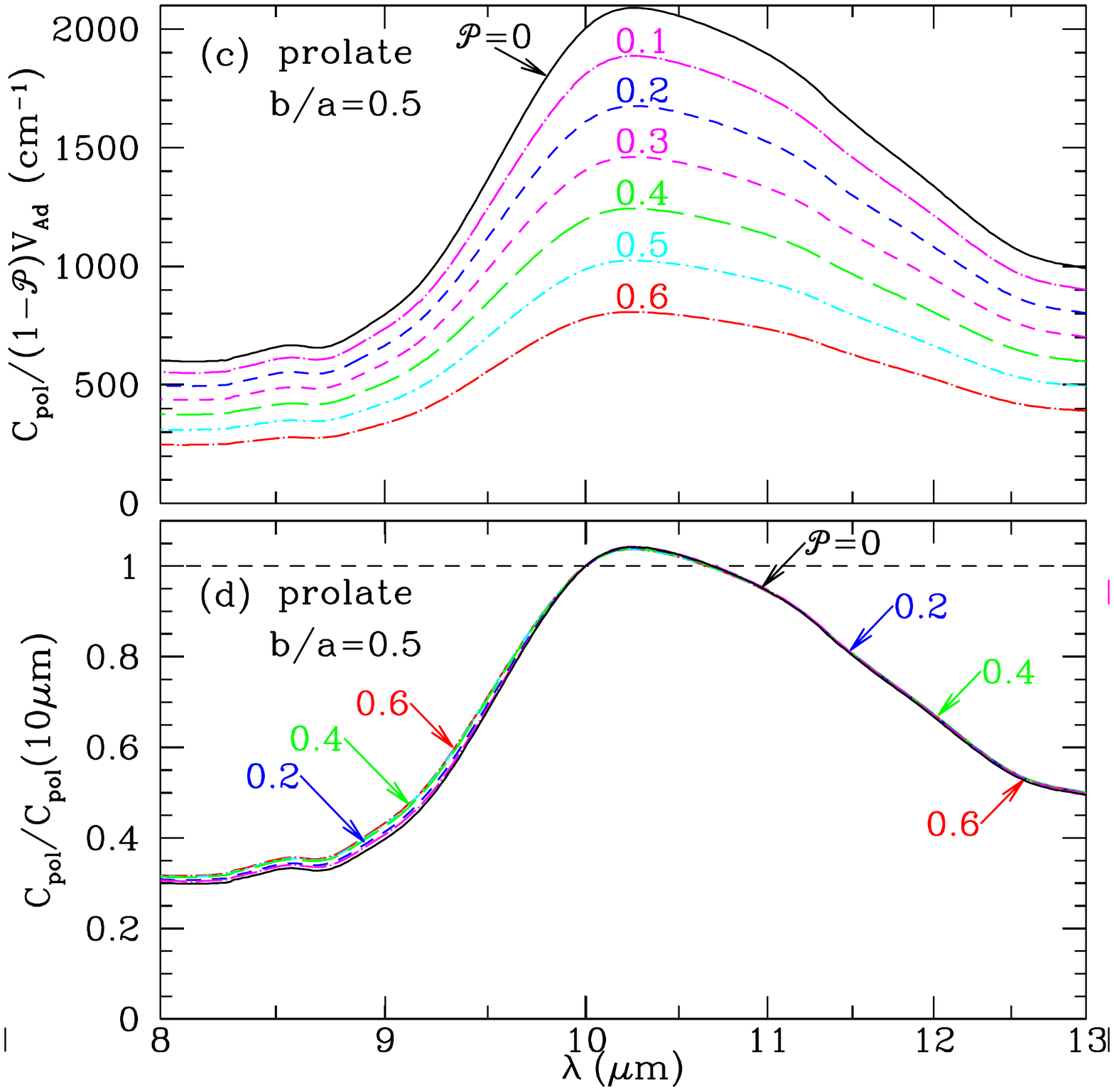}
\caption{\label{fig:Cpolz}\footnotesize
         Polarization cross section in the 8--13$\micron$ region
         for $b/a=3$ (oblate) and $b/a=0.333$ (prolate) spheroids
         for
         selected porosity $\poro$.
         The shape of the polarization profile
         depends only weakly on $\poro$.
         }
\end{center}
\vspace*{-0.3cm}
\end{figure}
Figure \ref{fig:Cpol}b also shows that the strength of the 20$\micron$
polarization relative to the $10\micron$ polarization is sensitive
to grain shape: the ratio
$\Cpol(19\micron)/\Cpol(10\micron)$ changes from $0.57$ to $0.70$
as $b/a$ varies from 3 to $1/3$.

\citet{Lee+Draine_1985} argued that 
the polarized 10$\micron$ 
feature toward the BN object was best fit by oblate spheroids.
However, their analysis was not self-consistent, because it
adopted a single dielectric function that had 
been ``derived'' from the extinction
assuming the grains to be spherical; this dielectric function was then used
to calculate extinction vs.\ $\lambda$ for nonspherical grains.
We now treat the problem self-consistently.

\begin{figure}[t]
\def\figw{10.0cm}
\begin{center}
\includegraphics[width=\figw,angle=0,clip=true,
                 trim=0.5cm 5.0cm 0.5cm 2.5cm]
{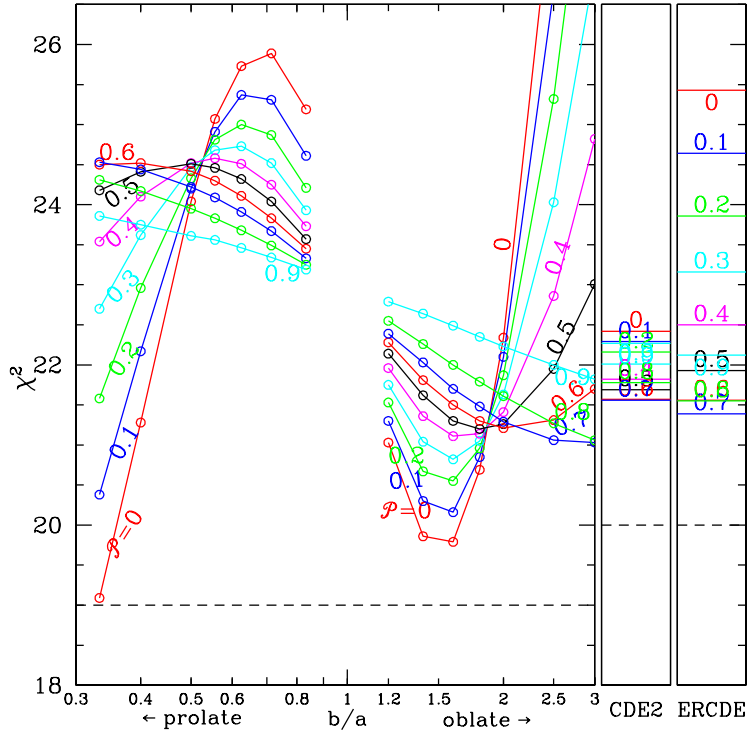}
\caption{\label{fig:chi2_vs_shape}\footnotesize
         Figure-of-merit $\chi^2$ (see Eq.\ \ref{eq:chi^2})
         for spheroids as a function
         of axis ratio $b/a$, for selected values of
         $\poro$.
         The dashed line shows the expected value of $\chi^2\approx19$
         if the model were perfect, and errors are random and correctly
         estimated.
         The panels on the right show $\chi^2$ for the CDE2 and ERCDE
         shape distributions; with 2 degrees of freedom, we would
         expect $\chi^2=20$ (dashed line).
         The best fit is found for prolate
         spheroids with $a/b=3$ and $\poro=0$.
         }
\end{center}
\vspace*{-0.3cm}
\end{figure}

For each choice of grain shape and $\poro$ we have
a different dielectric function
$\epsilon(\lambda; b/a, \poro)$, constrained to reproduce the observed
absorption, but giving a distinct polarization profile.
By comparing the models with spectropolarimetric observations,
we hope to narrow the range of grain shapes consistent with observations.
Figure \ref{fig:chi2_vs_shape} 
shows a goodness-of-fit metric
\beq \label{eq:chi^2}
\chi^2 \equiv \sum_{j=1}^{N_{\rm dat}}
\frac{1}{\sigma_j^2} 
\left[
   \frac{p_{\rm obs}(\lambda_j)}{p_{\rm max}}-
   A
   \frac{\Cpol^{\rm(ed)}(\lambda_j)}{\Cpol^{\rm(ed)}(10\micron)}
\right]^2
~~~, 
\eeq 
for different axis ratios and selected $\poro$,
where the $N_{\rm dat}=22$ measurements $p_{\rm obs}(\lambda_j)$ 
and uncertainties $\sigma_j$ (shown in Figure \ref{fig:polprof})
are
from \citet{Wright+Aitken+Smith+etal_2002}, and the scale factor $A$
is adjusted to minimize $\chi^2$ for each model.
Figure \ref{fig:chi2_vs_shape} also shows $\chi^2$ for selected $\poro$
for the CDE2 and ERCDE continuous distributions of ellipsoidal shapes.

We have 3 adjustable parameters:
$b/a$,
$\poro$,
and the factor $A$ in Eq.\ (\ref{eq:chi^2}).
We have no {\it a priori} constraint on axis ratio $b/a$, other than that
it be large enough to be able to reproduce the observed polarization
of starlight;
allowed values of $b/a$ are delineated by \citet{Draine+Hensley_2020c} for
different $\poro$.
We have no {\it a priori} constraint on porosity, other than
$0\leq\poro<1$.
If the measurement uncertainties $\sigma_j$ are independent
and correctly estimated, we would
expect to have a minimum $\chi^2\gtsim 22-3=19$
(the dashed line in Figures \ref{fig:chi2_vs_shape}a-d).
Recognizing that the errors may have been under- or over-estimated, 
we are not concerned if the minimum of $\chi^2$ differs somewhat from 19.

It is evident from Figure \ref{fig:chi2_vs_shape}
that oblate and prolate spheroids
can both provide acceptable fits,
depending on $\poro$.
The best fit is found for $a/b=3$ and $\poro=0$.

We favor modest axis ratios, e.g., $a/b\approx 2$, or $b/a\approx 1.6$,
for several reasons:
\begin{enumerate}
\item The distribution of grain sizes and shapes may be due in part to
fragmentation.  Fragmentation produces
fragments with typical axis ratios $\sim 1\!:\!\sqrt{2}\!:\!2$,
at least for larger (cm-sized) bodies 
\citep{Fujiwara+Kamimoto+Tsukamoto_1978}.
\item For extreme axis ratios, the observed polarization of starlight
would require only a small fractional alignment of the
$a\gtsim0.1\micron$ grains
\citep[see discussion in][]{Draine+Hensley_2020c}.  
Although the physics of grain
alignment remains uncertain, some analyses suggest that
$a\gtsim0.15\micron$ grains will be rotating suprathermally
\citep{Purcell_1979,Draine+Weingartner_1996} and may be expected to
have high alignment fractions.  If so, the axis ratios should be
more modest, perhaps $b/a\approx 1.4-1.8$,
or $a/b\approx 1.6-2.5$.
\item The observed ratio of polarized submm emission
to starlight polarization
is sensitive to grain shape.
For the present model (with both starlight extinction and FIR-submm 
emission dominated by a single grain type -- astrodust) 
we favor modest axis ratios 
$b/a\approx 0.5$
\citep{Draine+Hensley_2020c,Hensley+Draine_2020c}.
\end{enumerate}

It should be kept in mind that $\chi^2$ in Fig.\ \ref{fig:chi2_vs_shape}
is based on difficult
observations of linear polarization (Fig.\ \ref{fig:polprof})
made on only 2 stars.

\subsection{Porosity}

Figure \ref{fig:Cpolz} shows an expanded view of
the 10$\micron$ polarization profile, for different
values of $\poro$.  
We see that for fixed shape $b/a=2$, varying $\poro$ has less of
an effect on the polarization profile than does varying the grain shape
(compare Figures \ref{fig:Cpol}d and \ref{fig:Cpolz}d).

Figure \ref{fig:chi2_vs_shape} shows $\chi^2$ vs.\ $b/a$ 
for spheroids  with selected porosities $\poro$, 
and $\chi^2$ 
for ellipsoids with the CDE2 or ERCDE shape distributions and
different porosities.
While the best fits to the observed polarization profiles 
are found for the more extreme prolate spheroids
($a/b=3$) and
$\poro=0$, the fits appear acceptable for all of the
prolate cases, and for many of the oblate spheroids as well.
Only strongly oblate shapes ($b/a>2$) 
with low porosities ($\poro< 0.5$) are strongly disfavored.
For the CDE2 and ERCDE shape distributions, the best fits are found for
the highest porosities.

\begin{figure}[t]
\begin{center}
\includegraphics[width=10.0cm,angle=270,clip=true,
                 trim=0.5cm 0.5cm 0.5cm 0.5cm]
{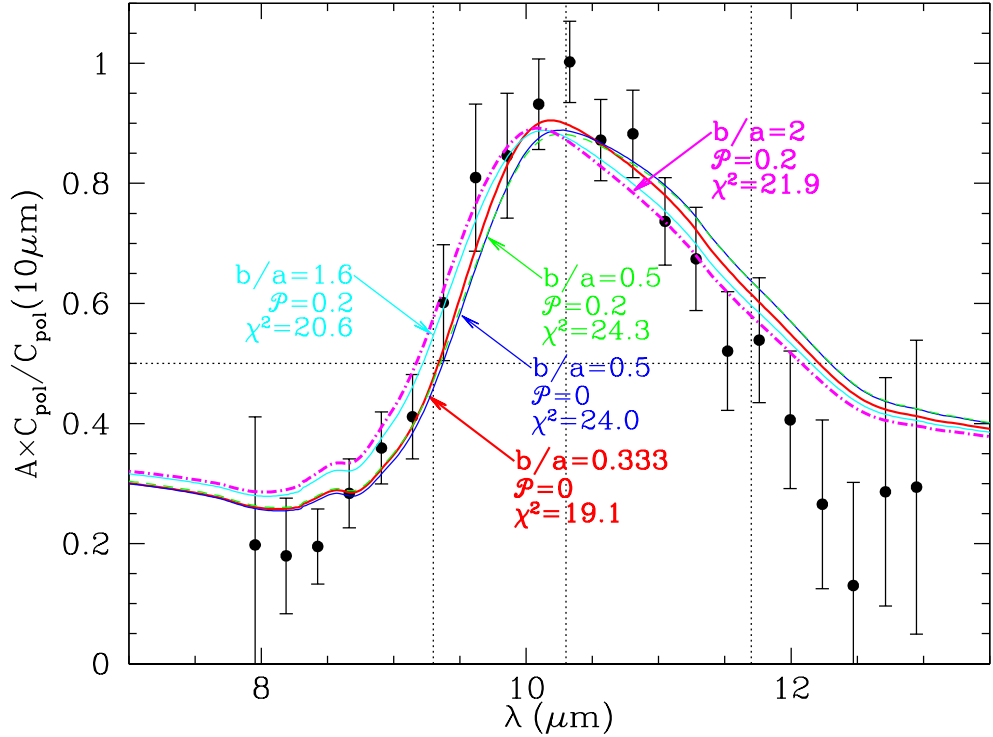}
\caption{\label{fig:Cpol_zoom}\footnotesize
         Data points: average polarization profile 
         $p_{\rm obs}(\lambda)/p_{\rm max}$ measured
         for WR48A and WR112, two sightlines through the diffuse ISM
         \citep{Wright+Aitken+Smith+etal_2002}.
         Curves: $A\times C_{\rm pol}/C_{\rm pol}(10\micron)$
         for selected models,
         with the factor $A$ adjusted to minimize $\chi^2$
         [see Eq.\ (\ref{eq:chi^2})].
         The best fit shown (red) is for 
         $b/a=0.333$ prolate spheroids with $\poro=0$. 
         Other examples of good fits are
         $b/a=0.5$ prolate spheroids with $\poro=0$
         and $0.2$,
         and $b/a=1.6$ and $2$ oblate spheroids with
         $\poro=0.2$.
         }
\end{center}
\vspace*{-0.3cm}
\end{figure}

\subsection{Polarization Profile: Examples}

Figure \ref{fig:Cpol_zoom} shows the
polarization measured toward WR48A and WR112
\citep{Wright+Aitken+Smith+etal_2002} together with selected 
models that are
in good agreement with the observations.
The best-fitting model uses
prolate spheroids with
porosity $\poro=0$, and axis ratio
$a/b=3$, but fits that are almost as good are provided by
$a/b=2$  and $\poro=0$ or $\poro=0.2$.
Also shown are oblate models with $b/a=1.6$ and porosities
$\poro=0$ and $0.4$.

\begin{figure}[t]
\begin{center}
\includegraphics[width=8.5cm,angle=0,clip=true,
                 trim=0.5cm 5.0cm 0.5cm 0.5cm]
{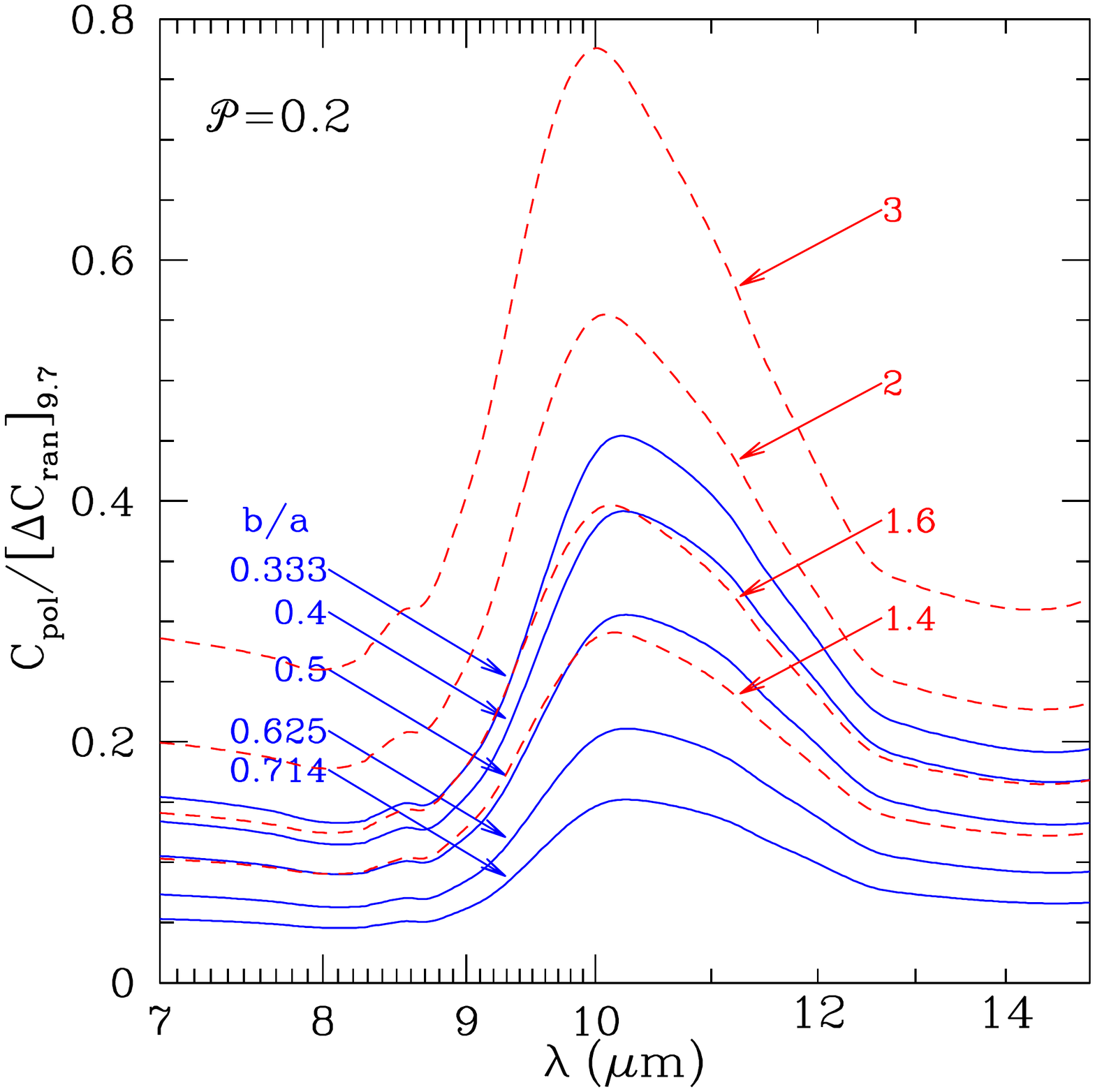}
\caption{\label{fig:cpol/cran_max}\footnotesize
         $\Cpol/[\Delta\Cran]_{9.7\mu{\rm m}}$ for oblate (red)
         and prolate (blue)
         spheroids with varying axis ratio $b/a$,
         for porosity $\poro=0.2$.
         }
\end{center}
\vspace*{-0.3cm}
\end{figure}
\begin{figure}[ht]
\begin{center}
\includegraphics[angle=0,width=8.0cm,
                 clip=true,trim=0.5cm 0.5cm 0.5cm 0.5cm]
{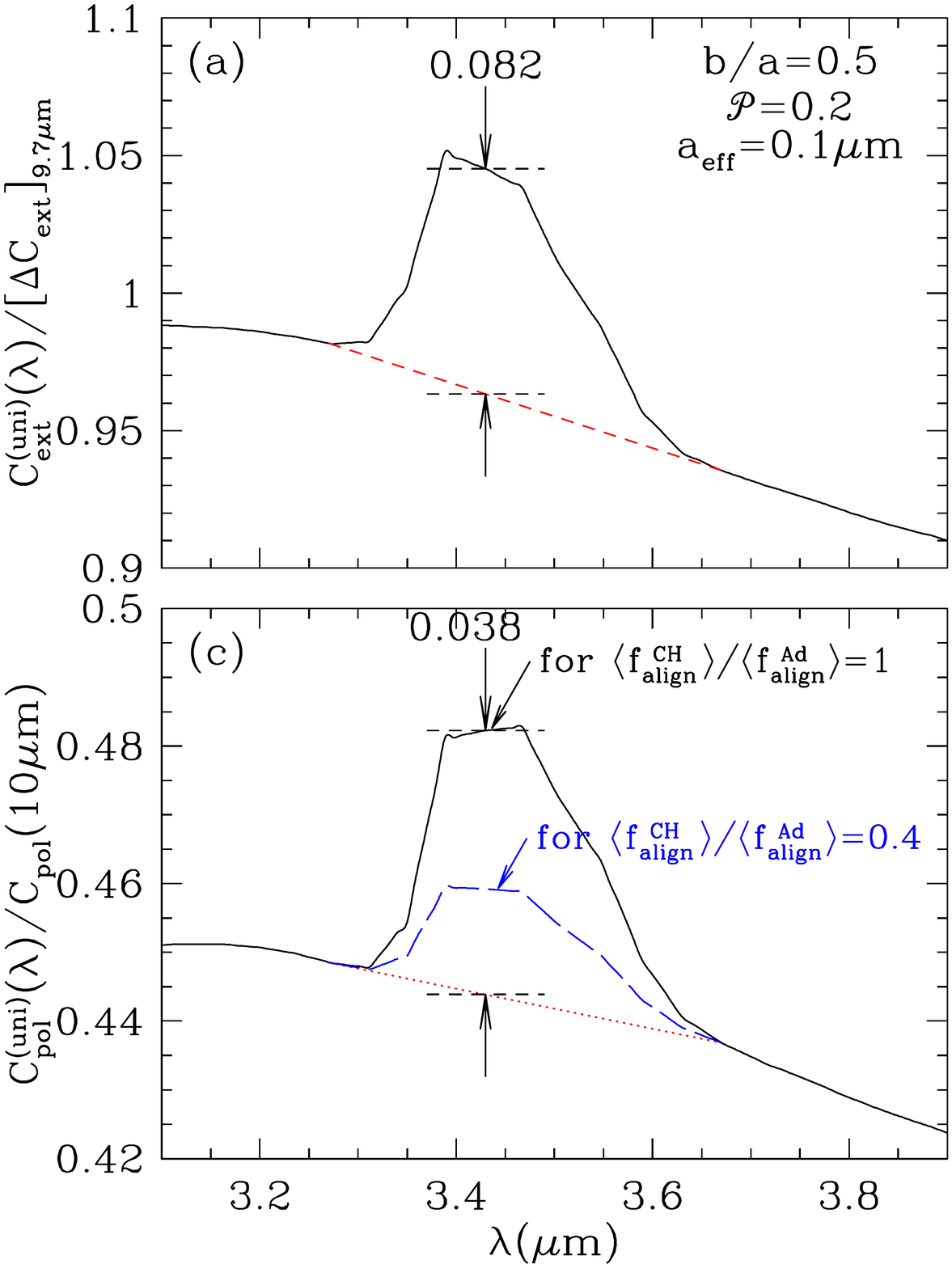}
\includegraphics[angle=0,width=8.0cm,
                 clip=true,trim=0.5cm 0.5cm 0.5cm 0.5cm]
{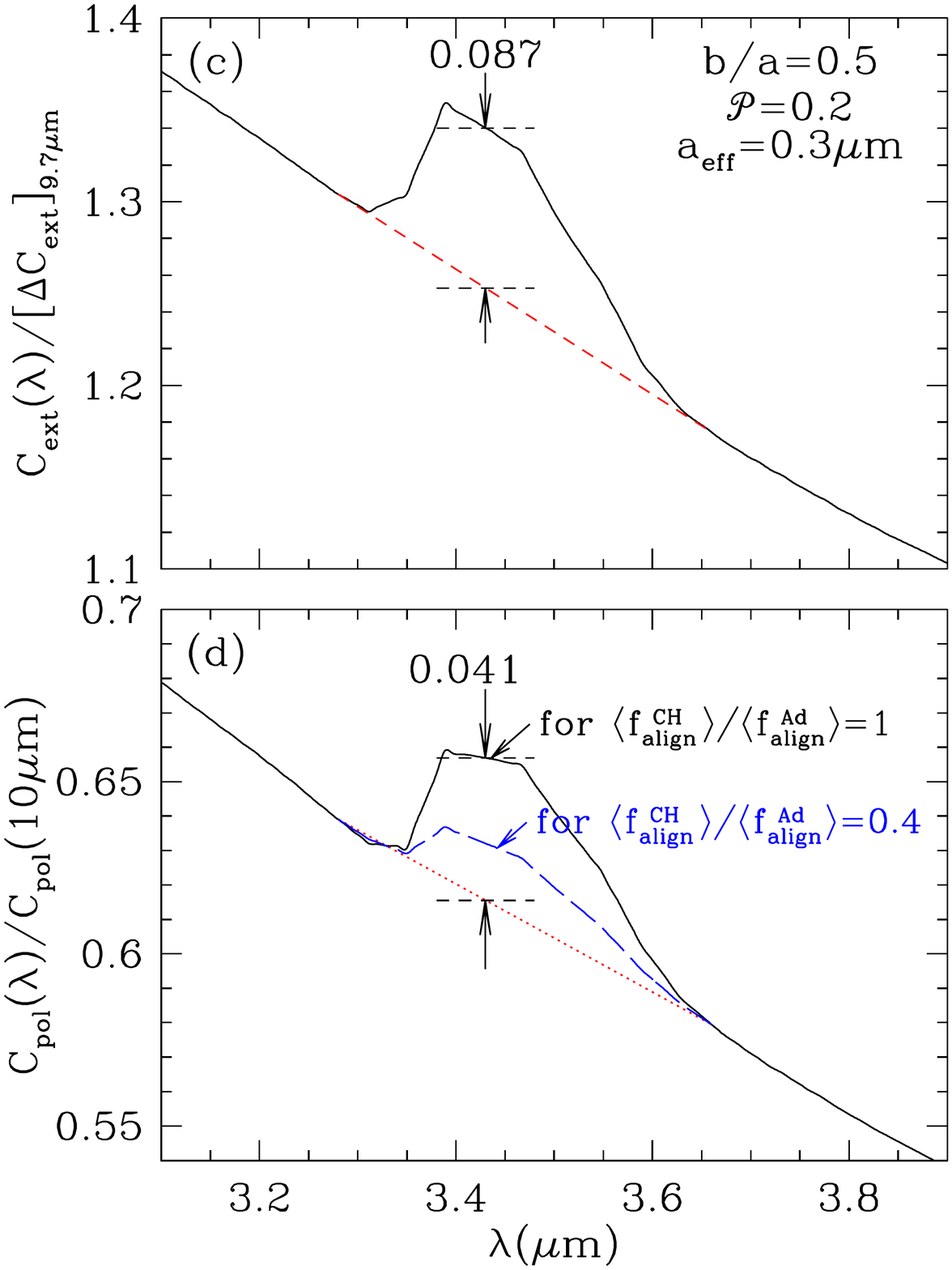}
\caption{\footnotesize\label{fig:CH polarization}
CH feature in extinction and polarization, calculated for
prolate \astrodust\ grains with $a/b=2$, porosity
$\poro=0$.
(a) and (c): Extinction relative to extinction in the $9.7\micron$ silicate
feature, for $\aeff=0.1\micron$ and $0.3\micron$.
For $\aeff=0.3\micron$, scattering increases the extinction by $\sim$40\%.
(b) and (d): Polarization relative to polarization at $10\micron$,
for $\aeff=0.1\micron$ and $0.3\micron$.
Solid black curve: assuming the CH absorber is
uniformly mixed with the silicate material in \astrodust.
Dashed curve: estimated polarization for 
$\langle \falign^{\rm CH}\rangle/\langle \falign^\Ad\rangle=0.4$
(see text), which could correspond to the CH absorber being
concentrated in a surface layer with thickness
$\Delta\approx0.01\micron$
(see Fig.\ \ref{fig:falign CH}).
}
\end{center}
\end{figure}


\section{\label{sec:alignment}
         Infrared and Submm Polarization}

The objective of this section is to relate the polarization in the $10\micron$
feature to polarization at other wavelengths, as a model prediction.
Of particular interest are 
(1) polarization in the far-infrared and submm thermal continuum, and
(2) polarization in the $3.4\micron$ extinction
feature produced by CH stretching modes in aliphatic hydrocarbons. 


Consider a population of grains, each assumed to
be spinning around its principal axis $\bahat_1$.
Let $a$ be the spherical-volume-equivalent grain radius.
The fractional alignment of grains of size $a$ with the local
magnetic field $\bB_0$ is 
\beq
\falign(a)\equiv
\frac{3}{2}
\left[\langle(\bahat_1\cdot\bBhat_0)^2\rangle-\frac{1}{3}\right]
~~~,
\eeq
with $\falign=0$ for random orientations, and $\falign=1$
for perfect alignment of $\bahat_1$ with $\bB_0$.
The mass-weighted alignment of the \astrodust\ material is
\beq \label{eq:falign}
\langle \falign^\Ad \rangle \equiv
\frac{\int_{a_{\rm min}}^{a_{\rm max}} da
\falign(a)a^3 (dn/da)}
{\int_{a_{\rm min}}^{a_{\rm max}} da
\, a^3 (dn/da)}
~,
\eeq
where
$n(a)$ is the number of grains smaller than $a$.

Let $\gamma$ be the angle between the magnetic field $\bB_0$ and
the line of sight $\bzhat$.
Let $\bxhat$ and $\byhat$ be $\perp$ and $\parallel$ to
the projection of $\bB_0$ on the plane of the sky, and let
$\tau_x$, $\tau_y$ be the optical depths for $\bE\parallel\bxhat$,
$\bE\parallel\byhat$.
Provided the grains are in the Rayleigh limit, with 
$\Cpol \propto V$, 
the fractional polarization $p$ per unit $10\micron$ feature depth
$\deltataupk$ is just
\beq \label{eq:pol/tau}
\frac{p_{\rm ext}(\lambda)}{\deltataupk} = \frac{1}{\deltataupk}
\left[
\frac{e^{-\tau_y}-e^{-\tau_x}}{e^{-\tau_x}+e^{-\tau_y}}
\right]
\approx \frac{\tau_x-\tau_y}{2\deltataupk} \approx 
\frac{\Cpol(\lambda)}{[\Delta\Cran]_{9.7\mu{\rm m}}}
\langle\falign^\Ad\rangle\sin^2\gamma
~~~.
\eeq
The polarizing ability of \astrodust\ grains, relative to extinction in
the 9.7$\micron$ feature, is
shown in Figure \ref{fig:cpol/cran_max} for 
$\poro=0.2$
and selected shapes.
For $b/a=0.5$ and $\poro=0.2$ we have 
\beq \label{eq:Cpol10/Cran for b/a=1.4}
\frac{\Cpol^{\rm(ed)}(10\micron)}{[\Delta \Cran^{\rm(ed)}]_{9.7\mu{\rm m}}}
\approx 0.29
~.
\eeq

\subsection{\label{sec:3.4um polarization}
            3.4$\mu$m Feature Polarization}

CH absorption will produce a
polarization feature at $3.4\micron$.
We define an effective alignment fraction for the CH:
\beq
\langle f_{\rm align}^{\rm CH}\rangle
\equiv
\frac{\int da f_{\rm align}(a) f_{\rm CH}(a) a^3 (dn/da)}
{\int da f_{\rm CH}(a) a^3 (dn/da)}
\eeq
where $f_{\rm CH}(a)$ is the fraction of the volume of grains of
radius $a$
occupied by the $3.4\micron$ absorber.
If the CH absorber is uniformly distributed through the astrodust, then
$f_{\rm CH}(a)$ is independent of $a$, and 
$\langle f_{\rm align}^{\rm CH}\rangle = 
\langle f_{\rm align}^{\rm Ad}\rangle$.

At 3.4$\micron$, $a\gtsim0.1\micron$ grains are large enough
that scattering begins to be important, 
and the extinction and polarization cross sections no 
longer vary linearly with grain mass.
Let $C_{\rm ext}^{\rm (uni)}(\lambda)$
and polarization $C_{\rm pol}^{\rm (uni)}(\lambda)$ 
(see Figure \ref{fig:CH polarization})
be extinction and
polarization cross sections calculated
for the case of
uniformly-distributed CH absorption (i.e., $f_{\rm CH}$ independent of
$a$).

For $0.1<a <0.3\micron$ grains, the 3.4$\micron$ feature
would have polarization, relative to $10\micron$ polarization,
\beq \label{eq:p34}
\frac{[\Delta p]_{3.4\mu{\rm m}}}{p(10\micron)} 
= \frac{\Delta\Cpol^{\rm (uni)}(3.41\micron)}{\Cpol(10\micron)}
\frac{\langle\falign^{\rm CH}\rangle}{\langle\falign^\Ad\rangle}
\approx 0.040
\frac{\langle\falign^{\rm CH}\rangle}{\langle\falign^\Ad\rangle}
~,
\eeq
where the value $0.040$ is 
intermediate between $0.038$ for $\aeff=0.1\micron$ and $0.041$ for
$0.3\micron$ (see Figures 
\ref{fig:CH polarization}c,d).
The
Quintuplet sources near the Galactic Center have
\citep{Chiar+Adamson+Whittet+etal_2006}
\beqa \label{eq:GCS3-II}
\frac{[\Delta p]_{3.4\micron}}{p(10\micron)} 
&=& 
\frac{(0.0006\pm0.0013)}{(0.090\pm0.003)} \approx 0.007 \pm 0.015
~~~{\rm for~GCS3-II}
\\ \label{eq:GCS3-IV}
&=&
\frac{(0.0015\pm0031)}{(0.102\pm0.015)} ~~~\approx 0.015 \pm 0.030
~~~{\rm for~GCS3-IV}
~;
\eeqa
the error bars are claimed to be 99\% confidence intervals.
If the CH and the silicate are in the same aligned grains, with
$\langle\falign^{\rm CH}\rangle/\langle\falign^\Ad\rangle=1$,
the predicted value (\ref{eq:p34}) exceeds the observed values by a factor
$\sim$3.  
Therefore, the CH absorber and the silicate
absorber cannot be identically distributed.  How can the \astrodust\ hypothesis
accomodate this?

We are postulating that \astrodust\ contains 
both silicate and carbonaceous material, in approximately constant ratios.  
However, it may be that only a fraction of the carbonaceous material produces
$3.4\micron$ absorption.
\citet{Mennella+Brucato+Colangeli+Palumbo_1999,
       Mennella+Brucato+Colangeli+Palumbo_2002} argue that the $3.4\micron$
CH absorption feature
is the result of exposure of carbonaceous material to H atoms.  
If so, the CH absorption may be concentrated in ``activated'' surface layers 
on the grains, where hydrogenation by inward-diffusing H atoms 
has produced the CH bonds responsible for the absorption feature.

The observed wavelength-dependence of starlight polarization 
requires that grains smaller than $\sim$$0.1\micron$ be minimally aligned
\citep{Kim+Martin_1995b,Draine+Fraisse_2009}.
For grain size distributions consistent with the observed
interstellar reddening, 
most of the grain surface area is provided by small grains.

Previous studies have considered the possibility that the $3.4\micron$
absorption arises
in hydrocarbon mantles with silicate cores,
concluding that the 3.4$\micron$ polarization,
relative to silicate feature polarization, is fairly insensitive to
whether the 3.4$\micron$ absorber is in a mantle or mixed
through the grain
\citep{Li+Greenberg_2002,Li+Liang+Li_2014}.
This is true for grains for which the ratio of $3.4\micron$ 
absorption to silicate
absorption is constant.
However, if the $3.4\micron$ 
absorption were concentrated in a thin surface layer with
thickness independent of grain size, then small grains would make a greater
contribution to $3.4\micron$ absorption (relative to silicate absorption) than
would larger grains.
Since small grains are relatively unaligned, and account for most of the
total grain surface area, the $3.4\micron$ 
absorption would be skewed toward the
unaligned grain population, resulting in smaller values of
$[\Delta p]_{3.4\micron}/p(10\micron)$ than would otherwise be expected.


\begin{figure}[ht]
\begin{center}
\includegraphics[angle=0,width=10.0cm,
                 clip=tru,trim=0.5cm 5.0cm 0.5cm 2.5cm]
{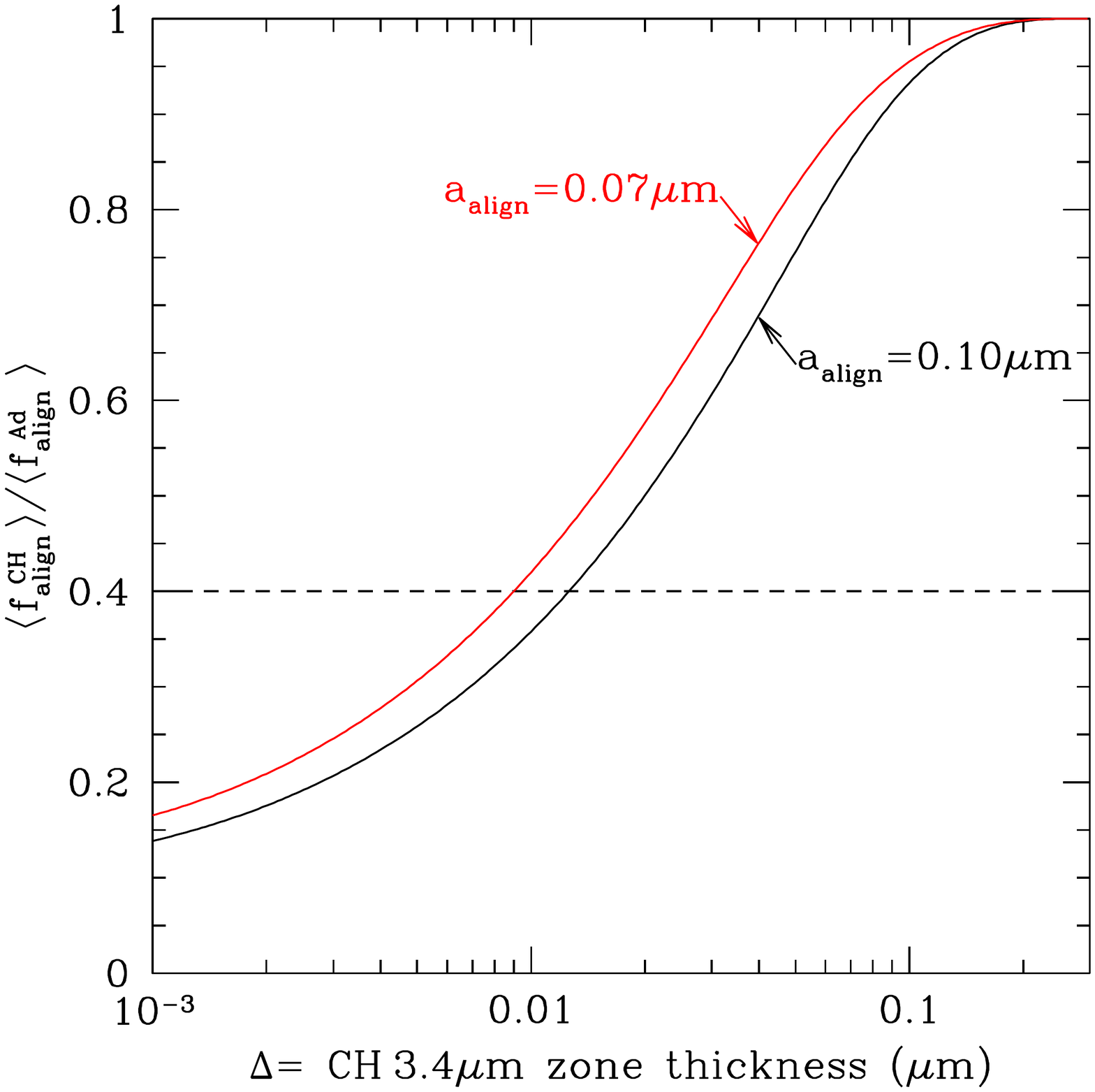}
\caption{\label{fig:falign CH}\footnotesize
Fractional alignment of CH $3.4\micron$ absorbers relative to fractional
alignment of the silicate absorbers as a function of the
thickness $\Delta$ of the CH $3.4\micron$ absorption layer (see text).
Upper limits on CH $3.4\micron$ polarization appear to require
$\langle\falign^{\rm CH}\rangle/\langle\falign^\Ad\rangle\ltsim 0.4$,
or $\Delta\ltsim 0.01\micron$ if the IR opacity is dominated by a single
grain type.
}
\end{center}
\end{figure}

If the CH $3.4\micron$ absorber is in a surface layer of thickness $\Delta$, 
then the aligned fraction for CH $3.4\micron$ absorbers is
\beq
\langle \falign^{\rm CH}\rangle \approx
\frac{\int da \falign(a)
\left[a^3-({\rm max}(a-\Delta,0))^3\right](dn/da)}
{\int da 
\left[a^3-({\rm max}(a-\Delta,0))^3\right](dn/da)}
~~.
\eeq


Figure \ref{fig:falign CH}
shows the ratio $\langle\falign^{\rm CH}\rangle/\langle\falign^\Ad\rangle$
for a MRN size distribution 
($dn/da\propto a^{-3.5}$, $0.001\micron < a < 0.30\micron$) 
and a step-function alignment fraction ($\falign=0$ for $a<a_{\rm align}$,
$\falign=const$ for $a>a_{\rm align}$).
If $a_{\rm align}\approx 0.1\micron$
and $\Delta= 0.01\micron$, the polarization in the CH $3.4\micron$
feature would be suppressed by
a factor $\sim$0.4 relative to the value if the CH $3.4\micron$ 
absorbers were uniformly
distributed in the \astrodust\ material, and \astrodust\ would have
\beq \label{eq:Deltap(3.4)/p(10) prediction}
\frac{[\Delta p]_{3.4\mu{\rm m}}}{p(10\micron)}
\approx 0.040\times 0.4 \approx 0.016
~~.
\eeq
This predicted $3.4\micron$ polarization relative to 10$\micron$ polarization
is marginally consistent
with Eqs.\ (\ref{eq:GCS3-II},\ref{eq:GCS3-IV}),
currently the best upper limits on 
$[\Delta p]_{3.4}/p(10\micron)$.

Further improvements in mid-IR spectropolarimetry should be
able to detect the 
predicted polarization (\ref{eq:Deltap(3.4)/p(10) prediction})
associated with the
$3.4\micron$ feature if it originates near the surfaces of astrodust
grains.
If the silicate and carbonaceous material are distributed in the same grains,
as in our \astrodust\ model, this would allow determination of the
thickness $\Delta$ of the CH ``activated'' surface layers.

\begin{figure}[t]
\begin{center}
\includegraphics[width=10.0cm,angle=0,clip=true,
                 trim=0.5cm 0.5cm 0.5cm 0.5cm]
{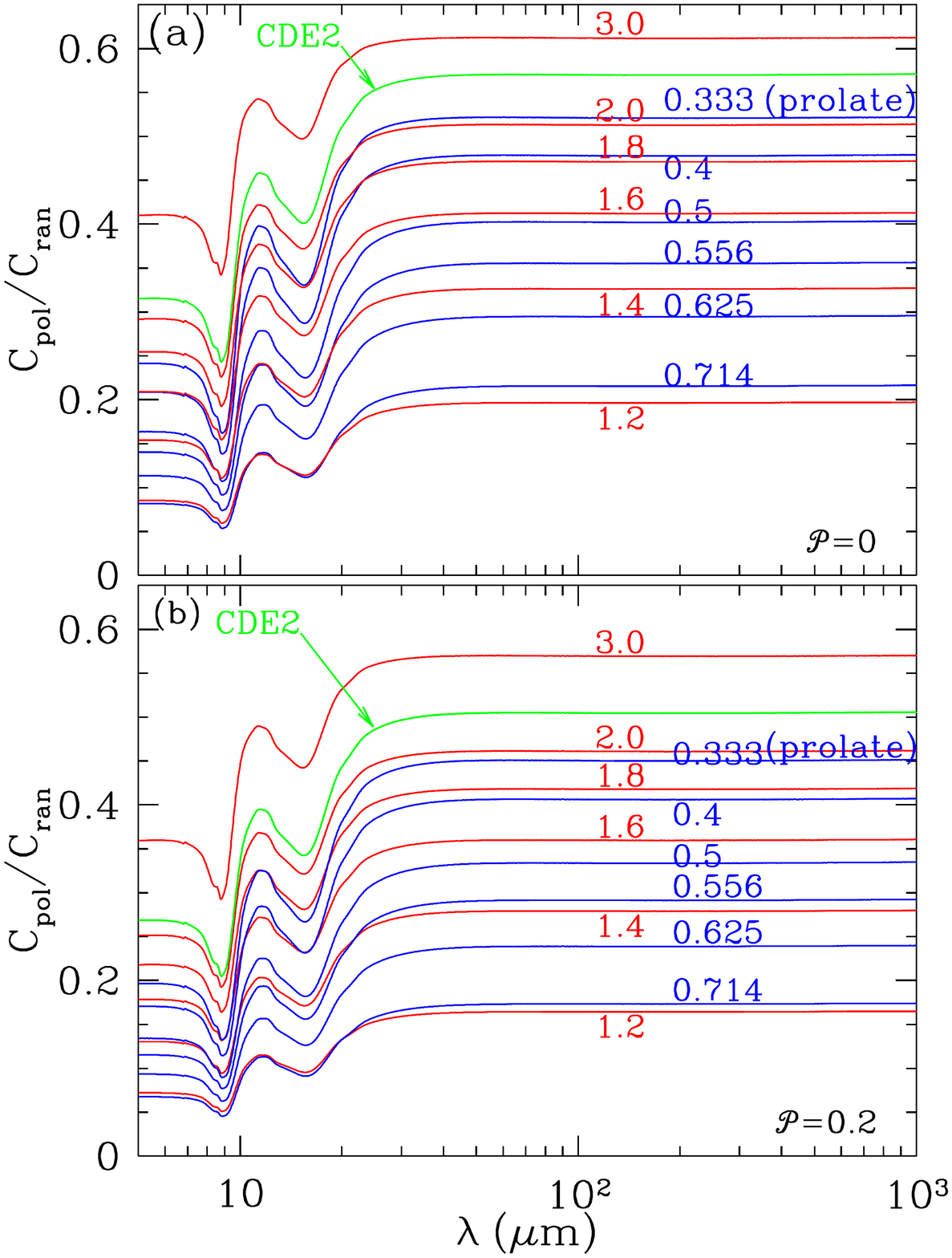}
\caption{\label{fig:cpol/cran}\footnotesize
         (a) $\Cpol/\Cran$ for oblate (red)
         and prolate (blue)
         spheroids with varying axis ratio $b/a$,
         for dielectric functions with porosity
         $\poro=0$ and $\fFe=0$.
         (b) Same as (a), but for $\poro=0.2$.
         For modest axis ratios (e.g., $0.5 \ltsim b/a \ltsim 1.6$),
         changing the porosity from $\poro=0$ to $0.2$ reduces
         $\Cpol/\Cran$ by $\sim$$20\%$.
         }
\end{center}
\vspace*{-0.3cm}
\end{figure}

\subsection{\label{sec:fir and submm polarization}
         Far-Infrared and Submm Polarization}

Optically-thin thermal emission from \astrodust\ grains has
polarization fraction \citep{Draine+Hensley_2020a}
\beqa \label{eq:optthinpol}
p_{{\rm em},\lambda}^\Ad 
&=& 
\left[\frac{C_x-C_y}{C_x+C_y}\right]_\lambda^\Ad =
\left[\frac{\Cpol}{\Cran}\right]_\lambda^\Ad
\times
\frac{\langle\falign^\Ad\rangle\sin^2\gamma}
{1 -\left[\Cpol/\Cran\right]_\lambda^\Ad\langle\falign^\Ad\rangle
\left(\sin^2\gamma-\frac{2}{3}\right)}
\\
&\approx&\left[\frac{\Cpol}{\Cran}\right]_\lambda^\Ad
\times
\langle\falign^\Ad\rangle\sin^2\gamma
~~~.
\eeqa
At long wavelengths ($\lambda > 50\micron$), the large values of $\epsilon_1$
(see Figs.\ \ref{fig:epsilon_vary_b/a}, \ref{fig:epsilon_vary_b/a_P0.20})
result in relatively large values of $[\Cpol/\Cran]_\lambda^\Ad$.
Fig.\ \ref{fig:cpol/cran} shows $[\Cpol/\Cran]_\lambda^\Ad$ 
from $\lambda=5\micron$ out to $1\mm$ for
different values of $b/a$, for $\poro=0$ and $0.2$.
We have seen above that the silicate polarization profile can be fit well by
prolate spheroids with $b/a\ltsim0.5$.
For $b/a=0.5$, Fig.\ \ref{fig:cpol/cran} shows that
such silicate-bearing grains have
$[\Cpol/\Cran]_\lambda^\Ad\approx 0.30$ at 
submm wavelengths (for $\poro=0.2$).

{\it Planck} has measured the polarization of the 
$850\micron$ emission from the diffuse ISM.
The highest fractional polarization observed (the 99.9th percentile) is
estimated to be $0.220_{-.014}^{+0.035}$ 
\citep{Planck_2018_XII}.
For optimal viewing geometry ($\sin^2\gamma=1$).
such high polarizations could be produced by $b/a=0.5$ prolate spheroids
with high degrees of alignment ($\langle\falign^\Ad\rangle\approx 0.7$).


\section{\label{sec:discuss}
         Discussion}

\subsection{Shape of the Silicate-Bearing Grains}

When this study was initiated, it was hoped that the shape of the
silicate polarization profile would strongly favor certain grain shapes and
porosities.  However, we find that, for self-consistent dielectric functions,
the {\it shape} of the
polarization profile depends only weakly on shape and porosity.
Existing observations 
of only the extinction and polarization of the 10$\micron$
silicate feature appear to allow both prolate and oblate shapes.
According to Figure \ref{fig:chi2_vs_shape},
the best-fitting shape for $\poro=0$ is an extreme
prolate shape, $b/a=3$, but we find that oblate shapes with
$b/a\approx1.4$ or $1.6$ provide fits that are almost as good.
Other polarization data -- the strength of 
starlight polarization at optical wavelengths, and
of polarized thermal emission at submm wavelengths -- 
provide additional constraints on grain shape and porosity.


The CDE2 and ERCDE shape distributions also 
give $10\micron$ polarization profiles
that are in good agreement with the observed profile.  However,
these shape distributions are not practical for modeling interstellar grains,
because
to model the polarization of starlight we require extinction
cross sections at 
wavelengths comparable to the grain size.
Such calculations for ellipsoids are time-consuming, and even for a single
grain mass and wavelength one would need to
sample many different ellipsoidal shapes 
in order to represent a continuous distribution
of ellipsoidal shapes such as CDE2 or ERCDE.

\subsection{Porosity}

As seen above (Fig.\ \ref{fig:chi2_vs_shape}), 
the porosity is only minimally constrained by the 
{\it shape} of the
$10\micron$ polarization profile.
\citet{Draine+Hensley_2020c} show that 
the strongest constraint on porosity is from the {\it strength} of the
polarization (both polarization of starlight, and polarization of
submm emission), which limit the porosity to a maximum value
that depends on grain shape.
Extreme porosities are ruled out.

\subsection{The Silicate Absorption Profile}

The present investigation was based upon the best available
observational determinations of the absorption by interstellar dust
over a wavelength range covering the 9.7$\micron$ and $18\micron$
silicate features.  The best data are from the Spitzer IRS instrument,
observing Cyg OB2-12
\citep{Ardila+VanDyk+Makowiecki+etal_2010,
       Fogerty+Forrest+Watson+etal_2016,
       Hensley+Draine_2020a}.
A weak broad absorption feature
at $11.1\pm0.10\micron$ is seen in the spectra of a number of heavily obscured
objects
\citep{Wright+Duy+Lawson_2016,Duy+Wright+Fujiyoshi+etal_2020},
which is interpreted as an Mg-rich
olivine, possibly the end-member forsterite Mg$_2$SiO$_4$.
Our determination of the silicate profile toward Cyg OB2-12 does not include
any obvious feature near $11.1\micron$, but this may be the result of
uncertainties in estimation of the underlying emission from Cyg OB2-12.

With the termination
of the cold Spitzer mission, spectrophotometry in this wavelength
range must be done through the 8--14$\micron$ atmospheric window until
the advent of 
the Mid Infrared Instrument (MIRI) on the James Webb Space Telescope
(JWST), which should provide
$\lambda/\Delta\lambda=5000$ spectroscopy over the 5--28$\micron$ range.

Spectrophotometry with MIRI is expected to 
significantly improve our knowledge of
the silicate absorption in the 5--28$\micron$ range.
If spectrophotometry with MIRI 
confirms the ``wiggles'' in the Spitzer IRS spectra
and shows them to be due to the interstellar extinction (rather than
the stellar atmospheres), these features will
provide clues to the composition and structure of the interstellar
amorphous silicate material, which laboratory synthesis could attempt
to replicate.

\subsection{Polarization in the Silicate Features}

For grain shapes ($a/b\approx0.5$) and likely degree of alignment
($\langle\falign^\Ad\rangle\approx0.7$) 
that appear 
to be consistent with (1) the silicate polarization profile
(2) the optical-UV polarization of starlight,
and
(3) typical levels of submm polarization observed by
{\it Planck},
we expect the silicate feature, observed in extinction, to produce
polarization.
For $b/a=0.5$ and $\poro=0.2$ we have
\beq
\frac{p(10\micron)}{\Delta\tau_{9.7}} \approx 
0.29\langle\falign^\Ad\rangle\sin^2\gamma
\eeq
[see eq.\ (\ref{eq:pol/tau},\ref{eq:Cpol10/Cran for b/a=1.4})].
If $\langle\falign^\Ad\rangle\approx 0.7$, we would expect to find values of
$p(10\micron)/\Delta\tau_{9.7}$ as large as $\sim0.20$ for the most favorable
geometries ($\sin^2\gamma\rightarrow 1$).

\citet{Smith+Wright+Aitken+etal_2000} present the results of 10$\micron$
spectroscopy and polarimetry of 55 infrared sources.
In their sample, 
the highest observed values of $p(10\micron)/\deltataupk$ are $0.038$
(the BN object in OMC1) and $0.035$ (Galactic Center source GCS\,IV
at $(\ell,b)=(0.16^\circ,-0.06^\circ)$).
Why have higher values of $p(10\micron)/\Delta\tau_{9.7}$ not been seen?

Most of the 55 sources in the \citet{Smith+Wright+Aitken+etal_2000} atlas
are infrared sources embedded in star-forming molecular clouds (e.g.,
the BN object) -- for which our diffuse ISM-based 
estimates of $b/a$ and $\langle\falign^\Ad\rangle$ may not apply.
The polarizing efficiency of grains in dark clouds is observed to
be reduced \citep{Whittet+Hough+Lazarian+Hoang_2008}.
The grain alignment in dark clouds may not achieve
values of $\langle\falign^\Ad\rangle$ 
as high as observed in the diffuse ISM; in addition, the grain
shapes within these clouds might conceivably be less elongated
than in the diffuse ISM.  Furthermore, the magnetic fields
in these turbulent clouds may have
an appreciable disordered component, which would lower the
degree of polarization.

The Galactic Center source GCS~IV is located in an area of the
sky where Planck found a relatively low fractional polarization at
$850\micron$ \citep{Planck_int_results_xix_2015} -- the $850\micron$
polarization appears to be only $\sim$$2\%$, small compared to the
highest values \citep[$\gtsim$20\%:][]{Planck_2018_XII}
observed by {\it Planck},
presumably
indicating some combination of low degrees of alignment and disordered and/or
unfavorable magnetic field geometry in the dust-containing regions
(i.e., low effective values of $\langle\falign^\Ad\rangle\sin^2\gamma$).
Given the low fractional ($\sim$$2\%$) polarization of the diffuse
850$\micron$ emission in this field,
we would expect a relatively low value of $p(10\micron)/\Delta\tau_{9.7}$
on the sightline to GCS\,IV, as observed.

It would be of great value to have measurements of
$p(10\micron)/\Delta\tau_{9.7}$ in regions where {\it Planck} finds
high values of $p(850\micron)$, ideally using sources where the bulk
of the extinction arises from the diffuse ISM (rather than being local
to the source).
For Cyg OB2-12 we predict $p(10\micron)\approx (2.2\pm0.3)$\%
\citep{Draine+Hensley_2020c}, which should be measurable.

This paper has stressed the value of mid-infrared
spectropolarimetry to constrain
grain geometry and the dielectric function of the silicate material.
The atlas of polarized spectra published by
\citet{Smith+Wright+Aitken+etal_2000} remains the ``state-of-the-art'',
but it is sobering to note in 2020
that all of the observations reported in
that paper were taken prior to February 1993.
It is regrettable that polarimetric capabilites
were not included in either the Spitzer or JWST instrument suites,
and disappointing that most
large telescopes that have come into operation
over the past three decades have not been equipped with
IR spectropolarimetric instruments.
The silicate polarization profile does depend on grain shape, but
the best available spectropolarimetry (toward WR-48a and WR-112=AFGL2104)
has estimated
observational uncertainties that permit both oblate and
prolate shapes.
Spectropolarimetry with factor of $\sim2$ 
higher signal/noise could remove this ambiguity, and reject the prolate
or oblate shapes (or perhaps both!).
The Canaricam instrument on the Gran Telescopio Canarias
\citep{Packham+Hough+Telesco_2005} appears to be the most
capable instrument currently available for mid-infrared spectropolarimetry,
and this instrument may be able to provide improved measurements of
the wavelength-dependence of the silicate polarization.

It would also be of great value to have spectropolarimety 
extending shortward and longward of the
atmospheric windows.  
Given the absence of polarimetry on JWST,
a mid-IR spectropolarimeter on a high altitude
balloon or on SOFIA \citep{Packham+Escuti+Boreman+etal_2008}
could provide valuable and unique constraints on models for the
silicate-bearing interstellar grain population.


\section{\label{sec:summary}
         Summary}

The principal findings of this study are as follows:
\begin{enumerate}
\item Observations of the far-infrared polarization fraction
from $250\micron$ to $3\mm$ favor a model for interstellar dust where the
opacity is dominated by a single grain material, which we term ``\astrodust.''
\item We use the observed 5--35$\micron$ extinction 
toward Cyg OB2-12
\citep{Hensley+Draine_2020a},
the opacity inferred from the observed far-infrared and
submm emission
\citep{Hensley+Draine_2020b}, 
plus other constraints (including depletion-based
estimates of grain mass and volume), to derive the
effective dielectric function $\epsilon(\lambda)$ for
\astrodust\ in the diffuse ISM.
\Astrodust\ material is $\sim$50\% amorphous silicate, but also
incorporates $\sim$$25\%$
of the carbon in diffuse 
clouds (an additional $\sim$$12\%$
of the carbon is in a population of PAH nanoparticles).
The hydrocarbon material in the \astrodust\ grains is
assumed to be responsible for the $3.4\micron$ absorption feature
in the diffuse ISM.
The PAHs are assumed to be randomly-oriented; all of the
polarized extinction and emission from the diffuse ISM is attributed to the
\astrodust\ grains.

The derived $\epsilon(\lambda)$ depends on assumptions including
the shape and porosity of the \astrodust\ grains.
The ``\astrodust'' dielectric functions obtained here will be made available
in computer-readable form at the time of publication.

\item For different grain shape and
$\poro$, 
we calculate the polarization cross section,
and predict the wavelength dependence of polarization from 
the near-IR (3.4$\micron$) to mm wavelengths.

\item Comparing the predicted $\Cpol(\lambda)/V$
with the $8$--$13\micron$ polarization observed toward two WR stars
\citep{Wright+Aitken+Smith+etal_2002}, we
conclude that:
\begin{itemize} 
   \item The silicate-bearing grains
    can be approximated by either oblate or prolate spheroids: the existing
    polarimetric data do not definitively 
    discriminate between oblate or prolate shapes.
   \item With existing spectropolarimetry, 
    the {\it shape} of the polarization profile is
    consistent with a wide range of porosities.
    Limits on porosity must come from considerations of
    the {\it strength} of the polarization, both the starlight polarization
    and the degree of polarization of the submm emission
    \citep{Draine+Hensley_2020c}.
   \item Future spectropolarimetry with improved
     signal/noise ratio has the potential
     to narrow the range of allowed shapes and porosities.
\end{itemize}
\item CH absorption in the aligned \astrodust\ grains will produce
a polarized extinction feature at 3.4$\micron$ [see Eq.\ (\ref{eq:p34})]:
\beq
\frac{[\Delta p]_{3.4\mu{\rm m}}}{p(10\micron)}
\approx 
0.040\times
\frac
{\langle\falign^{\rm CH}\rangle}
{\langle\falign^\Ad\rangle}
~~,
\eeq
where $\langle \falign^\Ad\rangle$ is the mass-averaged alignment of the
\astrodust\ grains, and $\langle \falign^{\rm CH}\rangle$ is the
3.4$\micron$ absorption feature-weighted fractional alignment.
If the $3.4\micron$ absorbers are
preferentially located near the grain surfaces, 
then the fact that small grains
are minimally aligned implies
$\langle \falign^{\rm CH}\rangle/\langle \falign^\Ad\rangle < 1$, with
the actual ratio
dependent on the degree to which the
$3.4\micron$ absorption is concentrated near
grain surfaces.
For 
$\langle \falign^{\rm CH}\rangle/\langle \falign^\Ad\rangle \ltsim 0.4$,
the predicted $3.4\micron$ feature polarization is consistent with the observed
upper limits on $3.4\micron$ feature polarization
\citep{Chiar+Adamson+Whittet+etal_2006}.
Such values would be consistent with the $3.4\micron$ absorption arising
in surface zones of thickness $\Delta\ltsim 0.01\micron$.
More sensitive 3.4$\micron$ polarimetry should be able to detect the
predicted polarization in the 3.4$\micron$ feature.
\end{enumerate}

\acknowledgments
This work was supported in part by NSF grants AST-1408723 and
AST-1908123, and
carried out in part at the Jet Propulsion Laboratory, California
Institute of Technology, under a contract with the National Aeronautics
and Space Administration.
We thank 
Shane Fogerty and Charles Poteet for providing data on Cyg OB2-12
in advance of publication.
We also thank
Megan Bedell,
Francois Boulanger,
Vincent Guillet,
Biwei Jiang,
Charles Telesco,
and
Chris Wright
for helpful discussions.

\appendix

\section{\label{app:mag dipole}
         Magnetic Dipole Absorption by Metallic Fe inclusions}

We assume that silicate-bearing grains contain ferromagnetic
metallic Fe inclusions, randomly-distributed throughout the volume,
with volume filling factor [see Eq.\ (\ref{eq:fvfe})]
\beq
f_{\rm fill} \approx \left(1-\poro\right)\frac{0.12\fFe}{1-0.22\fFe}
~~~.
\eeq
where $\fFe$ is the fraction of the iron in metallic inclusions.
For a spheroidal grain with principal axes $\bahat_j$,
the magnetic-dipole
contribution to the absorption is, for $\bH\parallel \bahat_j$,
\beq
C_{{\rm abs},j}^{\rm mag} =
\frac{\omega V}{c} 
{\rm Im} \left[\frac{\mu_{\rm eff} -1}{1+L_j(\mu_{\rm eff}-1)}\right]
~,
\eeq
where $\mu_{\rm eff}$, the effective-medium estimate for the magnetic
permeability of the grain material, is given by
\beq \label{eq:mu_eff and D_b}
\mu_{\rm eff} = 
1 + 4\pi 
\frac{f_{\rm fill}(\chi_++\chi_-)/3}
     {1-D_b f_{\rm fill}(\chi_++\chi_-)/3}
~,
\eeq
where 
\beq
\chi_\pm = \frac{\omega_{\rm M}}{\omega_0 - i\alpha_G\omega \mp \omega}
~,
\eeq
\citep{Draine+Hensley_2013}.
$D_b$ in Eq.\ (\ref{eq:mu_eff and D_b})
is the demagnetization factor perpendicular to the long
axis of a prolate inclusion; for 2:1 prolate inclusions,
$D_b = 5.193$.
$\alpha_G\approx 0.2$ is the Gilbert damping parameter,
and $\omega_{\rm M}/2\pi=4.91\GHz$ for Fe \citep{Draine+Hensley_2013}.
The parameter $\omega_0$ depends on
the shape of the ferromagnetic inclusions, 
varying from $\omega_0/2\pi=1.53\GHz$ for
spherical inclusions, to $\omega_0/2\pi=16.3\GHz$ for 2:1 prolate inclusions
\citep[see Table 2 of][]{Draine+Hensley_2013}.

\section{\label{app:optical-uv}
         Dielectric Function for \Astrodust\ Material from Optical to
         X-ray Wavelengths}

We seek a provisional dielectric function 
$\epsilon_{\rm mat}^{\rm ox}(\lambda)$ for \astrodust\ material
at wavelengths $\lambda < \lambda_0 \equiv 1\micron$ 
($\hbar\omega > \hbar\omega_0\equiv 1.24\eV$).
Let $\epsilon^{\rm ox}(\lambda)$ be the dielectric function for solid
\astrodust\ material with no voids and no Fe inclusions.
We proceed as follows:
\begin{enumerate}
\item Because we expect the silicate-bearing grains to account for
most of the far-infrared emission, these grains should be absorptive at the
energies $0.5\ltsim 5\eV$ where the interstellar starlight
energy density $u_\nu$ 
peaks [the \citet{Mathis+Mezger+Panagia_1983} estimate 
for the radiation field has $\nu u_\nu$ peaking at $h\nu=1.32\eV$].
We assume 
\beq
{\rm Im}(\epsilon_{\rm mat}^{\rm ox}) \approx 0.20 .
~~~{\rm for}~~ 1\ltsim E \ltsim 5\eV
~.
\eeq
This is about twice as large as was 
assumed for astrosilicate by \citet{Draine+Lee_1984,Li+Draine_2001b,
Draine_2003c,Draine+Li_2007}.
Adoption of a larger value of ${\rm Im}(\epsilon)$ at optical
wavelengths is motivated by the finding 
\citep{Planck_DL07_2016} that
the far-infrared power per unit starlight extinction 
is about twice as large as had been estimated for the
\citet{Draine+Li_2007} dust model and the MMP83 estimate for the
starlight intensity.
\item For computational convenience in obtaining
${\rm Re}(\epsilon_{\rm mat}^{\rm ox})$ using the Kramers-Kronig relations, 
we extend $\epsilon_{\rm mat}^{\rm ox}$
into the IR:
\beq \label{eq:ox extrap to ir}
{\rm Im}(\epsilon_{\rm mat}^{\rm ox}) = 0.20 (1\micron/\lambda)^2
~~~{\rm for}~\lambda > 1\micron
~.
\eeq
The assumed form of Eq.\ (\ref{eq:ox extrap to ir}) for $\lambda > 1\micron$
does not affect our final result, as we
subsequently use $\epsilon_{\rm mat}^{\rm ir}$ to add to (or subtract from)
$\epsilon_{\rm mat}^{\rm ox}$ at $\lambda > 1\micron$
to comply with observational constraints
[see Eq.\ (\ref{eq:oscillator model})].
\begin{figure}[ht]
\begin{center}
\includegraphics[angle=0,width=8.0cm,
                 clip=true,trim=0.5cm 0.5cm 0.5cm 0.5cm]
{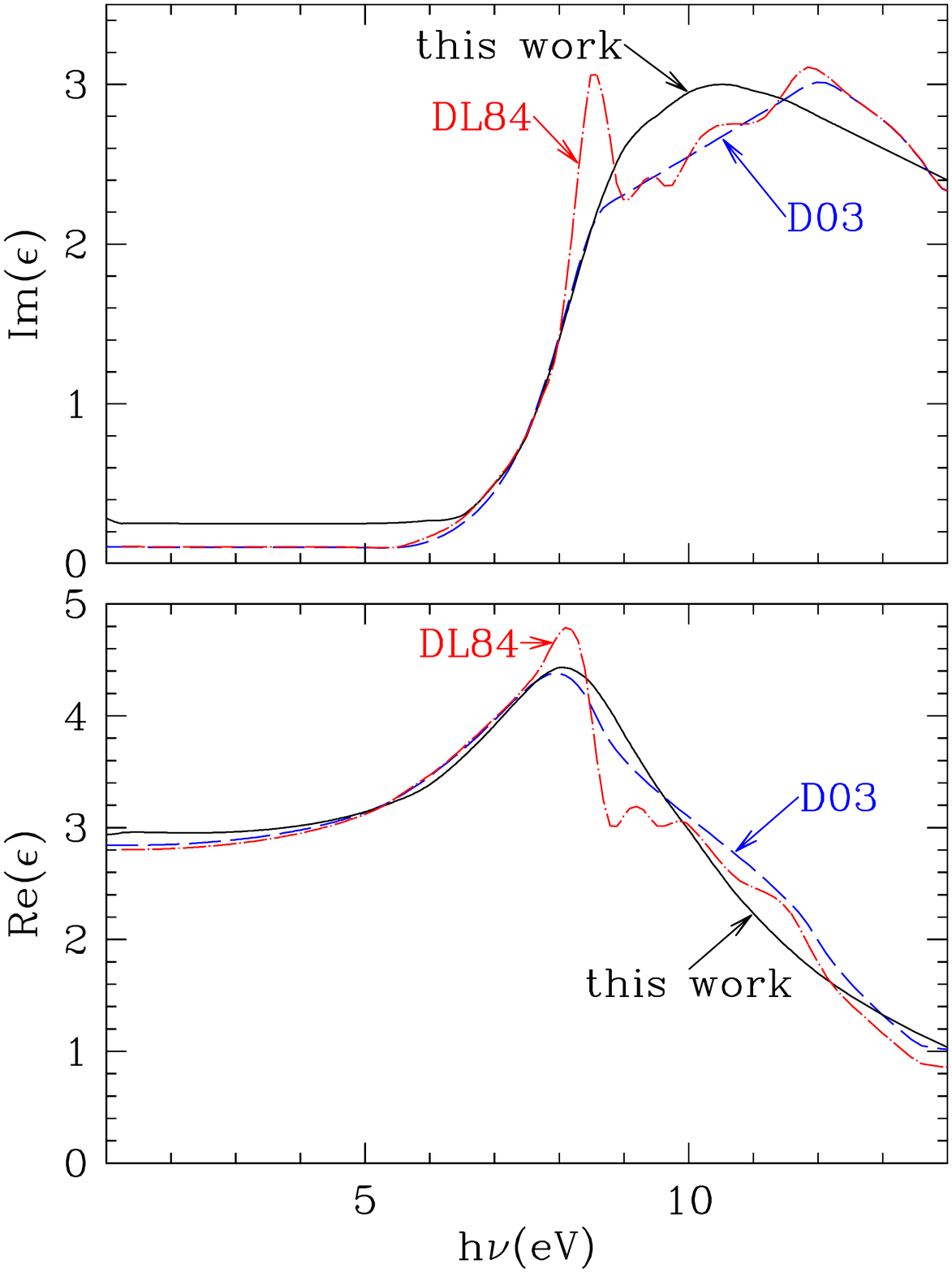}
\includegraphics[angle=0,width=8.0cm,
                 clip=true,trim=0.5cm 0.5cm 0.5cm 0.5cm]
{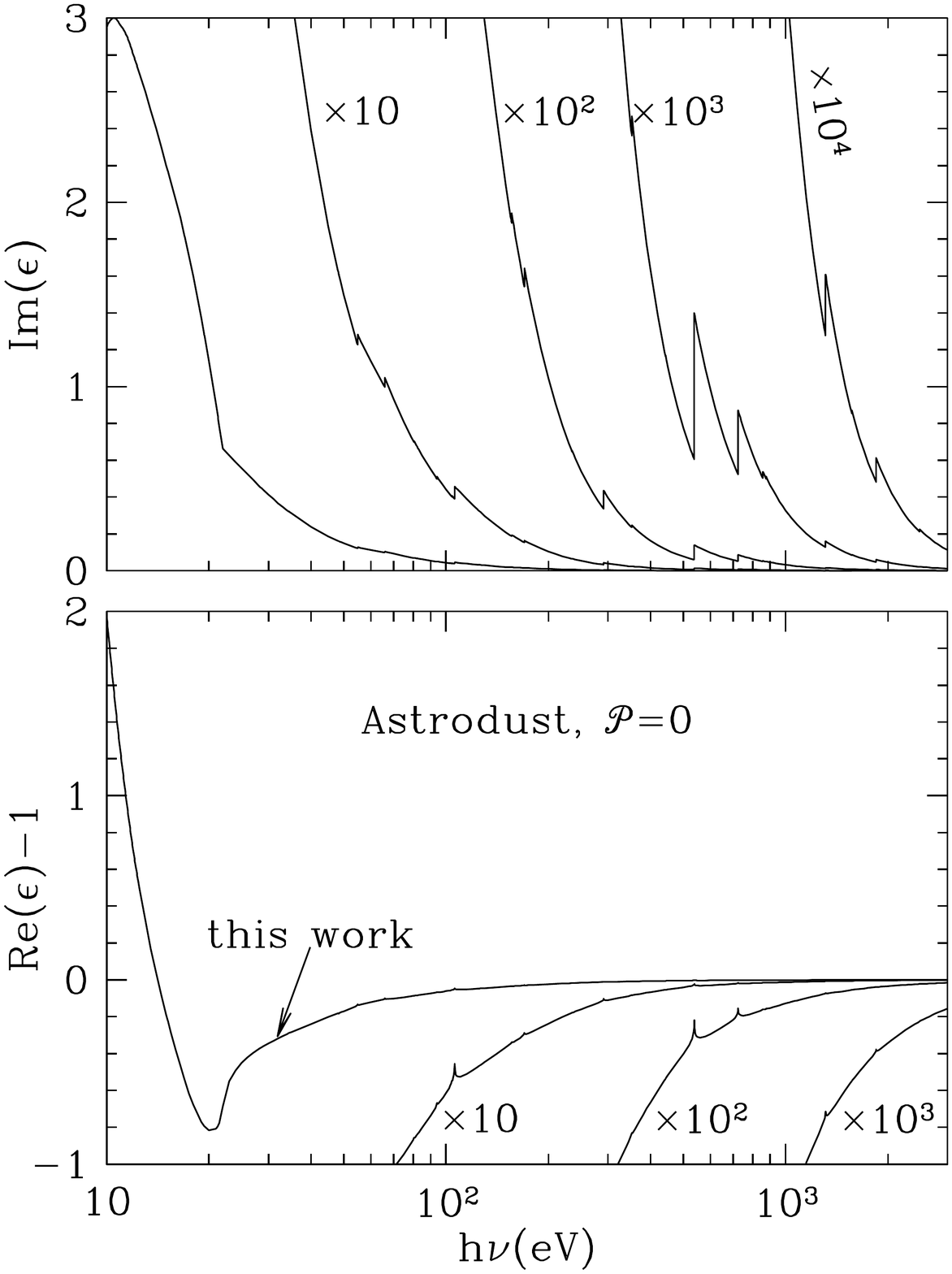}
\caption{\footnotesize \label{fig:uv}
Solid curves: real and imaginary part of 
$\epsilon$ for \astrodust\ with ${\poro} = 0$.
Dashed curves (D03): silicate dielectric function from
\citet{Draine_2003c}.
X-ray absorption edges are seen at 528\eV (O\,K), 700\eV (Fe\,L), ...
Dot-dashed curves (DL84): astrosilicate dielectric function from
\citet{Draine+Lee_1984}.}
\end{center}
\vspace*{-0.3cm}
\end{figure}

\item For $h\nu \gtsim 8\eV$, we adopt a dielectric function similar
to the ``astrosilicate'' dielectric function of \citep{Draine+Lee_1984,
Draine_2003c}, with a rapid
increase in absorption when the photon energy is large enough to excite
valence-band electrons to the (empty) conduction band.
\item Let $\alpha_{\rm mat}$ 
be the attenuation coefficient for radiation propagating
through the matrix material.  The attenuation coefficient is directly
related to the imaginary part of the refractive index $m$:
\beq
\alpha_{\rm mat} = \frac{4\pi}{\lambda} {\rm Im}(m)
~~~,
\eeq
where $\lambda$ is the wavelength {\it in vacuo}.
Because ${\rm Im}(\epsilon) = {\rm Im}(m^2) = 2 {\rm Re}(m){\rm Im}(m)$,
we have
\beq
{\rm Im}(\epsilon_{\rm mat}^{\rm ox}) =
\frac{\lambda}{2\pi}
{\rm Re}(m)
\alpha_{\rm mat}
~~~.
\eeq
At high energies
we expect ${\rm Re}(m)\approx 1$, and the
absorption per atom is expected to be well-approximated by the
mean photoelectric absorption cross sections $\sigma_{{\rm pe},j}$
of isolated atoms of element $j$,
\beq \label{eq:xray absorption}
\alpha_{\rm mat}^{\rm ox} \approx \sum_j n_j \sigma_{{\rm pe},j}(h\nu)
~~~,
\eeq
where $n_j$ is the number density of atoms $j$.
Therefore, at high energies ($E > 22\eV$) we take
\beq \label{eq:B5}
{\rm Im}(\epsilon_{\rm mat}^{\rm ox}) \approx 
\frac{\lambda}{2\pi} \sum_j n_j \sigma_{{\rm pe},j}
~.
\eeq
We approximate the photoelectric absorption cross sections
$\sigma_{{\rm pe},j}(h\nu)$ by the photoionization fitting
functions estimated for inner shell electrons
by \citet{Verner+Yakovlev_1995} and for outer-shell electrons
by \citet{Verner+Ferland+Korista+Yakovlev_1996},
implemented in the fortran code {\tt phfit2.f} 
written by D.A. Verner.

For $15<E<22\eV$, we adopt ${\rm Im}(\epsilon^{\rm ox})$ that connects 
our adopted $E<15\eV$ absorption (see Figure \ref{fig:uv}) to
the absorption (\ref{eq:xray absorption}) for $E>22\eV$.\footnote{%
   The adopted ${\rm Im}(\epsilon)$ for $E<22\eV$
   (see Figure \ref{fig:uv})
   is somewhat larger than would be found by evaluating
   Eq.\ (\ref{eq:xray absorption}),
   as expected because ${\rm Re}(m)>1$ over this range.}
\item The real part of the dielectric function
is obtained using the Kramers-Kronig relations:
\beq
{\rm Re}\left(\epsilon_{\rm mat}^{\rm ox}(\omega)\right) = 
1 + \frac{2}{\pi}\int_0^\infty \frac{{\rm Im}
\left(\epsilon_{\rm mat}^{\rm ox}(x)\right)}
{x^2 - \omega^2} ~x dx
~.
\eeq
\item For nonzero porosity $\poro$ and/or material with metallic Fe inclusions,
we apply Bruggeman effective medium theory, and take the effective
dielectric function $\epsilon_{\rm eff}$ to satisfy
\beq
0 = 
(1-\poro-\fvFe)
\left(
\frac{\epsilon_{\rm mat}-\epsilon_{\rm eff}}
     {\epsilon_{\rm mat}+2\epsilon_{\rm eff}}
\right)+
\poro
\left(
\frac{1-\epsilon_{\rm eff}}
     {1+2\epsilon_{\rm eff}}
\right)+
\fvFe
\left(\frac{\epsilon_{\rm Fe}-\epsilon_{\rm eff}}
     {\epsilon_{\rm Fe}+2\epsilon_{\rm eff}}
\right)
~,~~~
\eeq
where $\fvFe$ is the fraction of the solid volume contribued by Fe
inclusions [see Eq.\ (\ref{eq:fvfe})].
The Fe inclusions and vacuum pores are both taken to be spherical
\citep{Bohren+Huffman_1983}.

\item By construction, $\epsilon^{\rm ox}(\lambda)$
satisfies the Kramers-Kronig relations.
In addition, we verify that our final $\epsilon$ 
satisfies the ``{\it f}-sum rules''
\citep{Altarelli+Dexter+Nussenzveig+Smith_1972}
\beq
\frac{m_e}{2\pi^2 e^2}
\int_0^\infty \omega {\rm Im}(\epsilon)d\omega =
\sum_j n_j
\eeq
\beq
\frac{m_e}{\pi^2 e^2}
\int_0^\infty \omega {\rm Im}(\sqrt{\epsilon}) d\omega =
\sum_j n_j
~~~,
\eeq
where $n_j$ is the number density of electrons in atomic shell $j$.
Figure \ref{fig:neff} shows
\beq \label{eq:neff_a}
n_{\rm eff}(E) \equiv 
\frac{m_e}{2\pi^2e^2}
\int_0^{E/\hbar} \omega {\rm Im}(\epsilon) d\omega
\eeq
and
\beq \label{eq:neff_b}
n_{\rm eff}(E) \equiv 
\frac{m_e}{\pi^2e^2}
\int_0^{E/\hbar} \omega {\rm Im}(\sqrt{\epsilon}) d\omega
~~~.
\eeq
\end{enumerate}
\begin{figure}
\begin{center}
\includegraphics[angle=270,width=12.0cm,clip=true,
                 trim=0.0cm 0.0cm 0.0cm 0.0cm]
{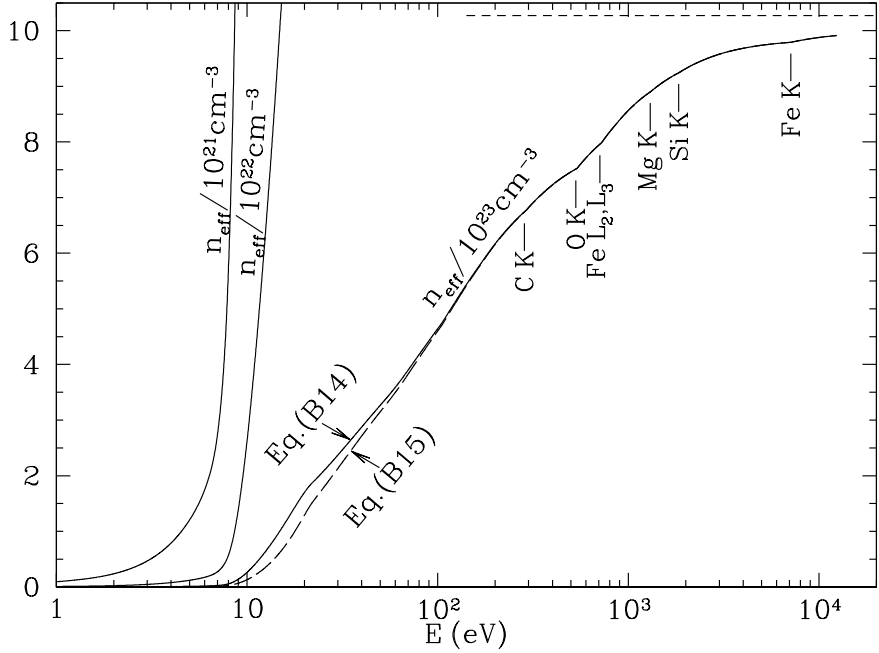}
\caption{\label{fig:neff} \footnotesize
$n_{\rm eff}(E)$ evaluated using Eq.\ (\ref{eq:neff_a}) and
(\ref{eq:neff_b}) using the \astrodust\ dielectric function for $\poro=0$.
The expected asymptotic limit $n_{\rm eff}=1.03\times10^{24}\cm^{-3}$
for the $F_\star=0.5$ composition in Table \ref{tab:composition}
is shown as a dashed line.}
\end{center}
\end{figure}

\section{\label{app:ferromagnetic}
         Dielectric Function for Grains with Ferromagnetic Inclusions}

\begin{figure}[h]
\begin{center}
\includegraphics[angle=270,width=12.0cm,clip=true,
                 trim=0.0cm 0.0cm 0.0cm 0.0cm]
{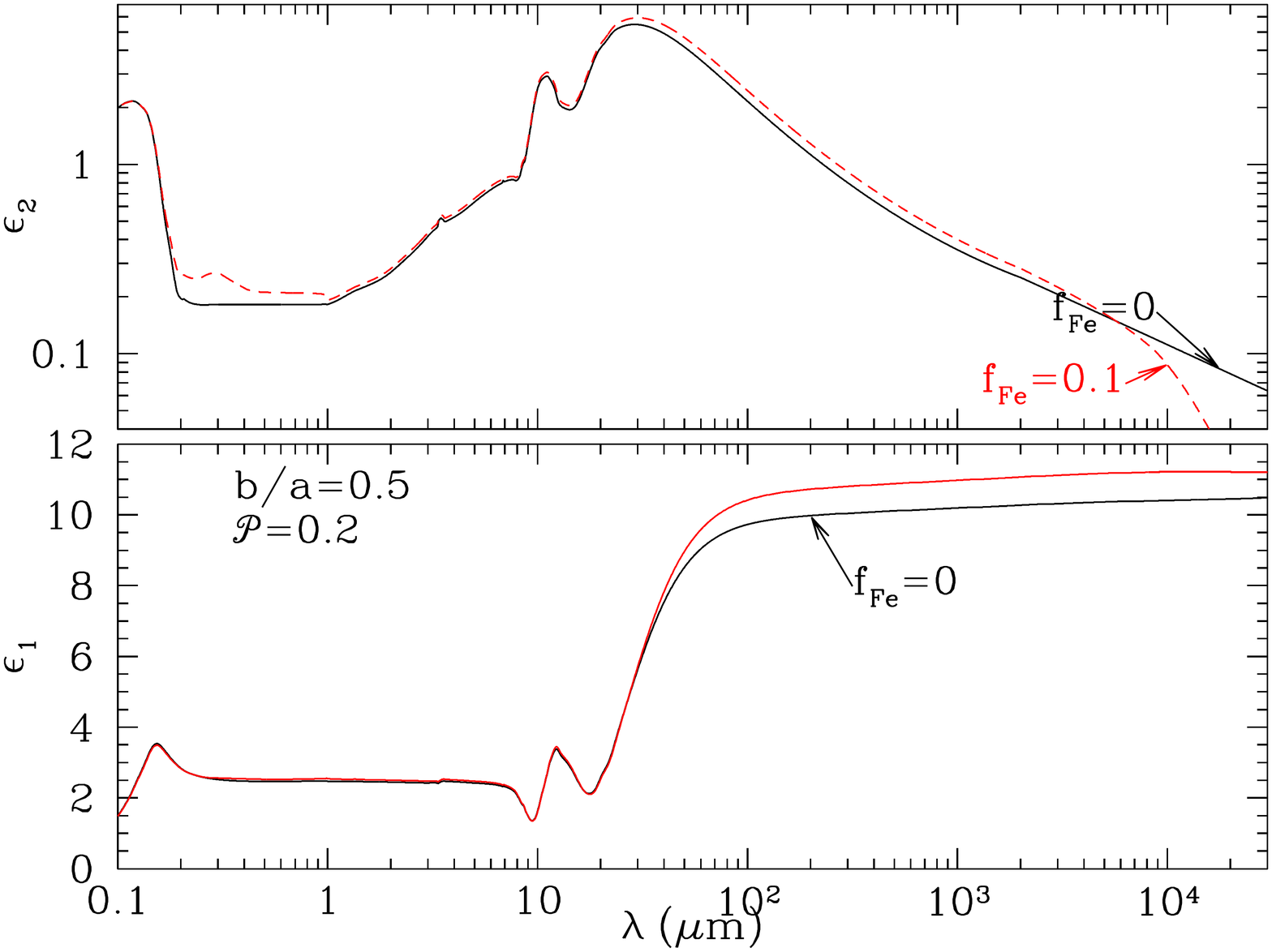}
\caption{\footnotesize \label{fig:vary fFe}
         Dielectric function $\epsilon$ for $b/a=0.5$ prolate spheroids,
         for $\poro=0.2$, and
         $\fFe=0$ and $0.1$, where $\fFe$ is the fraction of the Fe
         in
         metallic iron inclusions.
         The dielectric functions are essentially unchanged for
         $1\ltsim \lambda \ltsim 50\micron$, with only small
         changes for $50 \ltsim \lambda \ltsim 500\micron$.
         }
\end{center}
\vspace*{-0.3cm}
\end{figure}

The dielectric functions discussed above were calculated assuming
that any magnetic dipole absorption from ferromagnetic inclusions in the
silicate grains is negligible.
Figure \ref{fig:vary fFe}
show how the dielectric function would need to be modified if,
in fact, some fraction of the submm emission from silicate grains
is magnetic dipole radiation from ferromagnetic inclusions.
Fe inclusions increase the dust opacity in the visible-UV.
The magnetic effects become appreciable 
only at very long wavelengths,
$\lambda \gtsim 5\mm$ ($\nu < 60\GHz$).

\bibliography{/u/draine/work/libe/btdrefs}

\begin{thebibliography}{}
\expandafter\ifx\csname natexlab\endcsname\relax\def\natexlab#1{#1}\fi

\bibitem[{{Aitken} {et~al.}(1989){Aitken}, {Smith}, \&
  {Roche}}]{Aitken+Smith+Roche_1989}
{Aitken}, D.~K., {Smith}, C.~H., \& {Roche}, P.~F. 1989, \mnras, 236, 919

\bibitem[{{Alexander} \& {Ferguson}(1994)}]{Alexander+Ferguson_1994}
{Alexander}, D.~B., \& {Ferguson}, J.~W. 1994, in Lecture Notes in Physics,
  Berlin Springer Verlag, Vol. 428, IAU Colloq. 146: Molecules in the Stellar
  Environment, ed. U.~G. {Jorgensen}, 149

\bibitem[{{Altarelli} {et~al.}(1972){Altarelli}, {Dexter}, {Nussenzveig}, \&
  {Smith}}]{Altarelli+Dexter+Nussenzveig+Smith_1972}
{Altarelli}, M., {Dexter}, D.~L., {Nussenzveig}, H.~M., \& {Smith}, D.~Y. 1972,
  \prb, 6, 4502

\bibitem[{{Ardila} {et~al.}(2010){Ardila}, {Van Dyk}, {Makowiecki}, {Stauffer},
  {Song}, {Rho}, {Fajardo-Acosta}, {Hoard}, \&
  {Wachter}}]{Ardila+VanDyk+Makowiecki+etal_2010}
{Ardila}, D.~R., {Van Dyk}, S.~D., {Makowiecki}, W., {et~al.} 2010, \apjs, 191,
  301

\bibitem[{{Ashton} {et~al.}(2018){Ashton}, {Ade}, {Angil{\`e}}, {Benton},
  {Devlin}, {Dober}, {Fissel}, {Fukui}, {Galitzki}, {Gandilo}, {Klein},
  {Korotkov}, {Li}, {Martin}, {Matthews}, {Moncelsi}, {Nakamura},
  {Netterfield}, {Novak}, {Pascale}, {Poidevin}, {Santos}, {Savini}, {Scott},
  {Shariff}, {Soler}, {Thomas}, {Tucker}, {Tucker}, \&
  {Ward-Thompson}}]{Ashton+Ade+Angile+etal_2018}
{Ashton}, P.~C., {Ade}, P.~A.~R., {Angil{\`e}}, F.~E., {et~al.} 2018, \apj,
  857, 10

\bibitem[{{Bohlin} {et~al.}(1978){Bohlin}, {Savage}, \&
  {Drake}}]{Bohlin+Savage+Drake_1978}
{Bohlin}, R.~C., {Savage}, B.~D., \& {Drake}, J.~F. 1978, \apj, 224, 132

\bibitem[{Bohren \& Huffman(1983)}]{Bohren+Huffman_1983}
Bohren, C.~F., \& Huffman, D.~R. 1983, Absorption and Scattering of Light by
  Small Particles (New York: Wiley)

\bibitem[{{Chiar} \& {Tielens}(2006)}]{Chiar+Tielens_2006}
{Chiar}, J.~E., \& {Tielens}, A.~G.~G.~M. 2006, \apj, 637, 774

\bibitem[{{Chiar} {et~al.}(2013){Chiar}, {Tielens}, {Adamson}, \&
  {Ricca}}]{Chiar+Tielens+Adamson+Ricca_2013}
{Chiar}, J.~E., {Tielens}, A.~G.~G.~M., {Adamson}, A.~J., \& {Ricca}, A. 2013,
  \apj, 770, 78

\bibitem[{{Chiar} {et~al.}(2000){Chiar}, {Tielens}, {Whittet}, {Schutte},
  {Boogert}, {Lutz}, {van Dishoeck}, \&
  {Bernstein}}]{Chiar+Tielens+Whittet+etal_2000}
{Chiar}, J.~E., {Tielens}, A.~G.~G.~M., {Whittet}, D.~C.~B., {et~al.} 2000,
  \apj, 537, 749

\bibitem[{{Chiar} {et~al.}(2006){Chiar}, {Adamson}, {Whittet}, {Chrysostomou},
  {Hough}, {Kerr}, {Mason}, {Poche}, \&
  {Wright}}]{Chiar+Adamson+Whittet+etal_2006}
{Chiar}, J.~E., {Adamson}, A.~J., {Whittet}, D.~C.~B., {et~al.} 2006, \apj,
  651, 268

\bibitem[{{Clayton} {et~al.}(2003){Clayton}, {Wolff}, {Sofia}, {Gordon}, \&
  {Misselt}}]{Clayton+Wolff+Sofia+etal_2003}
{Clayton}, G.~C., {Wolff}, M.~J., {Sofia}, U.~J., {Gordon}, K.~D., \&
  {Misselt}, K.~A. 2003, \apj, 588, 871

\bibitem[{{Compi{\`e}gne} {et~al.}(2011){Compi{\`e}gne}, {Verstraete}, {Jones},
  {Bernard}, {Boulanger}, {Flagey}, {Le Bourlot}, {Paradis}, \&
  {Ysard}}]{Compiegne+Verstraete+Jones+etal_2011}
{Compi{\`e}gne}, M., {Verstraete}, L., {Jones}, A., {et~al.} 2011, \aap, 525,
  A103

\bibitem[{{Diplas} \& {Savage}(1994)}]{Diplas+Savage_1994}
{Diplas}, A., \& {Savage}, B.~D. 1994, \apj, 427, 274

\bibitem[{{Draine}(1989)}]{Draine_1989a}
{Draine}, B.~T. 1989, in IAU Symp. 135: Interstellar Dust, ed. L.~Allamandola
  \& A.~Tielens (Dordrecht: Kluwer), 313--327

\bibitem[{{Draine}(2003)}]{Draine_2003c}
{Draine}, B.~T. 2003, \apj, 598, 1026

\bibitem[{{Draine}(2009)}]{Draine_2009b}
{Draine}, B.~T. 2009, in \aspcs\ 414, Cosmic Dust -- Near and Far, ed.
  {T.~Henning, E.~Gr{\"u}n, \& J.~Steinacker}, 453--472

\bibitem[{{Draine} \& {Fraisse}(2009)}]{Draine+Fraisse_2009}
{Draine}, B.~T., \& {Fraisse}, A.~A. 2009, \apj, 696, 1

\bibitem[{{Draine} \& {Hensley}(2013)}]{Draine+Hensley_2013}
{Draine}, B.~T., \& {Hensley}, B. 2013, \apj, 765, 159

\bibitem[{{Draine} \& {Hensley}(2017)}]{Draine+Hensley_2020a}
{Draine}, B.~T., \& {Hensley}, B.~S. 2017, ArXiv:1710.08968

\bibitem[{{Draine} \& {Hensley}(2020)}]{Draine+Hensley_2020c}
---. 2020, Using the Starlight Polarization Efficiency Integral to Constrain
  Shapes and Porosities of Interstellar Grains (in preparation)

\bibitem[{{Draine} \& {Lee}(1984)}]{Draine+Lee_1984}
{Draine}, B.~T., \& {Lee}, H.~M. 1984, \apj, 285, 89

\bibitem[{{Draine} \& {Li}(2007)}]{Draine+Li_2007}
{Draine}, B.~T., \& {Li}, A. 2007, \apj, 657, 810

\bibitem[{{Draine} \& {Weingartner}(1996)}]{Draine+Weingartner_1996}
{Draine}, B.~T., \& {Weingartner}, J.~C. 1996, \apj, 470, 551

\bibitem[{{Duy} {et~al.}(2020){Duy}, {Wright}, {Fujiyoshi}, {Glasse},
  {Siebenmorgen}, {Smith}, {Stecklum}, \&
  {Sterzik}}]{Duy+Wright+Fujiyoshi+etal_2020}
{Duy}, T.~D., {Wright}, C.~M., {Fujiyoshi}, T., {et~al.} 2020, \mnras,
  doi:10.1093/mnras/staa396

\bibitem[{{Fabian} {et~al.}(2001){Fabian}, {Henning}, {J{\"a}ger}, {Mutschke},
  {Dorschner}, \& {Wehrhan}}]{Fabian+Henning+Jager+etal_2001}
{Fabian}, D., {Henning}, T., {J{\"a}ger}, C., {et~al.} 2001, \aap, 378, 228

\bibitem[{{Fanciullo} {et~al.}(2017){Fanciullo}, {Guillet}, {Boulanger}, \&
  {Jones}}]{Fanciullo+Guillet+Boulanger+Jones_2017}
{Fanciullo}, L., {Guillet}, V., {Boulanger}, F., \& {Jones}, A.~P. 2017, \aap,
  602, A7

\bibitem[{{Fitzpatrick} \& {Massa}(1986)}]{Fitzpatrick+Massa_1986}
{Fitzpatrick}, E.~L., \& {Massa}, D. 1986, \apj, 307, 286

\bibitem[{{Fogerty} {et~al.}(2016){Fogerty}, {Forrest}, {Watson}, {Sargent}, \&
  {Koch}}]{Fogerty+Forrest+Watson+etal_2016}
{Fogerty}, S., {Forrest}, W., {Watson}, D.~M., {Sargent}, B.~A., \& {Koch}, I.
  2016, \apj, 830, 71

\bibitem[{{Fujiwara} {et~al.}(1978){Fujiwara}, {Kamimoto}, \&
  {Tsukamoto}}]{Fujiwara+Kamimoto+Tsukamoto_1978}
{Fujiwara}, A., {Kamimoto}, G., \& {Tsukamoto}, A. 1978, \nat, 272, 602

\bibitem[{{Garbow} {et~al.}(1980){Garbow}, {Hillstrom}, \&
  {Mor\'e}}]{Garbow+Hillstrom+More_1980}
{Garbow}, B.~S., {Hillstrom}, K.~E., \& {Mor\'e}, J.~J. 1980, {minpack project,
  Argonne National Laboratory}, \url{http://www.netlib.org/minpack}

\bibitem[{{Guillet} {et~al.}(2018){Guillet}, {Fanciullo}, {Verstraete},
  {Boulanger}, {Jones}, {Miville-Desch{\^e}nes}, {Ysard}, {Levrier}, \&
  {Alves}}]{Guillet+Fanciullo+Verstraete+etal_2018}
{Guillet}, V., {Fanciullo}, L., {Verstraete}, L., {et~al.} 2018, \aap, 610, A16

\bibitem[{{Hensley} \& {Draine}(2020{\natexlab{a}})}]{Hensley+Draine_2020c}
{Hensley}, B.~S., \& {Draine}, B.~T. 2020{\natexlab{a}}, ``Unified Model of the
  Emission, Extinction, and Polarization by Dust in the Diffuse ISM" (in
  preparation)

\bibitem[{{Hensley} \& {Draine}(2020{\natexlab{b}})}]{Hensley+Draine_2020a}
---. 2020{\natexlab{b}}, \apj, 895, 38

\bibitem[{{Hensley} \& {Draine}(2020{\natexlab{c}})}]{Hensley+Draine_2020b}
---. 2020{\natexlab{c}}, ArXiv:2009.00018

\bibitem[{{Indebetouw} {et~al.}(2005){Indebetouw}, {Mathis}, {Babler}, {Meade},
  {Watson}, {Whitney}, {Wolff}, {Wolfire}, {Cohen}, {Bania}, {Benjamin},
  {Clemens}, {Dickey}, {Jackson}, {Kobulnicky}, {Marston}, {Mercer},
  {Stauffer}, {Stolovy}, \& {Churchwell}}]{Indebetouw+Mathis+Babler+etal_2005}
{Indebetouw}, R., {Mathis}, J.~S., {Babler}, B.~L., {et~al.} 2005, \apj, 619,
  931

\bibitem[{{Jenkins}(2009)}]{Jenkins_2009}
{Jenkins}, E.~B. 2009, \apj, 700, 1299

\bibitem[{{Jones} {et~al.}(2013){Jones}, {Fanciullo}, {K{\"o}hler},
  {Verstraete}, {Guillet}, {Bocchio}, \&
  {Ysard}}]{Jones+Fanciullo+Kohler+etal_2013}
{Jones}, A.~P., {Fanciullo}, L., {K{\"o}hler}, M., {et~al.} 2013, \aap, 558,
  A62

\bibitem[{{Kim} \& {Martin}(1995)}]{Kim+Martin_1995b}
{Kim}, S.-H., \& {Martin}, P.~G. 1995, \apj, 444, 293

\bibitem[{{K{\"o}hler} {et~al.}(2015){K{\"o}hler}, {Ysard}, \&
  {Jones}}]{Kohler+Ysard+Jones_2015}
{K{\"o}hler}, M., {Ysard}, N., \& {Jones}, A.~P. 2015, \aap, 579, A15

\bibitem[{{Lee} \& {Draine}(1985)}]{Lee+Draine_1985}
{Lee}, H.~M., \& {Draine}, B.~T. 1985, \apj, 290, 211

\bibitem[{{Lenz} {et~al.}(2017){Lenz}, {Hensley}, \&
  {Dor{\'e}}}]{Lenz+Hensley+Dore_2017}
{Lenz}, D., {Hensley}, B.~S., \& {Dor{\'e}}, O. 2017, \apj, 846, 38

\bibitem[{{Li} \& {Draine}(2001)}]{Li+Draine_2001b}
{Li}, A., \& {Draine}, B.~T. 2001, \apj, 554, 778

\bibitem[{{Li} \& {Greenberg}(2002)}]{Li+Greenberg_2002}
{Li}, A., \& {Greenberg}, J.~M. 2002, \apj, 577, 789

\bibitem[{{Li} {et~al.}(2014){Li}, {Liang}, \& {Li}}]{Li+Liang+Li_2014}
{Li}, Q., {Liang}, S.~L., \& {Li}, A. 2014, \mnras, 440, L56

\bibitem[{{Mathis} {et~al.}(1983){Mathis}, {Mezger}, \&
  {Panagia}}]{Mathis+Mezger+Panagia_1983}
{Mathis}, J.~S., {Mezger}, P.~G., \& {Panagia}, N. 1983, \aap, 128, 212

\bibitem[{{Mathis} {et~al.}(1977){Mathis}, {Rumpl}, \&
  {Nordsieck}}]{Mathis+Rumpl+Nordsieck_1977}
{Mathis}, J.~S., {Rumpl}, W., \& {Nordsieck}, K.~H. 1977, \apj, 217, 425

\bibitem[{{Mennella} {et~al.}(1999){Mennella}, {Brucato}, {Colangeli}, \&
  {Palumbo}}]{Mennella+Brucato+Colangeli+Palumbo_1999}
{Mennella}, V., {Brucato}, J.~R., {Colangeli}, L., \& {Palumbo}, P. 1999,
  \apjl, 524, L71

\bibitem[{{Mennella} {et~al.}(2002){Mennella}, {Brucato}, {Colangeli}, \&
  {Palumbo}}]{Mennella+Brucato+Colangeli+Palumbo_2002}
---. 2002, \apj, 569, 531

\bibitem[{{Min} {et~al.}(2003){Min}, {Hovenier}, \& {de
  Koter}}]{Min+Hovenier+deKoter_2003}
{Min}, M., {Hovenier}, J.~W., \& {de Koter}, A. 2003, \aap, 404, 35

\bibitem[{{Min} {et~al.}(2008){Min}, {Hovenier}, {Waters}, \& {de
  Koter}}]{Min+Hovenier+Waters+deKoter_2008}
{Min}, M., {Hovenier}, J.~W., {Waters}, L.~B.~F.~M., \& {de Koter}, A. 2008,
  \aap, 489, 135

\bibitem[{{Min} {et~al.}(2007){Min}, {Waters}, {de Koter}, {Hovenier},
  {Keller}, \& {Markwick-Kemper}}]{Min+Waters+deKoter+etal_2007}
{Min}, M., {Waters}, L.~B.~F.~M., {de Koter}, A., {et~al.} 2007, \aap, 462, 667

\bibitem[{{Nguyen} {et~al.}(2018){Nguyen}, {Dawson}, {Miville-Desch{\^e}nes},
  {Tang}, {Li}, {Heiles}, {Murray}, {Stanimirovi{\'c}}, {Gibson},
  {McClure-Griffiths}, {Troland}, {Bronfman}, \&
  {Finger}}]{Nguyen+Dawson+Miville-Deschenes+etal_2018}
{Nguyen}, H., {Dawson}, J.~R., {Miville-Desch{\^e}nes}, M.~A., {et~al.} 2018,
  \apj, 862, 49

\bibitem[{{Nittler} \& {Ciesla}(2016)}]{Nittler+Ciesla_2016}
{Nittler}, L.~R., \& {Ciesla}, F. 2016, \araa, 54, 53

\bibitem[{{Ossenkopf} {et~al.}(1992){Ossenkopf}, {Henning}, \&
  {Mathis}}]{Ossenkopf+Henning+Mathis_1992}
{Ossenkopf}, V., {Henning}, T., \& {Mathis}, J.~S. 1992, \aap, 261, 567

\bibitem[{{Packham} {et~al.}(2005){Packham}, {Hough}, \&
  {Telesco}}]{Packham+Hough+Telesco_2005}
{Packham}, C., {Hough}, J.~H., \& {Telesco}, C.~M. 2005, in Astronomical
  Society of the Pacific Conference Series, Vol. 343, Astronomical Polarimetry:
  Current Status and Future Directions, ed. A.~{Adamson}, C.~{Aspin},
  C.~{Davis}, \& T.~{Fujiyoshi}, 38

\bibitem[{{Packham} {et~al.}(2008){Packham}, {Escuti}, {Boreman}, {Quijano},
  {Ginn}, {Franklin}, {Axon}, {Hough}, {Jones}, {Roche}, {Tamura}, {Telesco},
  {Levenson}, {Rodgers}, \& {McGuire}}]{Packham+Escuti+Boreman+etal_2008}
{Packham}, C., {Escuti}, M., {Boreman}, G., {et~al.} 2008, in \procspie, Vol.
  7014, Ground-based and Airborne Instrumentation for Astronomy II, 70142H

\bibitem[{{Planck Collaboration} {et~al.}(2015{\natexlab{a}}){Planck
  Collaboration}, {Ade}, {Aghanim}, {Alina}, {Alves}, {Armitage-Caplan},
  {Arnaud}, {Arzoumanian}, {Ashdown}, {Atrio-Barandela}, \&
  et~al.}]{Planck_int_results_xix_2015}
{Planck Collaboration}, {Ade}, P.~A.~R., {Aghanim}, N., {et~al.}
  2015{\natexlab{a}}, \aap, 576, A104

\bibitem[{{Planck Collaboration} {et~al.}(2015{\natexlab{b}}){Planck
  Collaboration}, {Ade}, {Alves}, {Aniano}, {Armitage-Caplan}, {Arnaud},
  {Atrio-Barandela}, {Aumont}, {Baccigalupi}, {Banday}, {Barreiro}, {Battaner},
  {Benabed}, {Benoit-L{\'e}vy}, {Bernard}, {Bersanelli}, {Bielewicz}, {Bock},
  {Bond}, {Borrill}, {Bouchet}, {Boulanger}, {Burigana}, {Cardoso}, {Catalano},
  {Chamballu}, {Chiang}, {Colombo}, {Combet}, {Couchot}, {Coulais}, {Crill},
  {Curto}, {Cuttaia}, {Danese}, {Davies}, {Davis}, {de Bernardis}, {de Zotti},
  {Delabrouille}, {D{\'e}sert}, {Dickinson}, {Diego}, {Donzelli}, {Dor{\'e}},
  {Douspis}, {Dunkley}, {Dupac}, {En{\ss}lin}, {Eriksen}, {Falgarone},
  {Finelli}, {Forni}, {Frailis}, {Fraisse}, {Franceschi}, {Galeotta}, {Ganga},
  {Ghosh}, {Giard}, {Gonz{\'a}lez-Nuevo}, {G{\'o}rski}, {Gregorio}, {Gruppuso},
  {Guillet}, {Hansen}, {Harrison}, {Helou}, {Hern{\'a}ndez-Monteagudo},
  {Hildebrandt}, {Hivon}, {Hobson}, {Holmes}, {Hornstrup}, {Jaffe}, {Jaffe},
  {Jones}, {Keih{\"a}nen}, {Keskitalo}, {Kisner}, {Kneissl}, {Knoche}, {Kunz},
  {Kurki-Suonio}, {Lagache}, {Lamarre}, {Lasenby}, {Lawrence}, {Leahy},
  {Leonardi}, {Levrier}, {Liguori}, {Lilje}, {Linden-V{\o}rnle},
  {L{\'o}pez-Caniego}, {Lubin}, {Mac{\'{\i}}as-P{\'e}rez}, {Maffei},
  {Magalh{\~a}es}, {Maino}, {Mandolesi}, {Maris}, {Marshall}, {Martin},
  {Mart{\'{\i}}nez-Gonz{\'a}lez}, {Masi}, {Matarrese}, {Mazzotta},
  {Melchiorri}, {Mendes}, {Mennella}, {Migliaccio}, {Miville-Desch{\^e}nes},
  {Moneti}, {Montier}, {Morgante}, {Mortlock}, {Munshi}, {Murphy}, {Naselsky},
  {Nati}, {Natoli}, {Netterfield}, {Noviello}, {Novikov}, {Novikov},
  {Oppermann}, {Oxborrow}, {Pagano}, {Pajot}, {Paoletti}, {Pasian},
  {Perdereau}, {Perotto}, {Perrotta}, {Piacentini}, {Pietrobon},
  {Plaszczynski}, {Pointecouteau}, {Polenta}, {Popa}, {Pratt}, {Rachen},
  {Reach}, {Reinecke}, {Remazeilles}, {Renault}, {Ricciardi}, {Riller},
  {Ristorcelli}, {Rocha}, {Rosset}, {Roudier}, {Rubi{\~n}o-Mart{\'{\i}}n},
  {Rusholme}, {Salerno}, {Sandri}, {Savini}, {Scott}, {Spencer}, {Stolyarov},
  {Stompor}, {Sudiwala}, {Sutton}, {Suur-Uski}, {Sygnet}, {Tauber}, {Terenzi},
  {Toffolatti}, {Tomasi}, {Tristram}, {Tucci}, {Valenziano}, {Valiviita}, {Van
  Tent}, {Vielva}, {Villa}, {Wandelt}, {Zacchei}, \&
  {Zonca}}]{Planck_int_results_xxii_2015}
{Planck Collaboration}, {Ade}, P.~A.~R., {Alves}, M.~I.~R., {et~al.}
  2015{\natexlab{b}}, \aap, 576, A107

\bibitem[{{Planck Collaboration} {et~al.}(2016){Planck Collaboration}, {Ade},
  {Aghanim}, {Alves}, {Aniano}, {Arnaud}, {Ashdown}, {Aumont}, {Baccigalupi},
  {Banday}, {Barreiro}, {Bartolo}, {Battaner}, {Benabed}, {Benoit-Levy},
  {Bernard}, {Bersanelli}, {Bielewicz}, {Bonaldi}, {Bonavera}, {Bond},
  {Borrill}, {Bouchet}, {Boulanger}, {Burigana}, {Butler}, {Calabrese},
  {Cardoso}, {Catalano}, {Chamballu}, {Chiang}, {Christensen}, {Clements},
  {Colombi}, {Colombo}, {Couchot}, {Crill}, {Curto}, {Cuttaia}, {Danese},
  {Davies}, {Davis}, {de Bernardis}, {de Rosa}, {de Zotti}, {Delabrouille},
  {Dickinson}, {Diego}, {Dole}, {Donzelli}, {Dore}, {Douspis}, {Draine},
  {Ducout}, {Dupac}, {Efstathiou}, {Elsner}, {Ensslin}, {Eriksen}, {Falgarone},
  {Finelli}, {Forni}, {Frailis}, {Fraisse}, {Franceschi}, {Frejsel},
  {Galeotta}, {Galli}, {Ganga}, {Ghosh}, {Giard}, {Gjerlow}, {Gonzalez-Nuevo},
  {Gorski}, {Gregorio}, {Gruppuso}, {Guillet}, {Hansen}, {Hanson}, {Harrison},
  {Henrot-Versille}, {Hernandez-Monteagudo}, {Herranz}, {Hildebrandt}, {Hivon},
  {Holmes}, {Hovest}, {Huffenberger}, {Hurier}, {Jaffe}, {Jaffe}, {Jones},
  {Keihanen}, {Keskitalo}, {Kisner}, {Kneissl}, {Knoche}, {Kunz},
  {Kurki-Suonio}, {Lagache}, {Lamarre}, {Lasenby}, {Lattanzi}, {Lawrence},
  {Leonardi}, {Levrier}, {Liguori}, {Lilje}, {Linden-Vornle}, {Lopez-Caniego},
  {Lubin}, {Macias-Perez}, {Maffei}, {Maino}, {Mandolesi}, {Maris}, {Marshall},
  {Martin}, {Martinez-Gonzalez}, {Masi}, {Matarrese}, {Mazzotta}, {Melchiorri},
  {Mendes}, {Mennella}, {Migliaccio}, {Miville-Deschenes}, {Moneti}, {Montier},
  {Morgante}, {Mortlock}, {Munshi}, {Murphy}, {Naselsky}, {Natoli},
  {Norgaard-Nielsen}, {Novikov}, {Novikov}, {Oxborrow}, {Pagano}, {Pajot},
  {Paladini}, {Paoletti}, {Pasian}, {Perdereau}, {Perotto}, {Perrotta},
  {Pettorino}, {Piacentini}, {Piat}, {Plaszczynski}, {Pointecouteau},
  {Polenta}, {Ponthieu}, {Popa}, {Pratt}, {Prunet}, {Puget}, {Rachen}, {Reach},
  {Rebolo}, {Reinecke}, {Remazeilles}, {Renault}, {Ristorcelli}, {Rocha},
  {Roudier}, {Rubio-Martin}, {Rusholme}, {Sandri}, {Santos}, {Scott},
  {Spencer}, {Stolyarov}, {Sudiwala}, {Sunyaev}, {Sutton}, {Suur-Uski},
  {Sygnet}, {Tauber}, {Terenzi}, {Toffolatti}, {Tomasi}, {Tristram}, {Tucci},
  {Umana}, {Valenziano}, {Valiviita}, {Van Tent}, {Vielva}, {Villa}, {Wade},
  {Wandelt}, {Wehus}, {Ysard}, {Yvon}, {Zacchei}, \&
  {Zonca}}]{Planck_DL07_2016}
{Planck Collaboration}, {Ade}, P.~A.~R., {Aghanim}, N., {et~al.} 2016, \aap,
  586, A132

\bibitem[{{Planck Collaboration} {et~al.}(2018){Planck Collaboration},
  {Aghanim}, {Akrami}, {Alves}, {Ashdown}, {Aumont}, {Baccigalupi},
  {Ballardini}, {Banday}, {Barreiro}, {Bartolo}, {Basak}, {Benabed}, {Bernard},
  {Bersanelli}, {Bielewicz}, {Bock}, {Bond}, {Borrill}, {Bouchet}, {Boulanger},
  {Bracco}, {Bucher}, {Burigana}, {Calabrese}, {Cardoso}, {Carron}, {Chary},
  {Chiang}, {Colombo}, {Combet}, {Crill}, {Cuttaia}, {de Bernardis}, {de
  Zotti}, {Delabrouille}, {Delouis}, {Di Valentino}, {Dickinson}, {Diego},
  {Dor{\'e}}, {Douspis}, {Ducout}, {Dupac}, {Efstathiou}, {Elsner},
  {En{\ss}lin}, {Eriksen}, {Fantaye}, {Fernandez-Cobos}, {Ferri{\`e}re},
  {Forastieri}, {Frailis}, {Fraisse}, {Franceschi}, {Frolov}, {Galeotta},
  {Galli}, {Ganga}, {G{\'e}nova-Santos}, {Gerbino}, {Ghosh},
  {Gonz{\'a}lez-Nuevo}, {G{\'o}rski}, {Gratton}, {Green}, {Gruppuso},
  {Gudmundsson}, {Guillet}, {Handley}, {Hansen}, {Helou}, {Herranz}, {Hivon},
  {Huang}, {Jaffe}, {Jones}, {Keih{\"a}nen}, {Keskitalo}, {Kiiveri}, {Kim},
  {Krachmalnicoff}, {Kunz}, {Kurki-Suonio}, {Lagache}, {Lamarre}, {Lasenby},
  {Lattanzi}, {Lawrence}, {Le Jeune}, {Levrier}, {Liguori}, {Lilje},
  {Lindholm}, {L{\'o}pez-Caniego}, {Lubin}, {Ma}, {Mac{\'{\i}}as-P{\'e}rez},
  {Maggio}, {Maino}, {Mandolesi}, {Mangilli}, {Marcos-Caballero}, {Maris},
  {Martin}, {Mart{\'{\i}}nez-Gonz{\'a}lez}, {Matarrese}, {Mauri}, {McEwen},
  {Melchiorri}, {Mennella}, {Migliaccio}, {Miville-Desch{\^e}nes}, {Molinari},
  {Moneti}, {Montier}, {Morgante}, {Moss}, {Natoli}, {Pagano}, {Paoletti},
  {Patanchon}, {Perrotta}, {Pettorino}, {Piacentini}, {Polastri}, {Polenta},
  {Puget}, {Rachen}, {Reinecke}, {Remazeilles}, {Renzi}, {Ristorcelli},
  {Rocha}, {Rosset}, {Roudier}, {Rubi{\~n}o-Mart{\'{\i}}n}, {Ruiz-Granados},
  {Salvati}, {Sandri}, {Savelainen}, {Scott}, {Sirignano}, {Sunyaev},
  {Suur-Uski}, {Tauber}, {Tavagnacco}, {Tenti}, {Toffolatti}, {Tomasi},
  {Trombetti}, {Valiviita}, {Van Tent}, {Vielva}, {Villa}, {Vittorio},
  {Wandelt}, {Wehus}, {Zacchei}, \& {Zonca}}]{Planck_2018_XII}
{Planck Collaboration}, {Aghanim}, N., {Akrami}, Y., {et~al.} 2018,
  ArXiv:1807.06212

\bibitem[{{Poteet} {et~al.}(2015){Poteet}, {Whittet}, \&
  {Draine}}]{Poteet+Whittet+Draine_2015}
{Poteet}, C.~A., {Whittet}, D.~C.~B., \& {Draine}, B.~T. 2015, \apj, 801, 110

\bibitem[{{Purcell}(1979)}]{Purcell_1979}
{Purcell}, E.~M. 1979, \apj, 231, 404

\bibitem[{{Rouleau} \& {Martin}(1991)}]{Rouleau+Martin_1991}
{Rouleau}, F., \& {Martin}, P.~G. 1991, \apj, 377, 526

\bibitem[{{Schlafly} {et~al.}(2016){Schlafly}, {Meisner}, {Stutz},
  {Kainulainen}, {Peek}, {Tchernyshyov}, {Rix}, {Finkbeiner}, {Covey}, {Green},
  {Bell}, {Burgett}, {Chambers}, {Draper}, {Flewelling}, {Hodapp}, {Kaiser},
  {Magnier}, {Martin}, {Metcalfe}, {Wainscoat}, \&
  {Waters}}]{Schlafly+Meisner+Stutz+etal_2016}
{Schlafly}, E.~F., {Meisner}, A.~M., {Stutz}, A.~M., {et~al.} 2016, \apj, 821,
  78

\bibitem[{{Schutte} {et~al.}(1998){Schutte}, {van der Hucht}, {Whittet},
  {Boogert}, {Tielens}, {Morris}, {Greenberg}, {Williams}, {van Dishoeck},
  {Chiar}, \& {de Graauw}}]{Schutte+vanderHucht+Whittet+etal_1998}
{Schutte}, W.~A., {van der Hucht}, K.~A., {Whittet}, D.~C.~B., {et~al.} 1998,
  \aap, 337, 261

\bibitem[{{Shariff} {et~al.}(2019){Shariff}, {Ade}, {Angil{\`e}}, {Ashton},
  {Benton}, {Devlin}, {Dober}, {Fissel}, {Fukui}, {Galitzki}, {Gandilo},
  {Klein}, {Korotkov}, {Li}, {Martin}, {Matthews}, {Moncelsi}, {Nakamura},
  {Netterfield}, {Novak}, {Pascale}, {Poidevin}, {Santos}, {Savini}, {Scott},
  {Diego Soler}, {Thomas}, {Tucker}, {Tucker}, \&
  {Ward-Thompson}}]{Shariff+Ade+Angile+etal_2019}
{Shariff}, J.~A., {Ade}, P.~A.~R., {Angil{\`e}}, F.~E., {et~al.} 2019, \apj,
  872, 197

\bibitem[{{Siebenmorgen} {et~al.}(2014){Siebenmorgen}, {Voshchinnikov}, \&
  {Bagnulo}}]{Siebenmorgen+Voshchinnikov+Bagnulo_2014}
{Siebenmorgen}, R., {Voshchinnikov}, N.~V., \& {Bagnulo}, S. 2014, \aap, 561,
  A82

\bibitem[{{Siebenmorgen} {et~al.}(2017){Siebenmorgen}, {Voshchinnikov},
  {Bagnulo}, \& {Cox}}]{Siebenmorgen+Voshchinnikov+Bagnulo+Cox_2017}
{Siebenmorgen}, R., {Voshchinnikov}, N.~V., {Bagnulo}, S., \& {Cox}, N.~L.
  2017, \planss, 149, 64

\bibitem[{{Smith} {et~al.}(2000){Smith}, {Wright}, {Aitken}, {Roche}, \&
  {Hough}}]{Smith+Wright+Aitken+etal_2000}
{Smith}, C.~H., {Wright}, C.~M., {Aitken}, D.~K., {Roche}, P.~F., \& {Hough},
  J.~H. 2000, \mnras, 312, 327

\bibitem[{{Steinacker} {et~al.}(2015){Steinacker}, {Andersen}, {Thi},
  {Paladini}, {Juvela}, {Bacmann}, {Pelkonen}, {Pagani}, {Lef{\`e}vre},
  {Henning}, \& {Noriega-Crespo}}]{Steinacker+Andersen+Thi+etal_2015}
{Steinacker}, J., {Andersen}, M., {Thi}, W.-F., {et~al.} 2015, \aap, 582, A70

\bibitem[{{Verner} {et~al.}(1996){Verner}, {Ferland}, {Korista}, \&
  {Yakovlev}}]{Verner+Ferland+Korista+Yakovlev_1996}
{Verner}, D.~A., {Ferland}, G.~J., {Korista}, K.~T., \& {Yakovlev}, D.~G. 1996,
  \apj, 465, 487

\bibitem[{{Verner} \& {Yakovlev}(1995)}]{Verner+Yakovlev_1995}
{Verner}, D.~A., \& {Yakovlev}, D.~G. 1995, \aaps, 109, 125

\bibitem[{{Weingartner} \& {Draine}(2001)}]{Weingartner+Draine_2001a}
{Weingartner}, J.~C., \& {Draine}, B.~T. 2001, \apj, 548, 296

\bibitem[{{Welty} {et~al.}(1999){Welty}, {Hobbs}, {Lauroesch}, {Morton},
  {Spitzer}, \& {York}}]{Welty+Hobbs+Lauroesch+etal_1999}
{Welty}, D.~E., {Hobbs}, L.~M., {Lauroesch}, J.~T., {et~al.} 1999, \apjs, 124,
  465

\bibitem[{{Whittet}(2010)}]{Whittet_2010}
{Whittet}, D.~C.~B. 2010, \apj, 710, 1009

\bibitem[{{Whittet}(2011)}]{Whittet_2011}
{Whittet}, D.~C.~B. 2011, in \aspcs, Vol. 449, Science from Small to Large
  Telescopes, ed. P.~{Bastien}, N.~{Manset}, D.~P. {Clemens}, \& N.~{St-Louis},
  93

\bibitem[{{Whittet} {et~al.}(2008){Whittet}, {Hough}, {Lazarian}, \&
  {Hoang}}]{Whittet+Hough+Lazarian+Hoang_2008}
{Whittet}, D.~C.~B., {Hough}, J.~H., {Lazarian}, A., \& {Hoang}, T. 2008, \apj,
  674, 304

\bibitem[{{Wright} {et~al.}(2002){Wright}, {Aitken}, {Smith}, {Roche}, \&
  {Laureijs}}]{Wright+Aitken+Smith+etal_2002}
{Wright}, C.~M., {Aitken}, D.~K., {Smith}, C.~H., {Roche}, P.~F., \&
  {Laureijs}, R.~J. 2002, in The Origin of Stars and Planets: The VLT View, ed.
  J.~F. {Alves} \& M.~J. {McCaughrean}, 85

\bibitem[{{Wright} {et~al.}(2016){Wright}, {Duy}, \&
  {Lawson}}]{Wright+Duy+Lawson_2016}
{Wright}, C.~M., {Duy}, T.~D., \& {Lawson}, W. 2016, \mnras, 457, 1593

\bibitem[{{Young} \& {Frederikse}(1973)}]{Young+Frederikse_1973}
{Young}, K.~F., \& {Frederikse}, H.~P.~R. 1973, \jpcrd, 2, 313

\bibitem[{{Zubko} {et~al.}(2004){Zubko}, {Dwek}, \&
  {Arendt}}]{Zubko+Dwek+Arendt_2004}
{Zubko}, V., {Dwek}, E., \& {Arendt}, R.~G. 2004, \apjs, 152, 211

\bibitem[{{Zubko} {et~al.}(1996){Zubko}, {Mennella}, {Colangeli}, \&
  {Bussoletti}}]{Zubko+Mennella+Colangeli+Bussoletti_1996}
{Zubko}, V.~G., {Mennella}, V., {Colangeli}, L., \& {Bussoletti}, E. 1996,
  \mnras, 282, 1321

\end{thebibliography}

\end{document}